\documentclass[aps,prb,twocolumn,floatfix]{revtex4-2}
\usepackage{psfrag,epsfig,amsfonts,amssymb,amsmath,wasysym,bm}
\usepackage{bbold}
\usepackage{color}
\usepackage{hyperref}

\usepackage{xcolor}

\usepackage[inline]{enumitem} 

\renewcommand{\Re}{\operatorname{Re}}

\newcommand{\kB}{k_{\mathrm B}} 

\newcommand{\RR}{{\mathbb R}}
\newcommand{\CC}{{\mathbb C}}
\newcommand{\NN}{{\mathbb N}}
\newcommand{\ZZ}{{\mathbb Z}}

\newcommand{\tr}{\operatorname{tr}}

\newcommand{\ord}{{\cal O}}

\newcommand{\lmat}{\left( \begin{matrix}}	
\newcommand{\rmat}{\end{matrix} \right)}	

\providecommand{\opnorm}[1]{\|#1\|_{\!\!\; op}}

\newcommand{\xx}{b}

\begin{document}

\title{Absence of thermalization after a local quench and strong violation of the eigenstate thermalization hypothesis}

\author{Peter Reimann}
\author{Christian Eidecker-Dunkel}
\affiliation{Faculty of Physics, 
Bielefeld University, 
33615 Bielefeld, Germany}
\date{\today}

\begin{abstract}
Absence of thermalization after a {\em global} quantum quench is a 
well-established numerical observation in 
integrable many-body systems, 
and can be 
empirically related to
a violation of the eigenstate thermalization 
hypothesis (ETH) 
in such models.
Still, in many of those 
examples a weaker version of the
{conventional}
ETH (wETH) has been numerically reported or 
even rigorously proven.
In this paper we show analytically and illustrate numerically 
that the absence of thermalization is already possible after 
a {\em local} quench.
A closely related finding is a 
strong violation of the ETH,
meaning that not even the wETH is fulfilled anymore.
In our analytical explorations we focus on XX-spin-chain models
with open boundary conditions, where the local quench is generated
by initiating the system in thermal equilibrium and then suddenly
switching
on
(or slightly changing) a single-spin impurity either at the
end or in the center of the chain.
Numerically we observe qualitatively similar phenomena 
also for more general XXZ-models in the case of an
end-impurity, but not in the case of a central impurity.
\end{abstract}

\maketitle

\section{Introduction}
\label{s1}

The problem of thermalization is a long-standing and still not 
satisfactorily understood issue in statistical physics
\cite{gog16,mor18,tas24}.
Essentially, it amounts to the question of 
why 
many-body systems can usually be described so successfully 
by one of the standard canonical ensembles 
at sufficiently late times, i.e., after the initial relaxation 
processes have died out.

Considerable new insights have been recently gained 
by means of the eigenstate thermalization hypothesis (ETH)
\cite{deu91,sre94,rig08}, 
postulating that already the system's energy 
eigenstates reproduce very accurately its thermal 
equilibrium properties, 
at least for generic, so-called nonintegrable models
\cite{dal16}.
On the other hand, 
integrable models are generally
expected (or defined) to exhibit an extensive number of 
conserved quantities in the form of suitable sums of local 
operators \cite{ess16}.
As a consequence, they are furthermore expected to 
violate the ETH and hence {\em not} to thermalize at 
least after a {\em global} quantum quench 
\cite{gog16,mor18,dal16,bar70}, 
essentially meaning that
the initial condition exhibits some non-thermal features  
throughout the entire system.
While there is ample numerical evidence in support of 
{this very appealing general picture},
a more rigorous rationale has been provided until 
now at most in some special cases.

A weaker
{version}
of the 
{conventional} 
ETH (wETH) proposes that the
system's thermal equilibrium properties are well
captured by the vast majority of its energy eigenstates,
but still admits some 
{sufficiently small fraction of} 
exceptions \cite{weth}.
{In contrast to the 
conventional
(or ``strong'') ETH,
the wETH can even be rigorously proven for a large class of 
Hamiltonians,} the main prerequisites being that
the system is translationally invariant and exhibits
only 
short-range interactions,
while 
the issue of integrability
does not seem to play {any} role.
Furthermore, such systems have been analytically shown to exhibit
$\mbox{(re)}$thermalization after a {\em local} quantum quench
{for} 
instance in Refs.~\cite{far17,dab22,rei25},
and it has been argued that
such a behavior should be expected to 
occur very generally \cite{bre20,pc}.

The main message of our present paper is 
that things are actually more subtle.
Namely, we will provide 
analytical examples 
exhibiting 
the
absence of thermalization after a local quench,
accompanied by a violation 
even of the wETH, which we dub
a
{\em strong violation of the ETH.}
Note that 
this does not contradict the above mentioned
previous findings since our present examples 
are {\em not} translationally invariant.
It also does not contradict the numerical
observation of thermalization after 
a local quench for instance in 
Refs.~\cite{san04,tor14,bre20,pan20,san20,bar09}.
The reason is that in those latter examples the 
local quench breaks the integrability of the system, 
but not in our present examples.

Specifically, we will analytically explore  
the XX-spin-chain model with open boundary conditions 
after switching-on (or slightly changing)
an impurity either at the chain-end or in the middle of the chain.
Furthermore, we will numerically demonstrate that also the
more general XXZ model behaves qualitatively similarly
for impurities at the chain-end, but very differently
for central impurities.

The rest of the paper is organized as follow.
In Sec.~\ref{s2} we provide the details of the 
considered spin-chain models and 
local quenches, the notions of (non)thermalization and 
strong violation of the ETH,
and how they can be related
to temporal correlations at thermal equilibrium.
In Secs.~\ref{s3} and \ref{s4} we analytically explore 
and numerically illustrate
for the XX model with an end-impurity those 
correlations and their implications regarding
the issues of
(non-)thermalization and 
strong violation of the ETH.
In Sec.~\ref{s5} the same properties for the
XXZ model are recovered by numerical means.
In Sec.~\ref{s6} we provide similar analytical
and numerical explorations, but now for the case of 
a central- instead of an end-impurity, and resulting
in a very different behavior of the XX 
and the XXZ models with respect to 
(non)thermalization 
and the ETH.

\section{General Framework}
\label{s2}

\subsection{Setup}
\label{s21}

Our starting point is the spin-1/2 XXZ-chain Hamiltonian
with open boundary conditions 
\begin{eqnarray}
H_{\rm XXZ} =  
\sum_{l=1}^{L-1}  
J_\perp (s^x_{l+1}s^x_l + s^y_{l+1}s^y_l) + J_z s^z_{l+1}s^z_l   
\ ,
\label{1}
\end{eqnarray}
where the $s^{a}_l$
with $a\in\{x,y,z\}$  
are spin-1/2 operators
at the chain sites $l\in\{1,...,L\}$ and the 
coupling parameters $J_\perp$ and $J_z$
quantify the nearest neighbor interactions
\cite{foot1}.

Upon adding a single magnetic impurity at an arbitrary but 
fixed chain site $\nu\in\{1,...,L\}$ we obtain the 
modified Hamiltonian
\begin{eqnarray}
H_0 = H_{\rm XXZ} + \gamma s_\nu^z
\ ,
\label{2}
\end{eqnarray}
where $\gamma$ may be viewed as a localized transverse field.
In the first part of the paper (Secs.~\ref{s3} and \ref{s4}) we will focus on the
special case 
$J_\perp =1$ and $J_z=0$ (XX model). 
In the second part of the paper and in the rest of the present section 
we will deal with the more general XXZ model.

The initial state of the system is assumed to be
given by a canonical ensemble (Gibbs state)
associated with $H_0$ from Eq.~(\ref{2}),
\begin{eqnarray}
\rho(0) & := & e^{-\beta H_0}/\tr\{e^{-\beta H_0}\}
\, ,
\label{3}
\\
\beta & := & 1/\kB T
\ ,
\label{4}
\end{eqnarray}
where $T$ is the temperature and $\kB$ Boltzmann's constant.

The temporal evolution of this initial state is governed 
by a perturbed Hamiltonian of the form
\begin{eqnarray}
H & = &  H_0 + g V
\ ,
\label{5}
\end{eqnarray}
with a local perturbation operator $V$ and
a coupling parameter $g$.
Working in units with $\hbar=1$,
the system's state at time $t>0$ 
can thus be written as
\begin{eqnarray}
\rho(t) = e^{-iHt}\rho(0)\,e^{iHt}
\ ,
\label{6}
\end{eqnarray}
and the concomitant expectation value of
an observable $A$ (Hermitian operator) as
\begin{eqnarray}
\langle A\rangle_t:=\tr\{\rho(t) A\}
\ .
\label{7}
\end{eqnarray}

For generic perturbations in Eq.~(\ref{5}), the  Hamiltonian $H$,
which governs the dynamics, is different from $H_0$
and
the Gibbs state (\ref{3}) amounts to a 
nonequilibrium initial condition.
Put differently, the system is in thermal equilibrium with respect 
to the Hamiltonian $H_0$ for all times 
$t<0$, experiences an instantaneous parameter change 
at $t=0$ (quantum quench) and is governed by $H$
for $t>0$.

Throughout our present work, we will focus on 
perturbations which are identical to the
impurity in Eq.~(\ref{2}), i.e.,
\begin{eqnarray}
V=s_\nu^z
\ .
\label{8}
\end{eqnarray}
Of particular interest are cases with $\gamma=0$ in Eq.~(\ref{2}), 
meaning that an impurity is suddenly ``switched-on'' at $t=0$.
More generally, i.e. for $\gamma\not=0$, 
the impurity is already present at times $t<0$, and suddenly changes 
its strength at $t=0$.
We also remark that generalizations to different 
spins in Eqs.~(\ref{2}) and (\ref{8}), say $s_\mu^z$ instead of $s_\nu^z$ in Eq.~(\ref{8}),
would in principle be straightforward, but technically much more involved,
and moreover seem to us of relatively minor physical interest.

\subsection{Thermalization and sufficient conditions for nonthermalization}
\label{s22}

At the focus of our present work is the issue of (non)thermalization,
i.e., the question of whether or not
the expectation values in Eq.~(\ref{7}) 
approach for sufficiently large chain-lengths $L$ and after sufficiently 
long times $t$ their thermal equilibrium value 
\begin{eqnarray}
\langle A\rangle_{\!\rm th} :=\tr\{\rho_{can} A\}
\ ,
\label{9}
\end{eqnarray}
where $\rho_{can}$ is the canonical ensemble 
\begin{eqnarray}
\rho_{can} & := & e^{-\beta H}/\tr\{e^{-\beta H} \}
\label{10}
\end{eqnarray}
pertaining to the postquench Hamiltonian $H$,
{see also Eqs.~(\ref{5}) and (\ref{6})}.

The following four remarks will be of considerable
importance later on.
(i)
Note that we tacitly have chosen the same $\beta$ in Eqs.~(\ref{3}) and (\ref{10}).
The intuitive justification is that a local perturbation
[see Eqs.~(\ref{5}) and (\ref{8})] should not notably change 
the temperature for sufficiently large systems.
A more detailed {and rigorous} 
justification is provided in Appendix \ref{appA}.
(ii) 
Since our models amount to isolated systems,
it might appear more natural to work with a 
microcanonical instead of a canonical 
ensemble in Eq.~(\ref{9}).
We tacitly take for granted that both ensembles would
result in (practically) identical thermal expectation 
values at least for those Hamiltonians and observables 
we will consider in the following.
A more rigorous justification of this equivalence 
of ensembles is provided for instance in 
Refs.~\cite{tou15,bra15,tas18,kuw20}.
Analogous remarks apply to the initial condition in Eq.~(\ref{3}).
(iii) 
Usually, the expectation values in Eq.~(\ref{7}) are notably 
different from the thermal value in Eq.~(\ref{9}) during 
some initial time-interval (relaxation process), 
but also at certain, arbitrarily 
large times $t$ (quantum recurrences or revivals).
With increasing system size, those exceptional times are 
furthermore expected to become more and more rare.
In other words, a necessary (but possibly not sufficient)
condition for thermalization is that the long-time average
of the expectation values
in Eq.~(\ref{7}) approaches the thermal value in Eq.~(\ref{9}) as the
system size increases.
In the following, we will be mainly interested 
in examples which do {\em not}
exhibit thermalization. 
Accordingly, we will utilize as a
sufficient condition for nonthermalization 
the criterion that the difference between the long-time 
average and the thermal value remains finite
for arbitrarily large system sizes 
(thermodynamic limit).
(iv) 
In order to demonstrate nonthermalization, it is furthermore
sufficient that this criterion is fulfilled
at least for one physically relevant observable 
$A$.
With respect to our specific model from
Eq.~(\ref{5}), we will therefore mainly focus on 
particularly simple observables of the form
\begin{eqnarray}
A=s_\alpha^z
\ ,
\label{11}
\end{eqnarray}
where $\alpha\in\{1,...,L\}$ may be different or equal to
the index $\nu$ of the perturbation $V$ in Eq.~(\ref{8}).

\subsection{The role of the temporal correlations}
\label{s23}

Exploiting Eq.~(\ref{6}) and the cyclic invariance of the trace, 
the expectation values in Eq.~(\ref{7}) can 
be rewritten as
\begin{eqnarray}
\langle A\rangle_t & = & \tr\{\rho(0) A(t)\} 
\, ,
\label{12}
\end{eqnarray}
where
\begin{eqnarray}
A(t)
& := &
e^{iH t} A e^{-iHt}
\label{13}
\end{eqnarray}
is the observable at time $t$ in the
Heisenberg picture.
Similarly, the temporal correlation
(also called, among others, dynamic or 
two-point correlation function) of 
two Hermitian operators 
$V$ and $A$ at thermal equilibrium 
is defined as
\begin{eqnarray}
C_{\! V\!\!A}(t)
& := & 
\langle V\! A (t)\rangle_{\!\rm th}
-
\langle V\rangle_{\!\rm th}
\langle A\rangle_{\!\rm th}
\ .
\label{14}
\end{eqnarray}
While $V$ and $A$ in this definition are still arbitrary, 
{for our present purposes it is sufficient to}
focus on the situation that $V$ 
is the local perturbation from Eq.~(\ref{5})
and $A$ some local observable,
as exemplified by Eqs.~(\ref{8}) and (\ref{11}),
{respectively}.
{The reason is that in this way it becomes possible to}
utilize
the generalization of Onsager's regression hypothesis from Ref.~\cite{rei24},
consisting in the approximation
\cite{fn}
\begin{eqnarray}
 \langle A\rangle_t - \langle A\rangle _{\!\rm th}
 & = &  
g \beta  \, \tilde C_{V\!A}(t)
\ ,
\label{15}
\\
\tilde C_{V\!A}(t)
& := &
\sum_{k=0}^\infty  \frac{(i \beta)^k}{(k+1)!}\,  
\frac{d^k}{dt^k} C_{V\!A}(t)
\ .
\label{16}
\end{eqnarray}
Note that the correlations in Eq.~(\ref{14}) are generally complex-valued, 
and likewise for every single summand 
{on the right-hand side of}
Eq.~(\ref{16}),
while the entire sum is provably real-valued \cite{rei24}.

Roughly speaking,
the relative error of the approximation (\ref{15})
becomes arbitrarily small
for sufficiently small values of $g\beta$.
More precisely speaking, the neglected terms
can be rigorously shown \cite{rei24} to be 
upper bounded (in modulus) by
$g^2\beta^2 \opnorm{V}^2 \opnorm{A} q(\beta)$,
where  $\opnorm{V}$ is the operator norm of $V$ 
(largest eigenvalue in modulus) and thus 
$\opnorm{V}=\opnorm{A}=1/2$ for our standard examples from 
Eqs.~(\ref{8}) and (\ref{11}) (see also footnote \cite{foot1}).
Furthermore, the function
$q(\beta)$ is of order unity at least for small-to-moderate
values of $\beta J_\perp$, $\beta J_z$, and $\beta(g+\gamma)$
in the case of our 
Hamiltonians (\ref{5}) of the specific form (\ref{1}), (\ref{2}), and (\ref{8}).

These general analytical predictions regarding the accuracy of the 
approximation (\ref{15}) have been numerically confirmed by a wide 
variety of examples in Refs.~\cite{rei24,rei25}.
A few more such example will also be presented later.
For the rest, our present mindset and foremost objective will be different: 
Taking the relation (\ref{15}) for granted, we will deduce from it
without any further approximation that certain models of the form
(\ref{1})-(\ref{8}) do {\em not} exhibit thermalization 
after a local (and sufficiently weak) quantum quench.

From Eqs.~(\ref{2}) and (\ref{8}) it follows that the Hamiltonian
$H$ in Eq.~(\ref{5}) does not depend separately on $\gamma$ and $g$, 
but only on the combination
\begin{eqnarray}
p:=2(g+\gamma)
\ ,
\label{17}
\end{eqnarray}
where the extra factor $2$ has been introduced for later convenience.
The same property is inherited via Eq.~(\ref{10}) by the 
thermal equilibrium values in Eq.~(\ref{9}), 
the correlations in Eq.~(\ref{14}), 
and the quantity $\tilde C_{V\!A}(t)$ in Eq.~(\ref{16}).
Moreover, for observables of the specific form (\ref{11})
it has been shown in Ref.~\cite{eid23} 
(see Supplemental Material I.D therein) 
that the thermal equilibrium values in Eq.~(\ref{9})
are odd functions of $p$,
while the correlations in Eq.~(\ref{14}) 
are even functions of $p$.
It follows that also the quantity
$\tilde C_{V\!A}(t)$ in Eq.~(\ref{16}) must be
an even function of $p$, i.e., 
invariant under a sign change of $p$.
For the correlations in Eq.~(\ref{14}) and the quantity
$\tilde C_{V\!A}(t)$ in Eq.~(\ref{16}), exactly the same 
invariance can be shown to apply under a 
sign change of the coupling parameter $J_\perp$
in Eq.~(\ref{1}).
Moreover, the same symmetry properties can be
extended to arbitrary observables of the 
form $A=s_l^a$.
Analogous generalizations are also valid 
for the modulus of the thermal equilibrium 
values in Eq.~(\ref{9}), while the behavior of the 
sign is more complicated \cite{eid23}.

\subsection{Nonthermalization and strong violation of the ETH}
\label{s24}

Let us define the long-time average of an arbitrary function $f(t)$ as 
\begin{eqnarray}
\overline{f(t)}:=\lim_{\tau\to\infty}\frac{1}{\tau}\int_0^\tau dt\, f(t)
\ .
\label{18}
\end{eqnarray}
Under very weak assumptions regarding the long-time behavior of 
$C_{V\!A}(t)$, which we henceforth tacitly take for granted, it readily
follows that the long-time average of 
$d^k C_{V\!A}(t)/d^k t$
is zero for any 
$k\geq 1$.
We thus can conclude from Eq.~(\ref{16}) that
\begin{eqnarray}
\overline{\tilde C_{V\!A}(t)} =  \overline{C_{V\!A}(t)}
\ .
\label{19}
\end{eqnarray}
According to items (iii) and (iv) above Eq.~(\ref{11}), the difference 
on the left-hand side of Eq.~(\ref{15}) plays a key role with respect to 
the question of {(non-)}thermalization:
a sufficient condition for nonthermalization
is a nonvanishing long-time average of this difference 
in the thermodynamic limit,
which in turn can be written by means of
Eqs.~(\ref{15}) and (\ref{19}) as
\begin{eqnarray}
\overline{\langle A\rangle_t} - \langle A\rangle _{\!\rm th}
& = &
g\beta \,\overline{C_{V\!A}(t)}
\ .
\label{20}
\end{eqnarray}
In other words, a non-vanishing value of 
$\overline{C_{V\!A}(t)}$ for at least one local observable $A$
implies nonthermalization
[and given this approximation itself is sufficiently accurate, 
as discussed below Eq.~(\ref{16})].

Denoting by ${\cal E}_n$ and $|n\rangle$ the eigenvalues and eigenvectors 
of the Hamiltonian $H$, 
choosing $V=A$, and exploiting Eqs.~(\ref{9}) and (\ref{14})-(\ref{19}),
one can readily show \cite{uhr14,alh20,eid23} that
\begin{eqnarray}
\langle A \rangle_{\!\rm th} 
& = & \sum_{n=1}^N p_n \, \langle n|A|n\rangle
\ ,
\label{21}
\\
\overline{C_{\!A\!A}(t)}
& = &
\sum_{n=1}^N p_n\,  \big[\langle n|A|n\rangle - \langle A \rangle_{\!\rm th}\big]^2
\ ,
\label{22}
\end{eqnarray}
where 
$p_n:= \langle n | \rho_{can} | n\rangle=e^{-\beta {\cal E}_n}/\sum_{m=1}^N e^{-\beta {\cal E}_n}$ 
is the population of the energy level $|n\rangle$ in the 
canonical ensemble from Eq.~(\ref{11}), and where
-- in case that $H$ exhibits degeneracies 
-- the eigenstates $|n\rangle$ 
must be chosen so that $A$ is diagonal in the 
corresponding eigenspaces of $H$.

Taking for granted the equivalence 
of ensembles [see also items (ii) above Eq.~(\ref{11})], 
a non-negligible value of the sum in Eq.~(\ref{22}) 
is thus readily found \cite{alh20} to be tantamount 
to a violation of the wETH as defined in 
Refs.~\cite{weth}, and thus to a 
strong violation of the ETH, 
see also Secs.~\ref{s1} and \ref{s44}.
According to the discussion below 
Eq.~(\ref{20}),
both properties furthermore
imply the absence of thermalization after a {\em local} quench.
Note that this is similar but different from the well-established
(mainly numerical) finding that a violation of the 
conventional
(strong) ETH implies non-thermalization after a {\em global} 
quench \cite{dal16}.

We close with two noteworthy remarks.
(i)
Secstions.~\ref{s22}-\ref{s24}
and, in particular, the conclusions in the
previous paragraph,
are 
not restricted to the specific XXZ model
from Eqs.~(\ref{1}), (\ref{2}), and (\ref{5}) which we 
consider in the rest of the paper, 
but are valid more generally.
(ii) To prove the occurrence instead of the absence of thermalization
after a local quantum quench, it is {\em not} sufficient to show that
Eq.~(\ref{22}) approaches zero in the thermodynamic limit.
Instead, one has to show that the time-average on the right-hand side
of Eq.~(\ref{20}) approaches zero for {\em all} local observables $A$
[and the given perturbation $V$ under consideration in Eq.~(\ref{5})].
Moreover, it is not sufficient to show that the time-average 
$\overline{\langle A\rangle_t}$ on the
left-hand side of Eq.~(\ref{20}) approaches the thermal value
$ \langle A\rangle _{\!\rm th}$.
Instead, one has to show that the time-dependent deviations
$\langle A\rangle_t - \langle A\rangle _{\!\rm th}$
become negligibly small for all (sufficiently late) times $t$
[apart from negligibly rare exceptions (quantum revivals),
see also item (iii) below Eq.~(\ref{10})].
A more detailed investigation of this issue 
has been carried out in Ref. \cite{rei25}.

\section{The XX Model}
\label{s3}

The 
XX model amounts to a special case 
of Eq.~(\ref{1}) with parameters
\begin{eqnarray}
J_\perp=1\, , \ J_z=0
\ .
\label{23}
\end{eqnarray}
Exploiting Eqs.~(\ref{1}), (\ref{2}), (\ref{8}), and
(\ref{17}),
the Hamiltonian (\ref{5}) thus assumes the form \cite{foot1}
\begin{eqnarray}
H=
\sum_{l=1}^{L-1}  
\left(s^x_{l+1}s^x_l + s^y_{l+1}s^y_l \right)
+
 \frac{p}{2} s_\nu^z
 \ .
\label{24}
\end{eqnarray}

\subsection{Jordan-Wigner transformation}
\label{s31}

It is well-known that 
{the Hamiltonian (\ref{24})}
can be analytically 
diagonalized by mapping it via a 
Jordan-Wigner 
transformation to an equivalent model of free 
(noninteracting, spinless) fermions, and is 
therefore often denoted as a
noninteracting integrable model \cite{spo18}.
For a very incomplete selection of pertinent previous 
works, employing these well-established concepts
to similar models as in our present paper, we refer
to Refs.~\cite{cru81,sto92,sto95,der00,bar70,lie61} 
and further
references therein. 
Accordingly, we confine ourselves to a very brief
summary of the main steps, the full details of which
can be found for instance in Ref.~\cite{rei25}.
In a first step, 
the raising and lowering Pauli matrices $\sigma_l^\pm:=\sigma_l^x \pm i \sigma_l^y$
are employed to define the operators 
\begin{eqnarray}
c_l:=\sigma_l^- \, Z_l
\, ,\ 
Z_l:=\prod_{k=1}^{l-1}(-\sigma_k^z)
\ ,
\label{25}
\end{eqnarray}
and 
their adjoint counterparts
$c_l^\dagger$, which are then shown to fulfill
the anti-commutation relations characteristic of
fermionic annihilation and creation operators.
Moreover, 
Eq.
(\ref{24}) can be rewritten
 \cite{foot1} as
\begin{eqnarray}
H & = & 
\frac{1}{2}\sum_{l=1}^{L-1} 
 (c_{l+1}^\dagger c_{l}+c_l^\dagger c_{l+1})
+ \frac{p}{2} c_\nu^\dagger c_\nu
\label{26}
\end{eqnarray} 
apart from an irrelevant additive constant.
The second step is based on the observation 
that for any given unitary $L\times L$ matrix $U$
with matrix elements $U_{kl}$, the operators 
\begin{eqnarray}
f_k:=\sum_{l=1}^L U_{kl} c_l
\label{27}
\end{eqnarray}
and their adjoint counterparts $f_k^\dagger$
are again fermionic annihilation and creation operators.
Moreover, by suitably choosing the unitary $U$,
the correspondingly transformed Hamiltonian from Eq.~(\ref{26}) 
assumes the form 
\begin{eqnarray}
H=\sum_{k=1}^L E_k f_k^\dagger f_k
\label{28}
\end{eqnarray}
with real valued energies $E_k$
[not to be confused with the energies ${\cal E}_n$ 
above Eq.~(\ref{21})].
As shown in Appendix A of Ref.~\cite{rei25},
for our specific example from Eq.~(\ref{24}) 
the matrix elements $U_{kl}$ 
of this unitary $U$ and the energies $E_k$ 
must solve, for any given $k\in\{1,...,L\}$, the $L$
coupled linear equations 
\begin{eqnarray}
U^\ast_{k l+1} \bar \delta_{lL} +  U^\ast_{k l-1} \bar\delta_{l1} & = & (2 E_k - p\;\!\delta_{l\nu}) \, U^\ast_{kl}
 \ ,
 \label{29}
\end{eqnarray}
where $l$ runs from $1$ to $L$,
$\delta_{jk}$ is the Kronecker delta, and 
$\bar \delta_{jk}:= 1- \delta_{jk}$.

Let us abbreviate by $|0\rangle:=|\!\downarrow \cdots\downarrow\rangle$
the specific $L$-spin product state where every single spin is 
in the ``down'' state 
in the eigenbasis of $s_l^z$, i.e. \cite{foot1},
$\sigma_l^z |0\rangle=-|0\rangle$ 
for any $l\in\{1,...,L\}$.
It follows that $\sigma_l^-|0\rangle=0$,
hence $c_l|0\rangle = 0$, and finally $f_l|0\rangle = 0$
for all $l\in\{1,...,L\}$.
In other words, $| 0 \rangle$ plays the role of the  vacuum state
with respect to the fermionic models in Eqs.~(\ref{26}) and (\ref{28}).

Denoting by 
$\vec \xx :=(\xx_1,...,\xx_L)$ a vector 
with $L$ ``binary'' components $\xx_k\in\{0,1\}$, 
we define 
\begin{eqnarray}
|\vec \xx\rangle := 
( f_1^\dagger)^{\xx_1}
( f_2^\dagger)^{\xx_2}
\cdots 
( f_L^\dagger)^{\xx_L}
\,  |0\rangle
\ .
\label{30}
\end{eqnarray}
One thus can infer from Eq.~(\ref{28}) that
\begin{eqnarray}
H \,|\vec \xx\rangle & = & E(\vec \xx)\, |\vec \xx\rangle
\ ,
\label{31}
\\
E(\vec \xx) & := & \sum_{k=1}^L \xx_k\, E_k
\ .
\label{32}
\end{eqnarray}
In other words, the $2^L$ different vectors $|\vec \xx\rangle$ 
are the eigenstates of $H$ and $E(\vec \xx)$ the 
corresponding eigenvalues.
[Hence, there must exist a one-to-one relation 
to the eigenvectors $|n\rangle$ and eigenvalues 
${\cal E}_n$ introduced above Eq.~(\ref{21}),
apart from some complications which possibly may arise
in the case of degeneracies, see below Eq.~(\ref{22}).]
Finally, in order to evaluate Eq.~(\ref{14}),
the above Jordan-Wigner
and unitary transformations must also be applied to the 
perturbations $V$ and observables $A$ of interest.
For our particular examples from Eqs.~(\ref{8}) and (\ref{11}) one thus 
may exploit Eq.~(\ref{25}) to infer \cite{rei25} the relation
\begin{eqnarray}
s_l^z = c_l^\dagger c_l -1/2
\ .
\label{33}
\end{eqnarray}
Finally, $c_l^\dagger c_l$ 
{on the right-hand side of Eq.~(\ref{33})}
must be expressed
in terms of the operators 
$f_k$ and $f_k^\dagger$ via 
the inversion of Eq.~(\ref{27}), i.e., 
\begin{eqnarray}
c_l=\sum_{k=1}^L U^\ast_{kl}f_k
\ ,
\label{34}
\end{eqnarray}
and likewise for $c^\dagger_l$.

\subsection{Previous results and remaining task}
\label{s32}

By means of this formalism it is possible to
evaluate the temporal correlations in Eq.~(\ref{14}) as well as the 
infinite sum on the right-hand side of Eq.~(\ref{16})
in closed analytical form without any further approximation.
Referring to Ref.~\cite{rei25} for the detailed calculations,
the so-obtained result for Eq.~(\ref{16}) can be written in the form
\begin{eqnarray}
\tilde C_{V\!A}(t)
& = & 
\sum_{j,k=1}^L 
U_{j\nu} U^\ast_{j\alpha} U^\ast_{k\nu} U_{k\alpha}
\, f_{jk} \, e_{jk}(t)
\ ,
\label{35}
\\
f_{jk}
& := & 
\frac{\tanh(\beta 
E_j/2)-\tanh(\beta 
E_k/2)}{2\beta(
E_j-
E_k)}
\ , \ \ \ \ \ \ 
\label{36}
\\
e_{jk}(t)
& := &
e^{-i(
E_{j}-
E_{k})t}
\ .
\label{37}
\end{eqnarray}
It immediately follows that Eq.~(\ref{36}) and thus 
(\ref{35}) are even functions of $\beta$.
By interchanging the summation indices $j$ and $k$ 
one furthermore sees that the right-hand side 
of Eq.~(\ref{35}) is equal to its complex conjugate,
i.e., we recover the general prediction  below Eq.~(\ref{16})
that $\tilde C_{V\!A}(t)$ must be a real-valued 
function of $t$.

The remaining task is the explicit solution of Eq.~(\ref{29}).
To simplify the notation, let us temporarily consider 
$k\in\{1,...,L\}$ as arbitrary but fixed, and employ the abbreviations
\begin{eqnarray}
E & := & E_k
\ ,
\label{38}
\\
a_l & := & \kappa \, 
U^\ast_{kl}
\ .
\label{39}
\end{eqnarray}
Moreover, while the $U_{kl}$ are normalized according to
$\sum_{l}|U_{kl}|^2=1$, we do not require such a 
normalization condition for the $a_l$'s.
This is accounted for by the still arbitrary proportionality
constant $\kappa\in \CC$ in Eq.~(\ref{39}).
Our only requirement is that $\kappa\not=0$, or equivalently, 
that at least one of the $a_l$'s must be non-zero.
Adopting these abbreviations, 
Eq.~(\ref{29}) amounts to 
\begin{eqnarray}
\bar \delta_{Ll} a_{l+1}+  p\;\!\delta_{\nu l} a_l  +\bar\delta_{1l}  a_{l-1}   & = & 2 E a_l 
 \ ,
 \label{40}
\end{eqnarray}
where $l$ runs from $1$ to $L$.

It is instructive to rewrite those $L$ equations in the form 
$K\vec a=E\vec a$, where
$\vec a$ is a $L$-dimensional column vector
with components $a_l$, 
and $K$ is a $L\times L$
matrix with elements 
$K_{lm}:=[\delta_{l+1m}+p\delta_{\nu l}\delta_{lm}+\delta_{l-1m}]/2$.
Observing that $K_{ml}=K_{lm}$ it follows that $K$ is 
self-adjoint,
i.e., we are actually dealing with a 
Hermitian
matrix  eigenvalue problem.
Moreover, 
we can conclude 
from Eqs.~(\ref{38}) and (\ref{39}) that the eigenvector 
$\vec a$ with eigenvalue $E=E_k$ can be identified 
(up to normalization) with the $k$-th column vector 
of the unitary matrix $U^\dagger$.

Altogether, we are thus left with the task to determine the
$L$ 
solutions of the eigenvalue problem (\ref{40}).
The energies $E_k$ and matrix elements $U_{kl}$ then
readily follow as detailed in and around 
Eqs.~(\ref{38}), (\ref{39}).

Before doing so, some general
properties of those solutions are noteworthy.
To begin with, we note that for any given solution $a_l$ of Eq.~(\ref{40}), 
also the real and imaginary parts of $a_l$ will be solutions.
It follows that, if it is convenient, we always may assume 
without loss of generality that the $a_l$ and hence 
the $U_{kl}$ in Eq.~(\ref{39}) are purely real quantities.
We thus can infer that Eq.~(\ref{35}) is symmetric with respect to the 
indices $\nu$ and $\alpha$, and hence with respect to the
perturbation $V$ in Eq.~(\ref{8}) and the observable 
$A$ in Eq.~(\ref{11}).
Moreover, one sees that a sign change of $t$  is 
tantamount to a complex conjugation of Eq.~(\ref{35}).
Recalling that Eq.~(\ref{35}) is real-valued [see below Eq.~(\ref{37})],
the result is an invariance under time-inversion.
Finally, given any solution of Eq.~(\ref{40}), let us define
$p':=-p$, $E':= -E$, and $a'_l:=(-1)^l a_l$.
It readily follows from Eq.~(\ref{40}) that
\begin{eqnarray}
\bar \delta_{Ll} a'_{l+1} +  p'\;\!\delta_{\nu l} a'_l  
+ \bar\delta_{1l}  a'_{l-1} =  2 E' a'_l
\ .
\label{41}
\end{eqnarray}
In other words, for every solution of the original problem 
we found a solution with an inverted sign of $p$.
Together with the discussion around Eqs.~(\ref{38}) and (\ref{39})
one thus can conclude that a sign change of $p$ implies 
a sign change of the energies $E_k$ (possibly after some suitable 
permutation of the labels $k$),
and that each matrix element $U_{kl}$ acquires a factor 
$(-1)^l$. 
Moreover, one can conclude that a sign change of $p$ 
is tantamount to interchanging the summation 
indices $j$ and $k$ on the right-hand side of Eq.~(\ref{35}).
Recalling the discussion below Eq.~(\ref{37}) it follows
that the sum is invariant under a sign change of $p$,
i.e., we recover the general prediction  below Eq.~(\ref{17})
that $\tilde C_{V\!A}(t)$ must be 
an even function of $p$.

\section{The XX-Model with an end-impurity}
\label{s4}

We focus on impurities and perturbations $s_\nu^z$
in Eqs.~(\ref{2}) and (\ref{8}) [and thus in Eq.~(\ref{24})] with
\begin{eqnarray}
\nu=L
\ ,
\label{42}
\end{eqnarray}
i.e., we may imagine the impurity to be located 
at the right chain-end [see also discussion below Eq.~(\ref{8})].
Accordingly, the $L$ equations (\ref{40}) can 
be rewritten as
\begin{eqnarray}
a_{l+1}+a_{l-1} & = & 2 E a_l \ \, \mbox{for $l=2,...,L-1\,$,}
\label{43}
\\
a_2 & = & 2 E a_1 
\ ,
\label{44}
\\
a_{L-1} + p\, a_L& = & 2E a_L 
\ .
\label{45}
\end{eqnarray}
A systematic solution of these equations is still very difficult. 
Hence, we will start out from an ansatz with some free parameter,
and then show that it solves Eqs.~(\ref{43})-(\ref{45}) by suitably 
choosing the parameter.
Moreover, we will show that the ansatz indeed generates
$L$ 
different such solutions 
[see below Eq.~(\ref{40})].
In other words, a complete solution of the original problem
is obtained, thus justifying the ansatz {\em a posteriori}.

\subsection{Ansatz}
\label{s41}

Our starting point is the observation that the quantities
\begin{eqnarray}
\alpha_l:= \sin(\varphi l)\ \, \mbox{for all $l\in\ZZ $}
\label{46}
\end{eqnarray}
satisfy for an arbitrary but fixed $\varphi\in\CC$ the relations
\begin{eqnarray}
\alpha_{l+1}+\alpha_{l-1}=2 \cos(\varphi)\, \alpha_l \ \, \mbox{for all $l\in\ZZ $}
\ .
\label{47}
\end{eqnarray}
In view of Eq.~(\ref{43}), this observation suggest the ansatz
\begin{eqnarray}
E & = & \cos(\varphi)
\ ,
\label{48}
\\
a_l & = & \alpha_l =\sin(\varphi l )\ \, \mbox{for $l=1,...,L$,}
\label{49}
\end{eqnarray}
with a generally {\em complex valued} parameter $\varphi$.
By exploiting Eqs.~(\ref{46}) and (\ref{47}),
one readily verifies that this ansatz indeed 
solves not only Eq.~(\ref{43}) but also Eq.~(\ref{44}).
We are thus left to solve the remaining equation
(\ref{45}) by suitably choosing the parameter $\varphi\in \CC$.
In fact, we have to show that there exist
$L$ different solutions of this equation
(see above). 

Introducing Eq.~(\ref{49}) into Eq.~(\ref{45}) yields
\begin{eqnarray}
2E\, a_L -p\, \alpha_L 
= \alpha_{L-1}
= 2 E\alpha_L - \alpha_{L+1}
\ ,
\label{50}
\end{eqnarray}
where we exploited Eqs.~(\ref{47}) and (\ref{48}) in the last step.
It follows that $p\alpha_L=\alpha_{L+1}$ and with Eq.~(\ref{46}) that
\begin{eqnarray}
\sin(\varphi(L+1))
& = & 
p \sin(\varphi L)
\ .
\label{51}
\end{eqnarray}

Altogether, considering the quantities $L$ and $p$ as given, we are 
left to determine $L$ solutions $\varphi\in\CC$ of Eq.~(\ref{51}).
More precisely speaking, we are actually not interested in the solutions $\varphi$
of Eq.~(\ref{51}) themselves, but rather in the concomitant 
eigenvectors and eigenvalues, 
encoded by $\{a_l\}_{l=1}^L$ and $E$,
which derive from $\varphi$ via Eqs.~(\ref{48}) and (\ref{49}).
For instance, one readily sees that $\varphi$ and $\varphi+2\pi$
give rise via Eqs.~(\ref{48}) and (\ref{49}) to exactly the same eigenvectors and 
eigenvalues, hence we can and will restrict ourselves to $\varphi$'s 
with the property $\mbox{Re}(\varphi)\in(-\pi,\pi]$.
Likewise, $\varphi$ gives rise via 
Eqs.~(\ref{48}) and (\ref{49}) to the same eigenvectors and eigenvalues
as $-\varphi$ (apart from an irrelevant
sign change of all $a_l$, see below Eq.~(\ref{39})).
Accordingly, we can and will restrict ourselves 
to $\varphi$'s which satisfy the extra condition 
\begin{eqnarray}
\mbox{Re}(\varphi)\in[0,\pi]\, .
\label{52}
\end{eqnarray}
Finally, in the special cases
$\varphi=0$ and $\varphi=\pi$
it follows from Eq.~(\ref{49})  that $a_l=0$ for
all $l=1,...,L$, which we excluded below Eq.~(\ref{39}).
Hence, Eq.~(\ref{52}) must be complemented by the
additional constraint
\begin{eqnarray}
\varphi\not \in\{0,\pi\}
\ .
\label{53}
\end{eqnarray}

As mentioned below Eq.~(\ref{28}) (see also below Eq.~(\ref{40})), 
only real eigenvalues $E$ are admissible in Eq.~(\ref{38}).
Writing $\varphi$ as $x+iy$ with $x,y\in\RR$, one readily sees that
the right hand side of Eq.~(\ref{48}) is real if and only if $\sin(x)\sinh(y)=0$,
which in turn is only possible in one of the following two cases.

Case A: $\sinh(y)=0$ with $y\in\RR$, implying $y=0$ and thus
$\varphi\in\RR$.
Together with Eqs.~(\ref{52}) and (\ref{53}) 
this  amounts to the requirement
\begin{eqnarray}
\mbox{case A: $\varphi \in(0,\pi)$.}
\label{54}
\end{eqnarray}

Case B: $\sin(x)=0$ with $x\in\RR$, implying $x/\pi\in\ZZ$
and thus $\varphi = iy+ n\pi$ with $y\in\RR$ and $n\in\ZZ$.
Similarly as above Eq.~(\ref{52}), one sees that it is actually
sufficient to focus on $y\in\RR_0^+$ and $n\in\{0,1\}$.
Together with Eqs.~(\ref{52})  and (\ref{53}) this amounts 
to the requirement
\begin{eqnarray}
\mbox{case B:}\, & & 
\varphi = iy+ n\pi\ \mbox{with $y\in\RR^+$ and $n\in\{0,1\}$.}
\ \ 
\label{55}
\end{eqnarray}

In the latter case B, it is for practical purposes
more convenient to replace $\varphi$ in Eq.~(\ref{49}) by 
$i\varphi$, yielding (apart from an irrelevant factor $i$, 
see below Eq.~(\ref{39}))
\begin{eqnarray}
a_l = \sinh(\varphi l)\ \, \mbox{for $l=1,...,L$}
\ .
\label{56}
\end{eqnarray}
Similarly, Eqs.~(\ref{48}) and (\ref{51}) now assume the form
\begin{eqnarray}
E & = & \cosh(\varphi)
\ ,
\label{57}
\\
\sinh(\varphi(L+1))
& = &
p\sinh(\varphi L)
\ .
\label{58}
\end{eqnarray}
Observing that above Eq.~(\ref{55}) we could equivalently replace 
the condition $y\in\RR_0^+$ by the condition $y\in\RR_0^-$,
the requirement (\ref{55}) now assumes the form
\begin{eqnarray}
\mbox{case B:}\, & & 
\varphi = y+ in\pi\ \mbox{with $y\in\RR^+$ and $n\in\{0,1\}$.}
\ \ 
\label{59}
\end{eqnarray}

\subsection{Analytical solution}
\label{s42}

In this section we work out the details of
the ansatz from the previous Sec. \ref{s41}.
In particular, we will treat the two cases 
in and above Eqs. (\ref{54}) 
and (\ref{59}) in two separate sections.
A summary and discussion of the so obtained 
solutions is provided in the subsequent  
Sec.~\ref{s43},
including the notions of (de)localized modes
appearing in the titles of the next sections.

\subsubsection{Case A: delocalized modes}
\label{s421}

According to Eqs.~(\ref{51}) and (\ref{54}) we have to solve
\begin{eqnarray}
\sin(\varphi (L+1)) & = &  p\, \sin(\varphi L)  \ \mbox{with $\varphi \in(0,\pi)$.}
\label{60}
\end{eqnarray}

To begin with, we show that there exists no solution of 
(\ref{60}) with the property $\sin(\varphi L)=0$
by means of a proof by contradiction.
Let us therefore assume that there exists a solution with
$\sin(\varphi L)=0$. 
The condition
$\varphi \in(0,\pi)$ in Eq.~(\ref{60}) implies that
$\varphi L=\pi n \ \mbox{with} \ \  n \in\{1,...,L-1\} \, . $
Similarly, we can conclude from Eq.~(\ref{60}) that 
$\sin(\varphi (L+1))=0$, and thus 
$\varphi (L+1)=\pi m\ \mbox{with} \ \  m\in\{1,...,L\} \, .$
In other words, we are looking for solutions of the equation
$L/(L+1)= n / m$
with $ n \in\{1,...,L-1\}$ and $ m\in\{1,...,L\}$.
One readily sees that no such solution exists:
Observing that $ n $ must be smaller than $L$ 
and smaller than $ m$, it follows that 
$n / m\leq  n /( n +1) < L/(L+1)$.

Accordingly, we can and will focus on
$\sin(\varphi L)\not\not =0$ in Eq.~(\ref{60}), and
our goal is to find solutions of
\begin{eqnarray}
f(\varphi) & = &  p \ \mbox{with $\varphi \in(0,\pi)$,}
\label{61}
\\
f(x) & := & \frac{\sin(x(L+1))}{\sin(xL)}
\nonumber
\\
& = &
\cos(x)+\sin(x)\,\cot(x L)
\ ,
\label{62}
\end{eqnarray}
where we exploited that
$\sin(x(L+1))=\sin(xL)\cos(x)+\cos(xL)\sin(x)$
in the last step.

\begin{figure}
\hspace*{-0.8cm}
\includegraphics[scale=0.95]{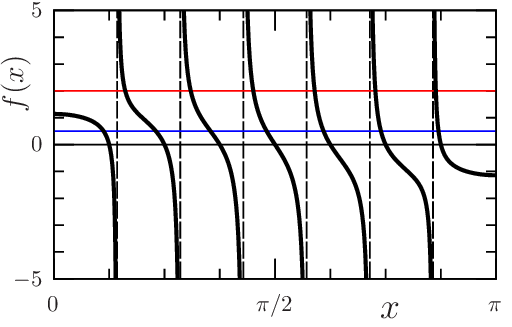}
\caption{
Black: The function $f(x)$ from Eq.~(\ref{62}) for $L=7$ and
$x\in(0,\pi)$ [see Eq.~(\ref{61})].
The dashed vertical lines are meant to indicate the $L-1$ poles
of $f(x)$.
The intersections of the black curve with the red horizontal line
exemplify the graphical solutions of Eq.~(\ref{61}) when 
choosing $p=2$.
Obviously, there are $L-1$ intersections.
Likewise the intersections with the blue line 
exemplify the $L$ solutions when $p=0.5$.
}
\label{fig1}
\end{figure}

Next we turn to a closer inspection of the function
$f(x)$ from Eq.~(\ref{62}) with $x\in(0,\pi)$ [see Eq.~(\ref{61})].
An illustration for $L=7$ is provided in Fig.~\ref{fig1}.
The following main properties of $f(x)$ can be
{readily verified analytically or by looking at}
Fig.~\ref{fig1}.
(i) 
It exhibits the symmetry $f(\pi-x)=-f(x)$ \cite{foot3}.
(ii)
It exhibits $L$ zeros at $x= n\pi/(L+1)$ with $ n\in\{1,...,L\}$.
(iii)
It exhibits $L-1$ simple poles at $x= m\pi/L$ with 
$ m\in\{1,...,L-1\}$, and is strictly monotonically 
decreasing for all other $x\in(0,\pi)$
(the analytical proof is provided in
Appendix \ref{appB}).
(iv) It approaches the value 
\begin{eqnarray}
p_c:=\frac{L+1}{L}
\label{63}
\end{eqnarray}
for $x\to 0$
and the value $-p_c$ for $x\to\pi$.
With these properties of $f(x)$ in mind, we are in 
the position to discuss the solutions of Eq.~(\ref{61}).

Let us first focus on the case $|p|>p_c$.
From the properties (iii) and (iv) or from Fig.~\ref{fig1} 
one can conclude that every pole of $f(x)$
gives rise to exactly one solution of Eq.~(\ref{61}).
Furthermore, each solution is a strictly monotonically 
decreasing function of $p$.
Finally, the distance of every given solution 
to the corresponding pole has the same sign as $p$,
and cannot exceed the distance between neighboring 
poles.
In other words, there are $L-1$ pairwise distinct solutions
of Eq.~(\ref{61}) of the form 
\begin{eqnarray}
\varphi_k \! &= & \! 
\frac{\pi k+r_k}{L}\ \mbox{with $k \in\{1,...,L\!-\!1\}$} 
,
\label{64}
\end{eqnarray}
where the 
``remainders''  $r_k$ exhibit the same sign as $p$,
are strictly monotonically decreasing functions of $p$,
 and are bounded
according to 
\begin{eqnarray}
|r_k|< \pi
\ .
\label{65}
\end{eqnarray}
Finally, by combining Eq.~(\ref{64}) with Eqs.~(\ref{38}), (\ref{39}), (\ref{48}), and (\ref{49})
we obtain
\begin{eqnarray}
E_k & = & 
\mbox{$\cos(\frac{\pi k+r_k}{L}) $}
\ ,
\label{66}
\\
U_{kl} & = & 
\kappa_k^{-1}
\mbox{$\sin(\frac{l\pi k+lr_k}{L}) $}
\ ,
\label{67}
\end{eqnarray}
for $k\in\{1,...,L-1\}$ and $l\in\{1,...,L\}$,
where
\begin{eqnarray}
\kappa_k := \left(
\sum_{l=1}^L
\mbox{$\sin^2(\frac{l\pi k+lr_k}{L}) $}
\right)^{1/2}
\label{68}
\end{eqnarray}
is a normalization constant.

Our next observation is that 
$f(x)$ in Eq.~(\ref{62}) can be approximated
by $1+ x\cot(x L)$ for small $x$.
For large $L$ this implies that $f(x)$ stays
very close to unity with the exception of
a very small neighborhood of all those
poles which are located near zero.
We thus can conclude that $|r_k|\ll \pi$
for all $k\ll L$.
[This can {\em not} be seen in Fig.~\ref{fig1} since
the value of $L=7$ is still too small.]
Analogous conclusions apply for $x$ close to $\pi$
and thus for all $k$ with $L-k\ll L$.
Finally, it is obvious from Fig.~\ref{fig1}
that all $r_k$ approach zero for 
asymptotically large $|p|$. 
In all the remaining cases, i.e. if neither of the conditions
$|p|\gg p_c$, $k\ll L$, or $L-k\ll L$ is fulfilled, the $|r_k|$
will not be much smaller than required by 
(\ref{65}).
A more quantitative illustration of those general features
is obtained 
by introducing Eq.~(\ref{64}) into Eqs.~(\ref{61}) and (\ref{62}),
yielding after a straightforward calculation 
\begin{eqnarray}
\sin(r_k) = p^{-1} 
\sin(p_c r_k + \pi k/L)
\ .
\label{69}
\end{eqnarray}
Together with the constraints in and above Eq.~(\ref{65}),
one thus can deduce for large $|p|$ and any $k\in\{1,...,L-1\}$
the asymptotics
\begin{eqnarray}
r_k = 
p^{-1} \mbox{$\sin(\frac{\pi k}{L}) 
\left(1+\frac{p_c}{p}\cos(\frac{\pi k}{L}) +\ord (p^{-2}) \right)$.}
\label{70}
\end{eqnarray}
Likewise, Eq.~(\ref{69}) implies for $|p|>p_c$ and sufficiently
small $k/L$ (which is only possible if $L\gg 1$)
that
\begin{eqnarray}
r_k = 
\frac{1}{p-p_c}
\frac{\pi k}{L}
\, (1+\ord (k^2/L^2))
\ ,
\label{71}
\end{eqnarray}
and analogously for sufficiently small 
$\bar k/L$ with $\bar k:=L-k$ that
\begin{eqnarray}
r_k = 
\frac{1}{p+p_c}
\frac{\pi \bar k}{L}
\, (1+\ord (\bar k^2/L^2))
\ .
\label{72}
\end{eqnarray}

Finally, for sufficiently large $L$ we can approximate $p_c$ in Eq.~(\ref{63})
by unity. After a few steps one thus obtains from Eq.~(\ref{69}) the large-$L$
asymptotics
\begin{eqnarray}
\tan(r_k) = \frac{p^{-1}\sin(\frac{\pi k}{L})}{1-p^{-1}\cos(\frac{\pi k}{L})}
\ .
\label{73}
\end{eqnarray}
For any given $k\in\{1,...,L-1\}$ there exists a unique 
solution $r_k$ of this equation together with the 
constraints in and above Eq.~(\ref{65}).
Since $|p^{-1}|<p^{-1}_c<1$ [see Eq.~(\ref{63})],
one can moreover conclude that
this solution must actually satisfy the even stronger 
bound $|r_k|<\pi/2$ (compared to Eq.~(\ref{65})), 
and therefore can be explicitly obtained by 
acting with the main branch of the function $\arctan$ 
on both sides of Eq.~(\ref{73}).
Finally,
the denominator in Eq.~(\ref{73}) can always be 
rewritten as a convergent geometric series.
Together with $\arctan(x)=x-x^3/3+x^5/5 - ...$ 
this yields a convenient expansion of $r_k$ 
in powers of $p^{-1}\sin(\pi k/L)$ and $p^{-1}\cos(\pi k/L)$,
thus systematically generalizing Eq.~(\ref{70}) 
in the special case $L\gg 1$ and thus
$p_c\simeq 1$.

Turning to the case $|p| < p_c$, a similar qualitative 
discussion as above Eq.~(\ref{63}) applies.
The main differences 
can be briefly summarized as follows.
The salient point is that, as $p$ crosses the critical  
value $p_c$ from above, an additional 
solution of Eq.~(\ref{61})
appears in the interval between $x=0$ and the 
first pole of $f(x)$ at $x=\pi/L$, see also Fig.~\ref{fig1}. 
Likewise, one solution disappears when $p$ 
crosses the value $-p_c$ from above.
Altogether, we thus always obtain $L$ 
solutions of Eq.~(\ref{61}) if $|p|<p_c$.
In the special case $p=0$, all those solutions 
must coincide with the zeros of $f(x)$, see item 
(ii) above.
As $p$ moves away from zero, the distance of 
each solution to the corresponding zero of $f(x)$ 
must decrease strictly monotonically with $p$,
must be opposite in sign to the sign of $p$,
and cannot exceed the distance between 
neighboring zeros (see  Fig.~\ref{fig1}).
In other words, we now obtain $L$ pairwise distinct 
solutions of Eq.~(\ref{61}) of the form
\begin{eqnarray}
\varphi_k \! &= & \! 
\frac{\pi k+r_k}{L+1}\ \mbox{with $k \in\{1,...,L \}$,} 
\label{74}
\end{eqnarray}
where the $r_k$ are again bounded 
as in Eq.~(\ref{65}) and 
are strictly monotonically decreasing functions of $p$,
while their sign 
must now be opposite to the sign of $p$.
Likewise, the counterparts of Eqs.~(\ref{66})-(\ref{68})
now assume the form
\begin{eqnarray}
E_k & = & 
\mbox{$\cos(\frac{\pi k+r_k}{L+1}) $}
\ ,
\label{75}
\\
U_{kl} & = & 
\kappa_k^{-1}
\mbox{$\sin(\frac{l\pi k+lr_k}{L+1}) $}
\ ,
\label{76}
\\
\kappa_k 
& := &
\left(
\sum_{l=1}^{L}
\mbox{$\sin^2(\frac{l\pi k+lr_k}{L+1}) $}
\right)^{1/2}
\label{77}
\end{eqnarray}
but now for all $k,l\in\{1,...,L\}$.
As before, the $r_k$ are furthermore 
seen to become even much smaller  than required by 
(\ref{65}) in the special cases
$|p|\ll 1$, $k\ll L$, or $L+1-k\ll L$
(which is only possible for $L\gg1$),
but not in any other case.
Finally, upon introducing Eq.~(\ref{74}) into 
(\ref{61}) one now finds after 
a straightforward calculation that
\begin{eqnarray}
\sin(r_k) = p\,\sin(p_c^{-1} r_k-\pi k/(L+1))
\ ,
\label{78}
\end{eqnarray}
complemented by the constraints below Eq.~(\ref{74}).
We thus recover the same equation as in (\ref{69}) upon replacing 
$p$ by $-p^{-1}$, $p_c$ by $-p_c^{-1}$, and $L$ by $L+1$.
The same correspondence is inherited by the results 
(\ref{70})-(\ref{72}).
For small $p$ and any $k\in\{1,...,L\}$
one thus obtains
\begin{eqnarray}
r_k = 
- p\, 
\mbox{$\sin(\frac{\pi k}{L+1}) 
\left(1+\frac{p}{p_c}\cos(\frac{\pi k}{L+1}) +\ord (p^2) \right)$.}
\ \ 
\label{79}
\end{eqnarray}
Likewise, one finds for $|p|<p_c$ and sufficiently
small $k/L$ that
\begin{eqnarray}
r_k = 
-\frac{p}{1-p/p_c}
\frac{\pi k}{L+1}
\, (1+\ord (k^2/L^2))
\ ,
\label{80}
\end{eqnarray}
and analogously for sufficiently small $\bar k/L$ with $\bar k:=L+1-k$ that
\begin{eqnarray}
r_k = 
-\frac{p}{1+p/p_c}
\frac{\pi \bar k}{L+1}
\, (1+\ord (\bar k^2/L^2))
\ .
\label{81}
\end{eqnarray}

Finally, 
for sufficiently large $L$ we can approximate
$p_c$ in Eq.~(\ref{63}) by unity.
Similarly as in Eq.~(\ref{73}), one thus
can deduce from Eq.~(\ref{78})
the large-$L$ asymptotics
\begin{eqnarray}
\tan(r_k) = -\frac{p\sin(\frac{\pi k}{L+1})}{1-p\cos(\frac{\pi k}{L+1})}
\ .
\label{82}
\end{eqnarray}
As long as $|p|<1$, the discussion of this result is 
analogous as below Eq.~(\ref{73}) and therefore not 
repeated here.
The case $|p|=1$ will be separately considered below.
The remaining case $|p|>1$ is of limited interest.
Indeed, any given $p$ with $|p|>1$ will violate
for sufficiently large $L$ our present overall assumption 
$|p| < p_c=(L+1)/L$.
We thus restrict ourselves to two brief remarks 
concerning this case $1<|p|<p_c$.
(i) The approximation (\ref{82})
now becomes questionable since we assumed
in its derivation that $p_c$ can be 
approximated by unity.
(ii) This is the only case
where the inequality $|r_k|< \pi/2$ (see below 
Eqs. (\ref{73}) and (\ref{82})) may possibly be violated.
A simple example
arises for $p\uparrow p_c$ and $k=1$,
where one can readily infer from Fig.~\ref{fig1} that 
$\varphi_1\to 0$ and hence from Eq.~(\ref{74})
that $r_1\to-\pi$.

Yet another special case which can be solved exactly is when $|p|=1$
(note that this still implies $|p|<p_c$). 
Namely, one can show  (see Appendix \ref{appC}) that
\begin{eqnarray}
r_k & = & - \pi \frac{L+1-k}{2L+1} \ \ \mbox{for \ $p=1$,}
\nonumber
\\
r_k & = & \pi \frac{k}{2L+1} \ \ \mbox{for \ $p=-1$}
\label{83}
\end{eqnarray}
and $k\in\{1,...,L\}$,
implying with Eq.~(\ref{74}) that
\begin{eqnarray}
\varphi_k & = & \pi\frac{k-1/2}{L+1/2}\ \ \mbox{for \ $p=1$,}
\nonumber
\\
\varphi_k & = & \pi\frac{k}{L+1/2}\ \ \mbox{for \ $p=-1$.}
\label{84}
\end{eqnarray}
By combining the monotonicity property below Eq.~(\ref{74}) 
with Eqs.~(\ref{83}) one readily obtains a further refinement 
of the previously obtained bound
$|r_k|< \pi/2$ [see below Eqs.~(\ref{73}) and (\ref{82})].
Incidentally, these results for $|p|=1$ are equivalent to those 
obtained by Tasaki in Sec. 3.1 of Ref. \cite{tas24b}.

Altogether, in the case $|p|<p_c$ 
we thus found
all the $L$ solutions required below 
Eqs.~(\ref{40}) and  (\ref{45}).
On the other hand, in the case $|p|>p_c$ 
we only found $L-1$ solutions 
i.e., one solution is still ``missing''.
The latter will be obtained in the next section.

\subsubsection{Case B: localized modes}
\label{s422}

\begin{figure}
\hspace*{-0.8cm}
\includegraphics[scale=0.95]{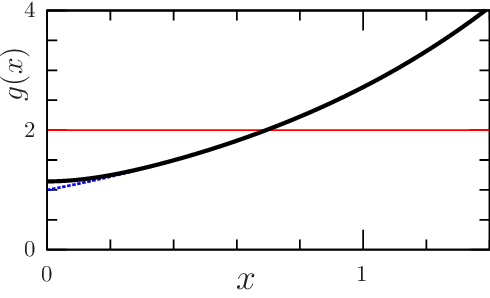}
\caption{
Black: The function $g(x)$ from Eq.~(\ref{86}) for $x>0$ (see Eq.~(\ref{85}))
and $L=7$. 
Similarly to Fig.~\ref{fig1}, 
the intersection of the black curve with the red horizontal line is the 
graphical solution of Eq.~(\ref{85}) when choosing $p = 2$.
Blue dashed 
curve: the approximation $g(x)\simeq e^x$ 
mentioned in the main text.
}
\label{fig2}
\end{figure}

According to Eqs.~(\ref{58}) and (\ref{59}) we have to solve
\begin{eqnarray}
g(y) & = & \pm p \ 
\mbox{with $y\in\RR^+$,}
\label{85}
\\
g(x) & := &\frac{\sinh(x(L+1))}{\sinh(xL)}
\ ,
\label{86}
\end{eqnarray}
where the plus sign arises when choosing $n=0$
in Eq.~(\ref{59}) and the minus sign when choosing 
$n=1$.

Restricting ourselves to $x\in\RR^+$ [see Eq.~(\ref{85})], 
the function $g(x)$ exhibit the following properties
(see also Fig.~\ref{fig2}).
(i)~It is strictly monotonically increasing.
(ii)~$g(x)\to p_c$ for $x\to 0$, where $p_c$ is defined in Eq.~(\ref{63}).
(iii)~$g(x)\sim e^x$ for $x\gg 1/L$ 
(see also the blue dashed curve in Fig.~\ref{fig2}).

From (i)-(iii) one can infer that there exists no solution 
of Eq.~(\ref{85}) if $|p|\leq p_c$, and exactly one solution
if $|p|>p_c$, in accordance with the discussion at the
end of the previous section.
Moreover, the choice $n=0$ applies in the case 
$p>p_c$ and $n=1$ in the case $p<-p_c$.
Denoting this solution for $|p|>p_c$ as $y_L$, and focusing on large 
$L$, we moreover can infer from Eqs.~(\ref{85}), (\ref{86}), and (iii)
the approximation
\begin{eqnarray}
y_L\simeq   
\ln(|p|)\ \mbox{if $\ln(|p|)\gg 1/L\,$.}
\label{87}
\end{eqnarray}
Note that for large $L$, 
the latter condition $\ln(|p|)\gg 1/L\,$ is equivalent to $|p|-p_c\gg 1/L$.
Finally, 
(\ref{59}) implies $\varphi_L=y_L$ if 
$p>p_c$
and $\varphi_L=y_L+i\pi$ if 
$p<-p_c$.
Together with Eqs.~(\ref{38}), (\ref{39}), (\ref{56}), (\ref{57}), 
(\ref{87}),
the so far still ``missing'' solution with index $k=L$ 
(see end of the previous section)
can thus be approximately rewritten for large $L$ 
and $|p|-p_c\gg 1/L$ 
as
\begin{eqnarray}
E_L
& = & 
(p+1/p)/2
\ ,
\label{88}
\\
U_{Ll} 
& = &
\sqrt{1-1/p^2}\, (1/p)^{\bar l}
\ ,
\label{89}
\end{eqnarray}
where
\begin{eqnarray}
\bar l := L - l
\label{90}
\end{eqnarray}
is the distance of the site $l$ from the right chain-end
(at the site $L$).
A more detailed analysis reveals that the relative error
of the approximation (\ref{88}) is on the order of $p^{-2L}$.
Regarding the approximation (\ref{89}), it is the absolute rather than the
relative error which is of foremost interest, and which is found to be
on the order of $p^{-(l+L)}$.

Observing that $|1/p|<1/p_c<1$ it follows that $U_{Ll}$ is maximal
(in modulus) for $l=L$ (chain end) and decreases exponentially 
with increasing distance $\bar l $ from the end.
Moreover, for $p>p_c$, the energy in Eq.~(\ref{87}) satisfies $E_L> 1$ 
and all $U_{Ll}$ in Eq.~(\ref{88}) are positive,
whereas for $p<-p_c$, the energy in Eq.~(\ref{87}) satisfies $E_L<-1$ 
and the $U_{Ll}$ in Eq.~(\ref{88}) exhibit alternating signs
\cite{foot3}.

Roughly speaking, that particular solution of type A 
which approaches zero and then disappears 
when $p$ crosses the critical value $p_c$
from below [see discussion above Eq.~(\ref{74})], 
may thus be imagined as actually ``turning from real to imaginary'' 
and then reappearing as a solution of type B, 
and likewise when $p$ crosses the value $-p_c$.
A more detailed discussion is 
provided in the next section.

\subsubsection{The case $|p|\simeq p_c$}
\label{s423}

If $|p|$ approaches $p_c$ from above, 
one readily confirms that there are still $L-1$ 
solutions of the form (\ref{66})-(\ref{68}), 
satisfying Eqs.~(\ref{65}) and (\ref{69}).
Concerning the remaining (or ``missing'')
solution, let us first focus on the case $p=p_c+\epsilon$
with $0<\epsilon\ll 1$. According to the properties 
(i) and (ii) below Eq.~(\ref{86}),
the solution $y$ of Eq.~(\ref{85}) must 
approach zero from above as $\epsilon \to 0$.
Moreover, the plus sign must be chosen in 
(\ref{85}) and thus $n=0$ in Eq.~(\ref{59}),
implying 
$\varphi=y$.
For sufficiently small $\epsilon$ it 
furthermore follows that $\varphi\ll L$ and with
(\ref{56}) that $a_l\simeq \varphi l$,
i.e., $a_l$ must be proportional to $l$.
Together with Eqs.~(\ref{38}), (\ref{39}), and (\ref{57})
this leads us to the conjecture that
the ``missing'' solution with $k=L$ 
assumes the form
\begin{eqnarray}
& & \mbox{$E_L=1$ and $U_{Ll}= \kappa_L^{-1}\, l$ if $p=p_c$\, ,}
\label{91}
\end{eqnarray}
where $\kappa_L:=[\sum_{l=1}^L l^2]^{1/2}=[L(L+1)(2L+1)/6]^{1/2}$.
This conjecture can be readily verified by noting
that Eq.~(\ref{91}) indeed satisfies Eqs.~(\ref{43})-(\ref{45}) 
[together with Eqs.~(\ref{38}) and (\ref{39})] even exactly.
Analogously one  finds the exact results 
\cite{foot3}
\begin{eqnarray}
& & \mbox{$E_L=-1$ and $U_{Ll}=\kappa_L^{-1}\,(-1)^l l$ if $p=-p_c$\, .}
\label{92}
\end{eqnarray}
Without going into the details we remark that
the very same 
findings
could also be 
deduced in the limit $|p|\uparrow p_c$ from the
results in Sec.~\ref{s421} (case A).
Moreover, for any given $L$ with $L\gg 1$ 
one can show that 
Eqs.~(\ref{91}) and (\ref{92})
still remain very good approximations 
as long  as $||p|-p_c|\ll 1/L$.
[We may also recall that the case $|p|=1$ corresponds to 
$||p|-p_c|=1/L$ and is solved exactly by Eqs.~(\ref{75})-(\ref{77})
with $r_k$ from 
Eq.~(\ref{83}).]

\subsection{Interim summary and discussion}
\label{s43}

As detailed below Eq.~(\ref{40}), our basic Eqs.~(\ref{29}) or
(\ref{40}) may be viewed as an $L$-dimensional Hermitian
matrix eigenvalue problem.
For $|p|<p_c$, the solutions ($L$ eigenvalues and eigenvectors)
are given by Eqs. 
(\ref{74})-(\ref{77}), 
where the $r_k$ are determined via Eqs.~(\ref{78}) and (\ref{65}).
Likewise, for $|p|\geq p_c$, there are 
$L-1$ solutions of the form 
(\ref{66})-(\ref{68}),
where the $r_k$ are determined via Eqs.~(\ref{69}) and (\ref{65}).
In addition, there is one more solution which is 
of the form (\ref{88}) and (\ref{89}) if $|p|>p_c$ and 
of the form (\ref{91}) and (\ref{92}) if $|p|=p_c$.

In view of Eq.~(\ref{65}), the quantities $r_k$ in Eqs.~(\ref{66}) 
and (\ref{75})
become negligible for sufficiently large $L$.
However, it is important to realize that those
$r_k$ are generally {\em not negligible} in Eqs.~(\ref{67}) 
and (\ref{76}) as 
soon as the indices $l$ are comparable in order of 
magnitude to $L$ [since the quantities $lr_k/L$ on 
the right-hand side of Eqs.~(\ref{67}) and (\ref{76})
are no longer small].
Exceptions are asymptotically large or small 
values of $|p|$ since all $r_k$ then approach zero 
according to Eqs.~(\ref{70}) and  (\ref{79}), respectively.
Further very important exceptions are
the normalization constants $\kappa_k$ from
(\ref{68}) and  (\ref{77}). 
Namely, as shown in Appendix \ref{appD}
they can be approximated for large $L$ as
\begin{eqnarray}
\kappa_k=\sqrt{L/2}
\ ,
\label{93}
\end{eqnarray}
where the omitted relative error 
{approaches zero for $L\to\infty$}. 

Put differently, the presence of an impurity ($p\not=0$) at the right
chain-end (at $l=L$) entails a marked asymmetry of the matrix elements
$U_{kl}$ with respect to the second index $l$
(while $k$ is arbitrary but fixed).
For instance, focusing on $|p|<p_c$ and
setting formally $l=0$ or $l=L+1$,
the matrix elements in Eq.~(\ref{76})
are zero in the former and nonzero in the latter case
[namely $U_{kL+1} \sim \sin(r_k)$].
An analogous asymmetry --  governed by $r_k$ --
arises when comparing
$l=1$ with $l=L$ or when $|p|>p_c$ [see Eq.~(\ref{67})].

Returning to the above mentioned $L$-dimensional Hermitian 
matrix eigenvalue problem, its solutions may also be
viewed as
$L$ eigenvalues
of the form (\ref{38}) and
$L$ eigenvectors $\vec a$ with components $a_l$
of the form (\ref{39}) [see also below Eq.~(\ref{40})].
From {Eqs.~(\ref{76}), (\ref{77}), and (\ref{93})}
one can infer that, for any given $k\in\{1,...,L\}$, 
most of those components assume values which are of 
comparable order of magnitude in the case
$|p|<p_c$. 
Henceforth, we will thus often employ the 
common
notion of  {\em delocalized} (or {\em extended})
modes (or eigenvectors)
for this type of solutions
(see Ref. \cite{mur19} and further references therein).
In the same vein, there are $L-1$ delocalized 
solutions in the case $|p| > p_c$ according 
to Eqs.~(\ref{67}), (\ref{68}), and (\ref{93}),
{and likewise for $|p|=p_c$ (see Sec.~\ref{s423}).}

On the other hand, the components of the $L$th
eigenvector in Eq.~(\ref{89}) are largest (in modulus)
at the right chain-end, corresponding to $\bar l =0$ 
in Eq.~(\ref{90}),
and decrease exponentially with increasing 
distance $\bar l $ from the chain-end.
Accordingly, we will often refer to such solutions as
{\em localized} modes \cite{mur19}.
More precisely speaking, Eq.~(\ref{89}) amounts to a strongly
(exponentially) localized solution, while the
corresponding solution for $|p|=p_c$ in
(\ref{92}) is weakly localized in the sense 
that the components only decrease algebraically 
with increasing distance from the right chain-end.

\subsection{Absence of thermalization and strong violation of the ETH for $|p|>p_c$}
\label{s44}

In this section we derive the first analytical key results of our present paper,
namely nonthermalization after a local quench with the property
$|p|>p_c$, and a concomitant strong ETH violation (see also Sec.~\ref{s1}).

As explained below Eq.~(\ref{19}), in order to show the absence of thermalization, 
it is sufficient that the long-time average of $\tilde C_{V\!A}(t)$ does not 
approach zero for large $L$.
The latter property is derived in the next section,
followed by a discussion of its implications in Sec.~\ref{s442}.

\subsubsection{Derivation}
\label{s441}

Focusing on $|p|>p_c$ and $L\gg 1$, we observed above Eq.~(\ref{64}) that the
quantities $\varphi_k$ in Eq.~(\ref{64}) are pairwise distinct and contained in the interval
$(0,\pi)$, implying that the $E_k$ in Eq.~(\ref{66}) are pairwise distinct
and contained in the interval $(-1,1)$, where $k\in\{1,...,L-1\}$.
Moreover, from Eq.~(\ref{88}) and the discussion above Eq.~(\ref{88})
it follows that $E_L$ is larger than unity in modulus.
Altogether, it follows that the $E_k$ are pairwise distinct for
all $k\in\{1,...,L\}$. As a consequence, the long-time average of
the quantities in Eq.~(\ref{37}) is given according to Eq.~(\ref{18}) 
by
$\overline{e_{jk}(t)}=\delta_{jk}$.
From Eq.~(\ref{35}) we thus can conclude that
\begin{eqnarray}
\overline{\tilde C_{V\!A}(t)}
= \sum_{k=1}^L |U_{k L}|^2 |U_{k \alpha}|^2 f_{kk}
\ , 
\label{94}
\end{eqnarray}
where we set $\nu=L$ according to Eq.~(\ref{42}).
With the help of the auxiliary quantities 
\begin{eqnarray}
R_{L,\alpha} & := &  \sum_{k=1}^{L-1} |U_{k L}|^2 |U_{k\alpha}|^2 f_{kk} \ ,
\label{95}
\\
S_{L,\alpha} & := & |U_{LL}|^2 |U_{L\alpha}|^2 f_{LL} \ ,
\label{96}
\end{eqnarray}
we can rewrite Eq.~(\ref{94}) as
\begin{eqnarray}
\overline{\tilde C_{V\!A}(t)} = R_{L,\alpha}+ S_{L,\alpha}
\ .
\label{97}
\end{eqnarray}
Moreover, by
employing l'Hopital's rule to evaluate the right-hand side
of Eq.~(\ref{36}) in the limit $E_j\to E_k$ we can conclude that
\begin{eqnarray}
f_{kk} = \frac{1}{4\cosh^2(\beta E_k/2)}
\ .
\label{98}
\end{eqnarray}
Our next goal is to derive upper and lower 
bounds for the quantity in Eq. (\ref{95}). 
In a first step, we can infer from Eq.~(\ref{98})
that $0<f_{kk}\leq 1/4$. 
Furthermore, by
observing Eq.~(\ref{67}) and the
large-$L$ approximation
(\ref{93}) it follows that $|U_{kl}|^2\leq 2/L$ 
for all $k\in\{1,...,L-1\}$ and $l\in\{1,...,L\}$.
Altogether, we thus can conclude from (\ref{95}) that
\begin{eqnarray}
R_{L,\alpha} \in [0,1/L]
\ .
\label{99}
\end{eqnarray}
Turning to the quantity in Eq. (\ref{96}),  
we can exploit Eqs.~(\ref{88}), (\ref{89}), and (\ref{98})
to infer
\begin{eqnarray}
S_{L,\alpha} :=  \left(1-\frac{1}{p^2}\right)^2
\left(\frac{1}{p^2}\right)^{\bar \alpha} 
\frac{1}{4\cosh^2(\beta \frac{p+p^{-1}}{4})}
\ ,
\ \ \ 
\label{100}
\end{eqnarray}
where $\bar \alpha:=L-\alpha$.
[Incidentally, the invariance of Eq. (\ref{100})
under a sign change of $p$ is in
accordance with the general symmetry 
considerations 
around Eq.~(\ref{41}).]

Observing Eq.~(\ref{99}) we thus can conclude that
the long-time average in Eq.~(\ref{97})
converges to $S_{L,\alpha}$ from Eq.~(\ref{100})
for large $L$ (and fixed $\bar\alpha$),
which in turn is generically non-zero.

\subsubsection{Discussion}
\label{s442}

The latter finding of a nonvanishing long-time 
average in Eq.~(\ref{97}) implies the
absence of thermalization for the
observable $A=s_\alpha^z$
[see below Eq.~(\ref{19})].
With Eqs.~(\ref{19}) and (\ref{20}), we furthermore obtain
the approximation
\begin{eqnarray}
\overline{\langle s_\alpha^z \rangle_t} - \langle s_\alpha^z\rangle _{\!\rm th}
& = &
g\beta \, S_{L,\alpha}
\ ,
\label{101}
\end{eqnarray}
where $S_{L,\alpha}$ is given by Eq.~(\ref{100}),
and where $|p|>p_c$ is tacitly taken for granted.
As expected, the 
nonthermalization effect is most pronounced
if the observable-site $\alpha$ coincides with the site $\nu=L$
of the perturbation [see Eqs.~(\ref{8}) and (\ref{42})],
and decreases exponentially with
increasing $\bar \alpha:=L-\alpha$.
 
We also remark that an arbitrarily small (nonzero) perturbation 
strength $g$ in (\ref{101}) is sufficient for nonthermalization provided
the prequench impurity strength $\gamma$ is notably larger 
(in modulus) than $1/2$, see also the discussion below Eq.~(\ref{8})
and the definitions (\ref{17}) and (\ref{63}).
Moreover, for small $g$ all our analytical approximations 
become asymptotically exact, as discussed below Eq. (\ref{16}).

Finally, these 
conclusions regarding nonvanishing long-time averages
in combination with the considerations 
in Secs.~\ref{s24} immediately imply that
our present model with $|p|>p_c$ must exhibit a 
strong violation of the ETH
in the sense that neither the strong nor the 
weak ETH is satisfied.
An alternative, more direct derivation of the same
result is provided in Appendix~\ref{appE}.

Obviously, the above finding of nonthermalization 
{and strong ETH violation}
for $|p|>p_c$
has its origin in the contribution of the localized mode (see Sec.~\ref{s43}) 
with index $k=L$ to the sum in Eq.~(\ref{94}).
Accordingly, the absence of such a mode in the case $|p|<p_c$ 
rules out a similar conclusion.
The same applies to the weakly localized case $|p|=p_c$, as can be inferred
from the results in Sec.~\ref{s423} or by taking the limit $|p|\to p_c=(L+1)/L$ in Eq.~(\ref{100}).

We finally mention that the thermal expectation values 
$\langle s_\alpha^z \rangle_{\rm th}$ do {\em not} exhibit any
kind of ``transition'' at $|p|=p_c$
(see Appendix \ref{appG} for the details).

\subsection{Numerical methods and results}
\label{s45}

The first purpose of this section is to illustrate and 
complement the analytical predictions and approximations
from the previous sections by quantitative numerical examples.
On the one hand, we therefore solved Eqs.~(\ref{1})-(\ref{11})
by means of standard, numerically exact diagonalization methods.
As usual, such computations are only practically
feasible for relatively small system sizes $L$.
[On the machines available to us, up to $L=24$ 
was doable with reasonable effort
and by taking advantage of dynamical
typicality concepts.]
Furthermore, we numerically evaluated the
analytical approximation (\ref{15}) by exploiting
the exact analytical result from Eqs.~(\ref{35})-(\ref{37}),
where the energies $E_k$ and the unitary matrix 
elements $U_{kl}$ were obtained by numerically solving 
Eq.~(\ref{29}) [or the equivalent Eqs. (\ref{43})-(\ref{45})].
Finally, we also found a way of how those energies $E_k$ and 
matrix elements $U_{kl}$ can be exploited to numerically
evaluate Eqs.~(\ref{1})-(\ref{11}) up to quite large
$L$ values  [for instance, $L=10^4$ turned out to be
still easily doable]. A more detailed account of this
numerical method is provided in Appendix \ref{appF}.

\begin{figure}
\hspace*{-0.8cm}
\includegraphics[scale=0.95]{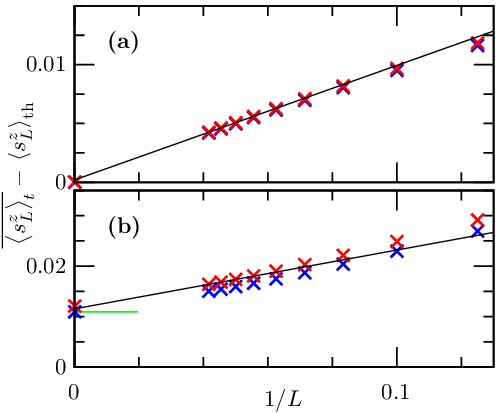}
\caption{
Red crosses: Numerically exact time-averaged expectation values
$\overline{\langle s^z_L\rangle_t} - \langle s^z_L\rangle _{\rm th}$
versus $1/L$ for the XX model from 
Eqs.~(\ref{1})-(\ref{11}) 
with parameter values
$J_\perp=1$, 
$J_z=0$,
$\nu=\alpha=L$,
$\beta=1$,
$\gamma=0$,
and $g=0.3$ in (a),
$g=0.6$ in (b).
Hence, $p=0.6<p_c$ in (a) and $p=1.2>p_c$ 
in (b), see Eqs. (\ref{17}) and (\ref{63}).
Blue crosses: 
Corresponding analytical approximations from 
Eq.~(\ref{15}), see also main text.
The leftmost crosses correspond to $L=10^4$,
followed by $L=24,22,...,10,8$.
The black lines are a guide to the eye, representing a linear
extrapolation towards $1/L=0$ 
based on the red crosses with 
$L\leq 24$.
The closeness of the lines to the red crosses at 
$L=10^4$ thus quantifies the reliability of 
such an extrapolation.
The green bar in (b) indicates the analytical approximation
for $L\to\infty$ from Eqs. (\ref{100}) and (\ref{101}).
}
\label{fig3}
\end{figure}

The second purpose of this section
is to ``gauge'' how well the behavior in the thermodynamic 
limit $L\to\infty$ can be extrapolated from the numerical
data up to $L=24$, i.e., by means of the above-mentioned
standard 
numerical
methods.
The underlying rationale is as follows:
In our present 
explorations of the XX model, we 
are able to compare the latter numerical data
with our approximate or even exact analytical findings,
as well as with the above mentioned more sophisticated
numerical methods for much larger $L$-values.
However, in our later explorations of the XXZ model,
the only remaining tool will be the standard numerical 
methods. Thus only data up to $L=24$ will be 
available for the extrapolation to the 
thermodynamic limit.
In doing so, it seems reasonable to expect 
and will be tacitly taken for granted that the 
quality of such an extrapolation is comparable 
to that in our present case of the XX model.

The red crosses in Fig.~\ref{fig3} exemplify the 
time-averaged expectation values
$\overline{\langle A\rangle_t} - \langle A\rangle _{\rm th}$
for $A=s_L^z$,
obtained by the above-mentioned numerical
solutions of the XX model
with parameter values as specified 
in the figure caption.
In other words, the prequench Hamiltonian $H_0$ appearing in
the initial Gibbs state (\ref{3}) is given by the basic XX-Hamiltonian 
from Eq.~(\ref{24}) with $p=0$ (no impurity, see also Eq.~(\ref{17})).
Similarly, the post´quench Hamiltonian is given by Eq.~(\ref{24}) 
with $p=0.6$ in (a) and $p=1.2$ in (b), i.e., it exhibits
an impurity at the right chain-end $(V=s_L^z$)
as specified in and around Eq.~(\ref{42}).
According to Eq.~(\ref{63}) this means that 
$p<p_c$ in (a) and $p>p_c$ in (b).

Comparing the blue and red crosses in Fig.~\ref{fig3}
illustrates that the approximation (\ref{15}) is very 
accurate in (a) and notably less accurate in (b),
in accordance with the general analytical
predictions below Eq. (\ref{16}). 
More precisely speaking, the value of $g\beta=0.3$ in (a) is still 
not really small, and all the more so for 
the value $g\beta=0.6$ in (b).
We have purposefully made this choice
in order to obtain some still clearly visible
differences between the blue and red crosses.
In other words, Fig. \ref{fig3} illustrates that the analytical 
error bound below Eq. (\ref{16}) is in general not very tight. 
We also validated (not shown) that the differences 
indeed become even much smaller upon decreasing 
$g\beta$, as predicted below Eq. (\ref{16}).
Another noteworthy observation is that the 
approximation (\ref{15}) obviously is 
(again as analytically predicted) 
not restricted to large system sizes $L$.

\begin{figure}
\hspace*{-0.8cm}
\includegraphics[scale=0.95]{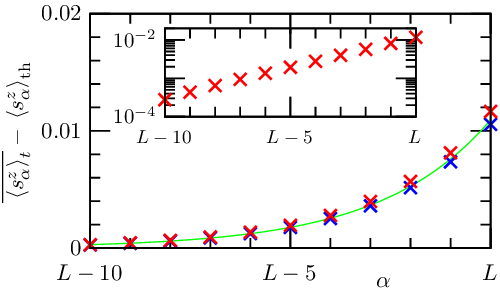}
\caption{
Time-averaged expectation values
$\overline{\langle s^z_\alpha\rangle_t} - \langle s^z_\alpha\rangle _{\rm th}$
vs $\alpha\in\{L-10,...,L\}$ for
the same XX model as in Fig.~\ref{fig3}(b).
Red crosses: Extrapolations of the numerically exact
results for $L=8,10,...,24$ to $L\to\infty$ by means of the
same procedure as indicated by the black lines in 
Fig.~\ref{fig3}.
In particular, the red cross for $\alpha=L$ corresponds to the
black line in Fig.~\ref{fig3}(b) at $1/L=0$.
Blue crosses: 
analogous extrapolations based on the 
approximation from Eq.~(\ref{15}).
Green curve: corresponding approximations from 
Eqs.~(\ref{100}) and (\ref{101}).
(Inset) Semilogarithmic plot of the same numerical 
results as in the main plot.
}
\label{fig4}
\end{figure}

The straight black lines in Fig.~\ref{fig3} depict the
extrapolation of the numerically exact results 
(red crosses) for $L\leq 24$ into the regime 
$L>24$.
The main point is that these extrapolations
are still quite close to the ``true'' numerical 
values at $L=10^4$ (leftmost red crosses).
We thus can conclude that already the numerically 
exact results up to $L=24$  
admit a quite faithful 
extrapolation of the behavior in the 
thermodynamic limit (see also second 
paragraph of this section).
In other words, the extrapolation would change
very little if also the leftmost cross at $L=10^4$
would be taken into account.
Moreover, such an even better extrapolation would
hit the origin even more closely in 
Fig.~\ref{fig3}(a) but definitely not in (b).
All this quite convincingly confirms absence of 
thermalization for $|p|>p_c$ but not for $|p|<p_c$
\cite{foot4},
as analytically predicted in the previous section.
Finally, the practically perfect agreement of the green bar
and the leftmost blue cross in Fig.~\ref{fig3}(b) illustrates the
very high accuracy of the analytical approximation 
from Eqs.~(\ref{100}) and (\ref{101}).

Figure~\ref{fig4} was obtained by similar
extrapolations to asymptotically large $L$ as in Fig.~\ref{fig3},
but now depicting results for more general observables 
$A=s_\alpha^z$ with $\alpha$ close to $L$ (right chain-end).
Moreover, we confined ourselves to the case
$g=0.6$, corresponding to panel (b) 
in Fig.~\ref{fig3} with $p=1.2>p_c$.
(Similarly to panel (a), $g=0.3$
yields identically vanishing results 
for $L\to\infty$).
Once again, the analytical approximations
from Eq. (\ref{15}) (blue) and from Eqs. (\ref{100}), (\ref{101})
(green) are seen to work quite well.
In particular, they faithfully reproduce the numerically 
observed ``localization'' of non-thermalization near
$\alpha=L$, 
as predicted below Eq.~(\ref{101}) and 
discussed in more detail in Sec.~\ref{s47}.
The inset as well as the green curve in the 
main plot corroborate an exponantial
decay of the numerical results
with $L-\alpha$, as analytically
predicted by Eq. (\ref{100}).

\subsection{Temporal relaxation behavior}
\label{s46}

While the focus in
the previous sections was on 
time-averaged expectation values,
the present section provides some
analytical approximations as well as numerical results 
regarding their time-dependence.

\subsubsection{Analytical approximations}
\label{s461}

Similarly as in Sec.~\ref{s44} [see also discussion below Eq.~(\ref{16})],
we take the approximation (\ref{15}) for granted and focus on
the behavior of the quantity $\tilde C_{V\!A}(t)$ from Eq.~(\ref{16}).
Moreover, the perturbations $V$ and observables $A$ 
are again assumed to be of the specific form (\ref{8}), 
(\ref{11}), and (\ref{42}), hence we can employ the
exact result (\ref{35}) as our starting point.

For simplicity, we focus on the most interesting case 
\begin{eqnarray}
\alpha=L
\ ,
\label{102}
\end{eqnarray}
meaning that both the perturbation $V$ and the observable $A$ are
given by the spin $s_L^z$ at the right chain-end.
Moreover, we focus on sufficiently high temperatures such that
\begin{eqnarray}
|\beta E_k|\ll 1\ \mbox{for all $k\in\{1,...,L\}$}
\ .
\label{103}
\end{eqnarray}
Therefore, all $f_{jk}$ in Eq.~(\ref{36}) can be approximated by $1/4$,
and the double sum in Eq.~(\ref{35}) can be rewritten
as the product of two simple sums.
In those cases where explicit analytical approximations
for $E_k$ and $U_{kl}$ are available (see Sec.~\ref{s42}),
and under the additional assumption that
\begin{eqnarray}
L\gg 1
\ ,
\label{104}
\end{eqnarray}
those sums can be furthermore approximated very well
in terms of Bessel functions
by means of a very similar line of reasoning as 
{for instance in Ref.~\cite{rei25}.}
Thus omitting the detailed calculations, the final results are
as follows.

For $|p|\ll p_c$ [where $p_c\simeq 1$ according to Eqs.~(\ref{63}) 
and (\ref{104})], we can exploit Eq.~(\ref{79}) to deduce the 
approximation
\begin{eqnarray}
\tilde C_{V\!A}(t) 
& \simeq & 
\frac{b(t) + p^2 c(t)}{4}\ \mbox{for  $|p|\ll 1$,}
\label{105}
\\
b(t)
& := &
[J_0(t)+J_2(t)]^2
\ ,
\label{106}
\\
c(t)
& := &
[J_1(t)+J_3(t)]^2 
\nonumber
\\
& & 
- 2\, [J_0(t)+J_2(t)]\,[J_2(t)+J_4(t)]
\ ,
\label{107}
\end{eqnarray}
{where} $J_n(t)$ are Bessel functions of the first kind.
Similarly,
by expoiting Eq.~(\ref{83}) 
one obtains
\begin{eqnarray}
\tilde C_{V\!A}(t) 
& \simeq & 
\frac{J^2_0(t) + J^2_1(t)}{4}\ \mbox{for  $|p|=1$.}
\label{108}
\end{eqnarray}
Finally, one can utilize Eq.~(\ref{69}) to infer
\begin{eqnarray}
\tilde C_{V\!A}(t) 
& \simeq & 
\frac{1-2\, p^{-2} d(t)}{4} \ \mbox{for  $|p|\gg 1$,}
\ \ \ \ \ \ \
\label{109}
\\
d(t)
& := &
1 - [J_0(t)+J_2(t)]\, \cos (E_Lt)
\ ,
\label{110}
\end{eqnarray}
{where} $E_L$ 
is approximately given by Eq.~(\ref{88}). 
Note that for large values of $|p|$, as required in Eq.~(\ref{109}),
also the energy $E_L$ in Eq.~(\ref{88}) becomes large (in modulus),
hence the condition (\ref{103}) becomes quite restrictive 
with respect to $\beta$.

The most important observation is that all Bessel functions $J_n(t)$ approach
zero for large $t$.
Hence, Eqs.~(\ref{105}) and (\ref{108}) approach zero for large $t$,
while the long time-limit in Eq.~(\ref{109}) agrees with the
long-time average from Eq.~(\ref{100}) [wherein $\bar\alpha=0$ and 
the last factor approaches $1/4$ according to Eqs.~(\ref{88}) and (\ref{103})].

Second, we recall that $\tilde C_{V\!A}(t)$ must be an even function of
$\beta$ and {of  $p$,} see 
below Eqs.~(\ref{37}) and (\ref{41}),
respectively.
Our present approximations are in accordance with these predictions.
In particular, they are independent of $\beta$, i.e., corrections of
order $\beta^2$ have been neglected [see also Eq.~(\ref{103})].
Moreover, the approximation (\ref{105}) includes terms up to second order in 
$p$, i.e., corrections of order $p^4$ have been neglected.
Likewise, corrections of order $1/p^4$ have been 
neglected in Eq.~(\ref{109}).
A quantification of the finite-$L$ corrections [cf. Eq.~(\ref{104})] 
is analytically less obvious (but numerically easy, see {next section}).

A more detailed discussion 
by exploiting the
well-known properties of the Bessel functions 
is straightforward but not of central interest 
with respect to the main issues of our present work.
Likewise, a generalization to other values of $\alpha$ 
than in Eq.~(\ref{102}) is readily feasible
in principle,
but the analytical 
expressions would become even more involved
than in the above findings.
Finally, going beyond the approximations (\ref{103}) and (\ref{104})
and beyond the $p$-values in Eqs.~(\ref{105}), (\ref{108}), and (\ref{109})
becomes very arduous by analytical means.

\subsubsection{Numerical examples}
\label{s462}

Here we illustrate the detailed temporal relaxation 
behavior by employing the same numerical 
methods as in Sec.~\ref{s45}.
Moreover, we compare the analytical
approximations from the previous section
with such numerically exact results.

\begin{figure}
\hspace*{-0.8cm}
\includegraphics[scale=0.95]{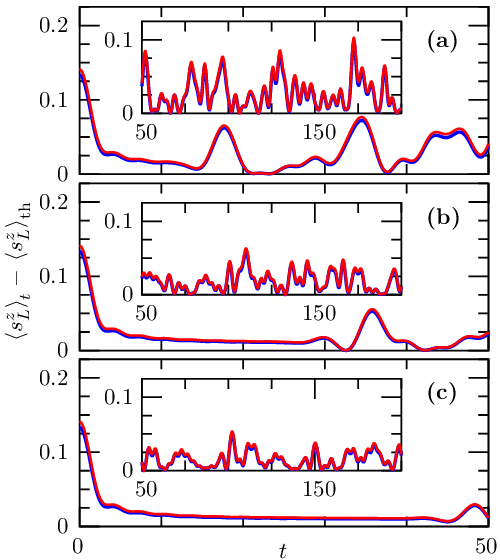}
\caption{
Red: numerically exact time-dependent expectation values
$\langle s^z_L\rangle_t - \langle s^z_L\rangle _{\rm th}$
for the same XX model 
as in Fig.~\ref{fig3}(b) with 
$L=8,\, 16,\,\textrm{and } 24$ in (a), (b), and (c), respectively.
Blue: corresponding analytical approximations from Eq.~(\ref{15}).
Main plots: initial relaxation behavior for $t\in[0,50]$.
(Insets) Long-time behavior for $t\in[50,200]$.
The blue curves are sometimes nearly covered by the red ones.
}
\label{fig5}
\end{figure}

The 
red
curves in Fig.~\ref{fig5} exemplify
the numerical expectation values 
$\langle A\rangle_t - \langle A\rangle _{\rm th}$
for $A=s_L^z$ and the same XX model as in
Fig.~\ref{fig3}(b).
Their very good agreement with the 
blue curves in Fig.~\ref{fig5} 
confirms once again (see also Sec.~\ref{s45})
that the approximation (\ref{15}) is very accurate, and works 
practically equally well for largely arbitrary system sizes $L$
and times $t$.
Likewise, we confirmed (not shown) that the agreement 
becomes even better for smaller values of $g\beta$,
as predicted below Eq.~(\ref{16}).

Another interesting observation in Fig.~\ref{fig5} is
that the curves seem to be (practically) independent of
$L$ as long as the time $t$ is smaller than about
$1.75\, L$ (in our present units), while exhibiting quite notable
and strongly $L$-dependent pseudorandom 
``fluctuations'' for even larger times $t$.
In other words, moderately large $L$ values
(such as in Fig.~\ref{fig5}) already emulate very 
well the behavior in the thermodynamic limit
$L\to\infty$ up to times
of about $1.75\, L$.
Indeed, our numerically exact results for $L=10^4$ would be 
indistinguishable for $t<1.75 L$ from the blue curves in Fig.~\ref{fig5}
and are therefore not shown.
The intuitive physical explanation is that the
system's behavior after a local quench 
at the right chain-end needs
some time (proportional to $L$) to ``notice'' that
the chain is of finite rather than infinite length.
A more detailed explanation and additional examples 
of these well-known general features can be found 
for instance in Ref.~\cite{rei25}.
 
The insets of Fig.~\ref{fig5} indicate
that the typical amplitudes of the above mentioned 
pseudorandom fluctuations for $t>1.75\, L$ 
decrease with increasing $L$.
(Incidentally, also the mean-value of those 
fluctuations depends on $L$, 
see also Fig.~\ref{fig3}.)
A much more detailed analytical and numerical 
investigation of this issue has been carried 
out in Ref.~ \cite{rei25} for a somewhat different 
model.
We found that largely the same observations
and conclusions are also recovered for our 
present model, and hence are not repeated 
here.
The main result is that the behavior
for a given, large but finite system size $L$ 
at very large times $t$ cannot be
easily related to the behavior at a fixed, 
large but finite time $t$  for very large 
systems sizes $L$.
Yet, if both $t$ and $L$ approach infinity,
the sequence of the limits becomes 
irrelevant.
Moreover, the behavior in this double-limit
can already be recovered in very good 
approximation for relatively small $L$
when focusing on times $t<1.75 L$.

We emphasize that our analytical demonstration of non-thermalization
in Sec.~\ref{s44} and its numerical confirmation in
Sec.~\ref{s45} remain entirely unaffected by all these 
quite subtle large-$t$ versus large-$L$ issues. 

\begin{figure}
\hspace*{-0.8cm}
\includegraphics[scale=0.95]{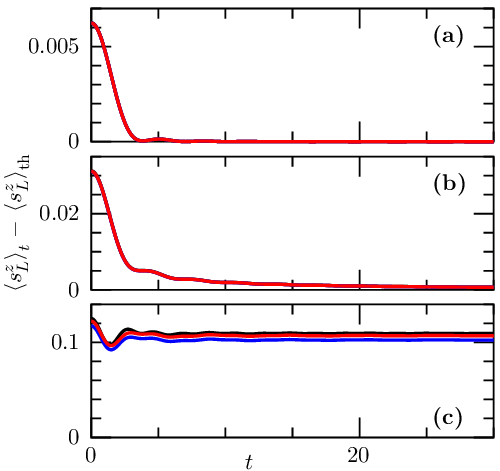}
\caption{
Red: numerically exact expectation values
{$\langle s^z_L\rangle_t - \langle s^z_L\rangle _{\rm th}$}
for the XX model from Eqs.~(\ref{1})-(\ref{11}) 
with parameter values
$J_\perp=1$, 
$J_z=0$,
$\nu=\alpha=L=10^4$,
$\beta=1/4$, 
$\gamma=0$
and 
$g=0.1,\, 0.5,\,{\textrm{and }} 2$ in (a), (b), and (c), respectively.
Blue: corresponding analytical predictions (\ref{15}). 
Black: corresponding approximations from 
Eqs.~(\ref{105}), (\ref{108}), and (\ref{109}) in (a), (b), and (c)
respectively.
The blue and black curves are (almost) covered by the red
ones in (a) and (b).
}
\label{fig6}
\end{figure}

Finally, we turn to the comparison of our 
analytical approximations from the previous 
section with numerical examples, see Fig.~\ref{fig6}.
Note that $p=0.2<p_c$ in (a),
$p=1\simeq p_c$ in (b),
and  $p=2>p_c $ in (c), 
where 
$p_c=(L+1)/L=1.0001$,
see also Eqs.~(\ref{17}) and (\ref{63}).
The 
black
curves in Fig.~\ref{fig6} exemplify the approximations
from Eqs.~(\ref{105})-(\ref{110}).
In order to fulfill the requirement (\ref{103}) in those approximations,
we have chosen in Fig.~\ref{fig6} a smaller value of $\beta$ than in
Fig.~\ref{fig5}.
While the agreement between the blue and black curves is very good in
(a) and (b), the deviations in (c) are somewhat larger.
The main reason seems to be that the condition $|p|\gg 1$ 
in Eq.~ (\ref{109}) is not
yet very well fulfilled for the choice of $p=2$ in Fig.~\ref{fig6}(c).
We numerically found that the agreement indeed becomes 
better upon further increasing $p$ (not shown). 
On the other hand, for larger $p$-values, 
correspondingly smaller $\beta$-values
must be chosen in order to still fulfill Eq.~(\ref{103}).
Moreover, upon increasing $p$ the overall temporal variations
become smaller and smaller, as can be most easily 
deduced from Eq.~ (\ref{109}).
In other words, the choice $p=2$ in Fig.~\ref{fig6}(c)
amounts to a compromise between 
a nearly trivial time-dependence and 
a breakdown of the analytical approximation
(\ref{109}).

Note that we restricted ourselves in Fig.~\ref{fig6} to the 
regime $t<1.75 \, L$ (see also the discussion above).
As expected, in the regime $t>1.75 \, L$ one finds 
(not shown) that the black curves will remain (practically) 
constant, 
while the red and blue curves will exhibit the usual 
pseudorandom fluctuations (see above), whose 
typical amplitudes are however already 
very small due to the quite large value of 
$L=10^4$.
As detailed in Ref.~\cite{rei25}, this is accompanied
by a breakdown of the approximation 
mentioned below Eq.~(\ref{104}).

Physically speaking, the case $p>p_c$ in Fig.~\ref{fig6}(c)
amounts to yet another example of non-thermalization
(nonzero long-time average).
Likewise, the case $p<p_c$ in Fig.~\ref{fig6}(a) gives rise
to a vanishing long-time average \cite{foot4}.
Finally, the case $p=p_c$ in Fig.~\ref{fig6}(b) may be viewed 
as a ``critical point,'' exhibiting 
a particularly slow relaxation towards zero.

\subsection{Signatures of localization}
\label{s47}

As detailed at the end of Sec.~\ref{s43}, for $|p|\geq p_c$ a so-called
localized mode (with index $k=L$) arises, whose characteristic signature
is that $U_{Ll}$ in Eq.~(\ref{89}) or Eq.~(\ref{91}) decreases (in modulus) with
increasing distance $\bar l:=L-l$ from the right chain-end
(exponentially for $|p|>p_c$ and algebraically for $|p|=p_c$).
Moreover, the appearance of such a localized mode is essential for
absence of thermalization, as observed at the end of Sec.~\ref{s442}.
It is however important to realize that 
this notion of localization
refers to the auxiliary
eigenvalue problem in and around Eq.~(\ref{40}).
The latter, in turn, arises by mapping the original 
XX model
(\ref{24}) by means of a Jordan-Wigner transformation 
to an equivalent model of free fermions, see Eq.~(\ref{28}).
Since the same kind of transformation will {\em not}
be possible for our later considered XXZ model, 
also such a notion of localization will be meaningless as it stands,
and thus needs to be adapted in some suitable way.

Therefore we first
address the question whether and how these localization 
features of the Jordan-Wigner transformed model 
manifest themselves already in the
original XX model
from Eq.~(\ref{24}).
Two particularly important such manifestations 
are obviously the absence of thermalization and the 
strong violation of the ETH
from Sec.~\ref{s44}.
But are there also some immediate signatures of 
``localization'' already in the original Hamiltonian (\ref{24})?

To begin with, let us consider
those eigenvectors $|\vec b\rangle$
of $H$ [see in and around Eq.~(\ref{31})]
for which the {general}
definition (\ref{30}) takes the particularly 
simple form of a ``one-particle excitation''
$| k \rangle := f_k^\dagger | 0 \rangle$
with an arbitrary but fixed $k\in\{1,...,L\}$.
Observing that $f^\dagger_k=\sum_{l=1}^L U^\ast_{kl} c_l^\dagger$ according to Eq.~(\ref{27}),
exploiting Eq.~(\ref{25}) to infer $c_l^\dagger=\sigma_l^+Z_l$,
and noting that $Z_l | 0 \rangle=| 0 \rangle$ [see below Eq.~(\ref{29})],
we thus can conclude 
that
\begin{eqnarray}
| k \rangle 
= \sum_{l=1}^L U^\ast_{kl}\,  \sigma^+_l | 0 \rangle
\ .
\label{111}
\end{eqnarray}
Recalling the definition of $| 0 \rangle$ below Eq.~(\ref{29}) 
it follows that  $\sigma^+_l | 0 \rangle$ may be viewed as a
``one-spin excitation'', i.e.,
a product state where the $l$-th spin points ``up'' and all other 
spins point ``down''.
Accordingly, {$| k \rangle$ in Eq.~(\ref{111}) thus} represents a linear 
combination of {all possible} one-spin excitations with 
coefficients $U^\ast_{kl}$.
Choosing $k=L$, the above mentioned localization properties
of $U_{Ll}$ then translate into a localization property of
$| L \rangle$ in the sense that one-spin excitations near
the right chain-end contribute with a much larger weight
$U_{Ll}$ than all the others.

Analogously, we may consider any eigenstate 
in Eq.~(\ref{30}) 
(apart from $| 0 \rangle$)
as some $n$-particle excitation of the general form
\begin{eqnarray}
|k_1,...,k_n\rangle :=  f_{k_1}^\dagger \cdots f_{k_n}^\dagger | 0 \rangle
\label{112}
\end{eqnarray}
with 
$k_1,...,k_n\in\{1,...,L\}$
and 
$k_1<k_2<...<k_n$.
Similarly as above, one thus can conclude 
(see Appendix A in Ref.~\cite{rei25}) that
\begin{eqnarray}
|k_1,...,k_n\rangle 
=  
\mbox{$\sum'$}
u(l_1,...,l_n)\,
\sigma^+_{l_1} \cdots \sigma^+_{l_n} | 0 \rangle
\, ,
\ \ \ \ \ \ 
\label{113}
\end{eqnarray}
where $\sum'$ indicates a sum over all  $l_1,...,l_n\in\{1,...,L\}$
with $l_1<l_2<...<l_n$ and thus consist of $\binom{L}{n}$ summands.
Furthermore,
\begin{eqnarray}
u(l_1,...,l_n) := \sum_\Pi 
\tau(\Pi({\bf l})) 
\, 
U^\ast_{k_1l_{\Pi(1)}} \cdots U^\ast_{k_nl_{\Pi(n)}}
\, , \ \ 
\label{114}
\end{eqnarray}
where the sum is meant to run over all permutations $\Pi$ of 
$\{1,...,n\}$ and thus consist of $n!$ summands,
and where $\tau({\bf l})$ is either $+1$ or $-1$,
but for the rest depends in a very complicated way on 
${\bf l}:=(l_1,...,l_n)$, ultimately
caused by the anti-commutation relations mentioned below Eq.~(\ref{25}).

Similarly as before, the last factors 
$\sigma^+_{l_1} \cdots \sigma^+_{l_n} | 0 \rangle$
in Eq.~(\ref{113}) may now be viewed as ``$n$-spin excitations.''
Moreover, 
one expects that the localization properties of $U_{Ll}$ 
must somehow enter the game if $k_n=L$.
However, the large number of summands in Eq.~(\ref{113})
and the complicated form of the coefficients in Eq.~(\ref{114}) 
make a more tangible physical interpretation quite difficult.
Therefore, we now turn to the signatures of localization
which do not manifest themselves directly in the system's energy
eigenstates, but rather in certain characteristic features
of some suitably chosen expectation values.

A first such signature arises for that specific eigenvector $|\vec b\rangle$
whose components are of the particular form $b_k=\delta_{kL}$ 
(and which thus coincides with the one-particle excitation $|L\rangle$
from above).
Namely,  its expectation values 
for the observables $A=s_\alpha^z+1/2$ [see also Eq.~(\ref{11})]
may be viewed as being localized near the right chain-end (at $\alpha=L$),
as can be inferred from Eq.~(\ref{e1}) in Appendix \ref{appE}.
However, this only concerns 
the above very special energy eigenstate $|\vec b\rangle$, 
whose energy is moreover located somewhere 
``in the middle of the spectrum of 
$H$'' according to Eq.~(\ref{32}). 
(In particular, it is far from the ground state.)
For more general eigenstates $|\vec b\rangle$,
i.e., many-particle excitations with $b_k=1$ not only for $k=L$, 
the traces of the 
localization properties of $U_{Ll}$ become less clean due to 
the additional ``pseudorandom'' summands 
$|U_{kl}|^2$ with $k<L$, 
see also Eqs.~(\ref{67}), (\ref{93}), and (\ref{e1}).

A second signature consist in the fact that the 
strong violation of the ETH
is most pronounced for small $\bar \alpha$ values, i.e., near the right chain-end,
as discussed in Appendix \ref{appE} around Eq.~(\ref{e4}).
A third signature arises by observing
that the absence of thermalization from 
Sec.~\ref{s44} may again be viewed as being localized near the right 
chain-end according to Eqs.~(\ref{100}) and (\ref{101}).
A quantitative numerical example is provided in Fig.~\ref{fig4}.

Note that the first two above signatures
are of rather theoretical character in the sense that 
properties of single {energy} eigenstates are
quite difficult to observe in an experimental many-body system.
The third one is experimentally unproblematic but
is of a rather ``indirect'' character
in the sense 
that it admits little insight
into the basic physical mechanism behind it.

In any case, the most fundamental signature of localization, 
namely the existence of a localized mode $U_{Ll}$, 
seems hardly seizable beyond the very special class of systems 
which can be mapped to an equivalent model of free fermions.
On the other hand, the second and third above signatures
are independent of whether or not such a mapping is available.
Moreover, they are in fact very closely related to each other
via our key relations (\ref{20}) and (\ref{22}).

Finally, it also seems noteworthy that the thermal 
expectation values $\langle s_\alpha^z \rangle_{\rm th}$ 
(see also Appendix \ref{appG})
do {\em not} exhibit any signature of the onset of localization 
when $p$ exceeds the critical value $p_c$.

\section{The XXZ Model with an end-impurity}
\label{s5}

The XXZ model may be viewed as a generalization of the XX model
from Eq.~(\ref{24}) when abandoning the requirement $J_z=0$ in Eq.~(\ref{23}),
i.e., the Hamiltonian is now of the form
\begin{eqnarray}
H= \sum_{l=1}^{L-1}  \left(s^x_{l+1}s^x_l + s^y_{l+1}s^y_l +J_z  s^z_{l+1}s^z_l \right)
+
 \frac{p}{2} s_\nu^z
 \ ,
 \ \ \ 
\label{115}
\end{eqnarray}
see also Eqs.~(\ref{1}), (\ref{2}), (\ref{5}), (\ref{8}), and (\ref{17}).
Moreover, we focus again on the case of an end-impurity
as specified in and around Eq.~(\ref{42}).
The main goal of this section is to show that qualitatively similar 
results as for the XX model are also recovered for the XXZ model.

To the best of our knowledge, a pertinent extension of the
general analytical framework 
from Sec. \ref{s3} is not available for our present XXZ model
with an impurity in Eq.~(\ref{115}).
Hence, we confine ourselves to purely 
numerical explorations, analogous to those in 
Secs.~\ref{s45} and \ref{s462}.

\begin{figure}
\hspace*{-0.8cm}
\includegraphics[scale=0.95]{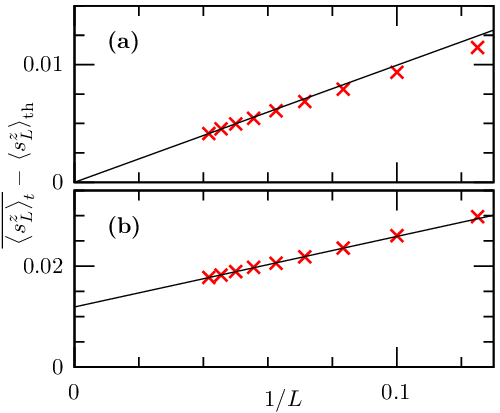}
\caption{
Same as Fig.~\ref{fig3} but now for the corresponding XXZ model with
$J_z=0.3$ in Eq.~(\ref{115}).
}
\label{fig7}
\end{figure}

In Fig.~\ref{fig7} we show the same type of numerically exact 
time-averaged expectation values as in Fig.~\ref{fig3},
but now for the XXZ model (\ref{115}) with $J_z=0.3$
instead of the XX model (corresponding to $J_z=0$).
Moreover, the numerical results for $L=10^4$ 
(leftmost crosses in Fig.~\ref{fig3})
as well as analytical approximations 
(blue crosses and green bar in Fig.~\ref{fig3})
are no longer available in Fig.~\ref{fig7} 
(see also previous paragraph and Sec.~\ref{s45}).
Apart from that, the numerical findings
in Fig.~\ref{fig7} are indeed quite similar to those
in Fig.~\ref{fig3}.
Finally, we recall the remarks in Sec. \ref{s45}
(paragraphs 2 and 4 therein)  concerning the reliability
of our large-$L$ extrapolations by means of the black lines.
Altogether, Fig.~\ref{fig7}(b) thus provides once again very convincing 
numerical evidence that the system does not exhibit thermalization
after a local quantum quench, see also the more detailed 
discussion of this issue in Sec. \ref{s45}.

\begin{figure}
\hspace*{-0.8cm}
\includegraphics[scale=0.95]{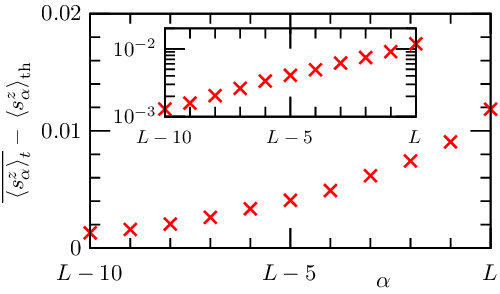}
\caption{
Same as Fig.~\ref{fig4} but now for the corresponding XXZ model with
$J_z=0.3$ in Eq.~(\ref{115}).
}
\label{fig8}
\end{figure}

As usual (see Sec.~\ref{s24}), once the absence
of thermalization has been established for a given system, 
we can conclude that the system also must exhibit a 
strong violation of the ETH.
Moreover, it seems reasonable to expect that the second 
and third main signature of localization from Sec.~\ref{s47} 
will be handed down from a nonthermalizing XX to the
corresponding XXZ model as well.
A numerical confirmation is provided by Fig.~\ref{fig8}.

\begin{figure}
\hspace*{-0.8cm}
\includegraphics[scale=0.95]{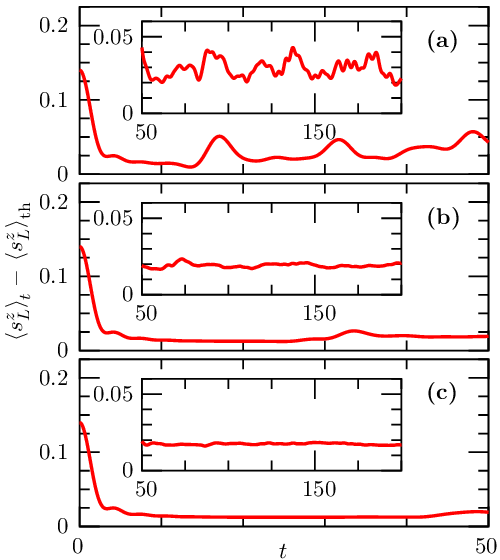}
\caption{
Same as Fig.~\ref{fig5} but now for the corresponding XXZ model with
$J_z=0.3$ in Eq.~(\ref{115}).
}
\label{fig9}
\end{figure}

Likewise, Fig.~\ref{fig9} is the counterpart of Fig.~\ref{fig5},
but now with $J_z=0.3$ instead of $J_z=0$,
and without any analytical approximations (blue and black curves in Fig.~\ref{fig5}).
For the rest, the numerically exact results (red curves) are again
qualitatively and even quantitatively quite similar in the two figures
for small-to moderate times with $t\leq 1.75 L$, see also Sec.~\ref{s462}.
On the other hand, the pseudorandom fluctuations for $t>1.75 L$
in Fig.~\ref{fig9} are still qualitatively similar as in Fig.~\ref{fig5},
but quantitatively much weaker \cite{foot6}.
Such quantitative differences have also been reported in related 
previous investigations.
Without going into the details, these differences 
can be traced back to the fact that
we are dealing with so-called interacting and noninteracting
integrable models in the case of our present 
XXZ and XX models, respectively \cite{spo18}.

A more detailed and systematic numerical exploration of how all 
these various features of the XXZ model depend on $J_z$,
and possibly also on $g$, $\gamma$, $L$, and $\beta$, 
is a quite interesting 
issue in its own right, which goes beyond the 
scope of our present work (see also Sec.~\ref{s7}).

\section{Inner impurities}
\label{s6}

\subsection{The XX model with an inner impurity}
\label{s61}

For the sake of simplicity, we will mainly 
focus on {\em odd chain lengths} $L$
and on impurities and perturbations $s_\nu^z$
in Eqs.~(\ref{2}) and (\ref{8}) [and thus in Eq.~(\ref{24})] with
\begin{eqnarray}
\nu=\frac{L+1}{2}
\ ,
\label{116}
\end{eqnarray}
i.e., we may imagine the impurity to be located exactly in the middle 
of the chain (henceforth denoted as central impurity),
see also below Eqs.~(\ref{8}) and (\ref{42}).
Generalizations to even $L$ and arbitrary inner
impurities 
are in principle straightforward,
but require more lengthy calculations and case differentiations,
and will therefore only be briefly 
summarized at the end of this section.
For the rest, 
the detailed calculations are largely analogous to those in the 
previous Sec.~\ref{s4}, and are therefore relegated to Appendix \ref{appH}.


Similarly as in Sec.~\ref{s43}, 
the first main implication of those analytical calculations
in Appendix \ref{appH} is 
that there are $L-1$ delocalized modes and one localized 
mode whenever $|p|$ exceeds the critical value 
\begin{eqnarray}
p_c := \frac{2}{\nu}=\frac{4}{L+1}
\ ,
\label{117}
\end{eqnarray}
where we utilized Eq.~(\ref{116}) in the last step.
Furthermore, one thus can infer -- similarly 
as in Sec.~\ref{s44} -- that the system exhibits
no thermalization after a local quench 
whenever $|p|>p_c$.
Quantitatively, by focusing on the most interesting regime 
$1/L\ll |p| \ll 1$,
the counterpart of Eqs.~(\ref{100}) and (\ref{101}) now assumes 
the form
\begin{eqnarray}
\overline{\langle s_\alpha^z \rangle_t} - \langle s_\alpha^z\rangle _{\!\rm th}
= g\beta \frac{p^2\, e^{-|p| |\nu-\alpha|}}{16 \, \cosh^2(\beta/2)}
\ .
\label{118}
\end{eqnarray}
Likewise, a strong violation of the ETH 
{immediately follows}
by similar arguments as in Sec.~\ref{s44},
and
the concomitant signatures of localization in the system's 
Hamiltonian (\ref{24}) are analogous to those in Sec.~\ref{s47}.

\begin{figure}
\hspace*{-0.8cm}
\includegraphics[scale=0.95]{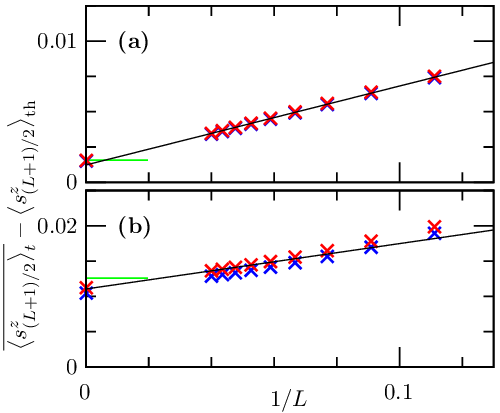}
\caption{
Same as in Fig.~\ref{fig3} but now for
$\nu=\alpha=(L+1)/2$ (central impurity),
$g=0.2$ in (a),
$g=0.4$ in (b),
and 
$L=10001,25,23,...,11,9$.
The green bars indicate the analytical approximation
for $L\to\infty$ from Eqs. (\ref{118}).
}
\label{fig10}
\end{figure}

The 
most important
difference compared to the end-impurities considered
in Sec.~\ref{s4} is that in the present case of a central impurity
the critical value in Eq.~(\ref{117}) 
approaches zero for large $L$.
In particular, a quench in the form of an arbitrarily weak 
central impurity thus always gives rise to nonthermalization 
in a sufficiently long spin-chain
[see also the discussion below Eq.~\ref{101})].
We also remark that the choice $\gamma=0$ is 
particularly natural in our present case [see below Eq.~(\ref{8})],
and implies together with Eq.~(\ref{17}) that the right-hand 
side of Eq~(\ref{118}) scales as $g^3$.
In other words, {the observable manifestations of}
nonthermalization are quantitatively 
quite weak for small $g$.

As an outlook we mention that for more general
impurities and perturbations $s_\nu^z$ in Eqs.~(\ref{2}) and (\ref{8})
[and thus in Eq.~(\ref{24})] with an arbitrary location $\nu\in\{1,...,L\}$
one finds (after similar but more involved calculations)
that the critical impurity strength is given by
\begin{eqnarray}
p_c = \frac{1}{\nu} + \frac{1}{L+1-\nu}
\ ,
\label{119}
\end{eqnarray}
and that the same conclusions as in the previous paragraph
apply whenever $|p|$ exceeds this critical value.
In particular, one readily recovers Eqs.~(\ref{63}) and (\ref{117})
in the special cases from Eqs.~(\ref{42}) and (\ref{116}), respectively.
Moreover, $p_c$ in Eq.~(\ref{119})
approaches zero whenever 
{the impurity site $\nu$ scales with the system size $L$
in such a way that both} $\nu$ and
$L-\nu$ diverge for large $L$.
On the other hand, {whenever either} $\nu$ or $L-\nu$ 
remains finite as $L\to\infty$,
the critical value in Eq.~(\ref{119})
approaches a non-vanishing limit.
Pictorially speaking, the impurity 
{may be said to be} located in the
``bulk'' of the spin-chain in the first case and in one
of the ``end-regions'' in the second case.

A numerical illustration of these predictions is presented in Fig.~\ref{fig10}
(see also Secs.~\ref{s45} and \ref{s5}).
As required at the beginning of this section, odd system sizes $L$
have been chosen,
and the values of $g=p/2$ [see Eq.~(\ref{63})] are still relatively small.
Overall, the main features and the discussion are largely the same as
in the case of an end-impurity, see Fig.~\ref{fig3} and Sec.~\ref{s45},
apart from one very prominent exception:
We now observe absence of thermalization
both in (a) and (b), 
in agreement with our above analytical predictions
[see in particular below Eq.~(\ref{118})].
Note that we have purposefully chosen in Fig.~\ref{fig10}
even smaller values of $g=p/2$ than in Fig.~\ref{fig3} 
in order to highlight 
this key difference between the two cases 
even more strikingly.
The fact that the deviations of the
green bars from the leftmost blue crosses 
are quantitatively larger in Fig.~\ref{fig10} 
than in Fig.~\ref{fig3} can be traced back to
the extra assumption $|p|\ll 1$ mentioned
above Eq.~(\ref{118}).
We also remark that our numerical results for even system sizes $L$
and $\nu=\alpha=L/2$ (not shown) are found to be very similar to those
in Fig.~\ref{fig10}, i.e., the restriction to odd $L$ and exactly centralized 
impurities in our above analytics indeed seems to be of minor importance.

\begin{figure}
\hspace*{-0.8cm}
\includegraphics[scale=0.95]{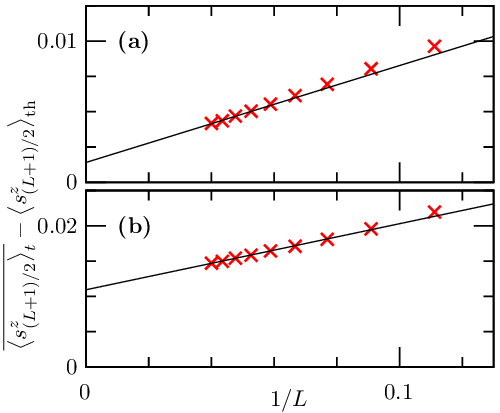}
\caption{
Same as in Fig.~\ref{fig10}, except that periodic instead of 
open boundary conditions have been employed.
For more details, see the main text.
}
\label{fig11}
\end{figure}

\subsubsection{Periodic boundary conditions}
\label{s611}

A further, quite minor and natural extension is to consider periodic
instead of open boundary conditions.
Formally, this amounts to replacing the upper summation limit
$L-1$ in Eq.~(\ref{24}) by $L$ and setting $s^a_{L+1}:=s^a_1$.
Obviously, in such a case the actual position $\nu$ of the impurity in
Eq.~(\ref{24}) becomes irrelevant.
Moreover, it seems reasonable to expect
that the system behaves similarly as an 
open system with a central impurity. 
A 
confirmation of this expectation follows
upon comparing the numerically exact findings in Fig.~\ref{fig11} 
with those in Fig.~\ref{fig10}.
We remark that also an analytical confirmation is in principle
possible by a pertinent generalization of the calculations
in Secs.~\ref{s3} and \ref{s4} (see also Appendix \ref{appH}).
We omit them here in view of the already quite considerable 
length of our paper as it stands.
Accordingly, also the leftmost red crosses, the blue crosses, 
and the green bars in Fig.~\ref{fig10} are no longer available.

\begin{figure}
\hspace*{-0.8cm}
\includegraphics[scale=0.95]{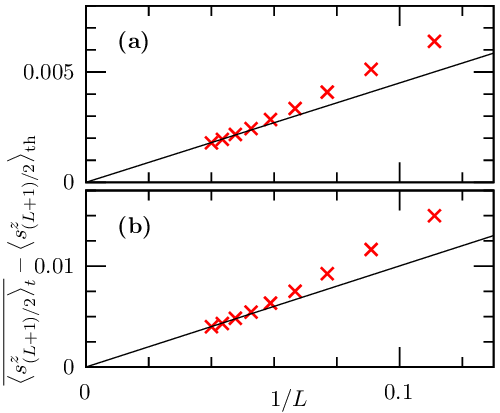}
\caption{
Same as Fig.~\ref{fig10} but now for the corresponding XXZ model with
$J_z=0.3$ in Eq.~(\ref{115}).
For more details see main text.
}
\label{fig12}
\end{figure}

\subsection{The XXZ model with a central impurity}
\label{s62}

As in Sec.~\ref{s5}, we consider the XXZ model from Eq.~(\ref{115}),
and we confine ourselves to numerical explorations.
As in the previous Sec.~\ref{s61}, we furthermore
focus on odd chain lengths $L$ and a central impurity
according to Eq.~(\ref{116}).

In Fig.~\ref{fig12} we show the same type of numerically exact 
time-averaged expectation values as in Fig.~\ref{fig10}, but now for the 
XXZ model (\ref{115}) with $J_z = 0.3$ instead of the XX model 
(corresponding to $J_z = 0$),
see also Fig.~\ref{fig7} and its discussion in Sec.~\ref{s5}.
Obviously, there is a very striking difference between Figs.~\ref{fig10}
and \ref{fig12}.
While the XX model exhibits absence of thermalization in
both Figs.~\ref{fig10}(a) and \ref{fig10}(b), the same is no longer 
the case for the XXZ model both in Figs.~\ref{fig12}(a) and \ref{fig12}(b)
\cite{foot4}.

Analogous results for the same or closely related
models have been reported previously for instance in 
Refs.~\cite{san04,tor14,bre20,pan20,san20,bar09}. 
In this sense,
the example from Fig.~\ref{fig12} has only been
included here in order to make 
the paper self-contained, while the observation of 
thermalization is in itself nothing new.
Likewise, the basic physical explanation of this
finding has been previously established 
and goes as follows:
As numerically shown in Refs.~\cite{san04,tor14,bre20,pan20,san20,bar09},
the well-known integrability of the XXZ model without any impurity
is broken by a central impurity.
More precisely speaking, it has been numerically found 
that the XXZ model with a central impurity satisfies the
conventional (strong) ETH (see also Sec.~\ref{s1})
and that its energy spectrum exhibits a so-called Wigner-Dyson 
level statistics, widely considered as a characteristic
signature of ``quantum chaos'' and thus of 
non-integrability.
Similar features are numerically 
seen 
also for XXZ models with periodic boundary conditions
in the presence of an impurity.
Moreover, the numerics in 
Refs.~\cite{san04,tor14,bre20,pan20,san20,bar09}
suggests that the integrability-breaking effect will
persist (in the thermodynamic limit)
even for arbitrary weak (but finite) impurities,
and likewise for arbitrarily strong (but finite) impurities.

Incidentally, the XXZ model with open boundary conditions
is well-known to remain integrable in the presence of an
end-impurity \cite{skl88,alc87,bei13},
and similarly for the XX model with any (local) impurity
[see below Eq.~(\ref{24})].
On the other hand, the XXZ model with periodic boundary 
conditions is provably nonintegrable in the presence of an 
impurity \cite{hok25}.
Finally, it is reasonable to expect that a central impurity
in a system with open boundary conditions results in similar
properties as an impurity in a system with periodic boundary 
conditions.
All our numerical findings are, of course, in perfect 
agreement with those general 
predictions.

\section{Conclusions}
\label{s7}

Our main finding is that already a local, and in principle arbitrarily 
weak quantum quench may be sufficient to 
prohibit a many-body system from exhibiting thermalization.
Specifically, we explored the behavior of some basic local observables 
in XX- and XXZ-spin-chain models,
initiated at thermal equilibrium (Gibbs state), and then
suddenly switching on (or slightly changing) a single-spin impurity
either at one of the chain's ends or in the middle of the chain.
We moreover showed that whenever a given 
system and observable exhibit such an 
absence of (re)thermalization,
this observable also 
exhibits a 
strong violation of the ETH,
meaning that it does not even
obey the weak ETH \cite{weth}.
Incidentally, this is somewhat reminiscent of the
connection between the absence of thermalization 
after a global quench and the violation
of the conventional (strong) ETH
in integrable systems,
which has been 
previously observed 
in numerous numerical 
examples
(see also Sec.~\ref{s1}).

In the case of the XX model, a fully analytical solution 
of the problem was possible.
After employing a standard Jordan-Wigner transformation, 
the main 
achievement
was to determine 
the eigenmodes and eigenvalues of the transformed 
Hamiltonian in the presence of an impurity in the 
original model, see Secs.~\ref{s41}, \ref{s42}, and Appendix~\ref{appH}.
We remark that somewhat similar results 
have been previously obtained for the transverse field Ising model 
with an end-impurity in Ref.~\cite{fra16}, but that no tangible 
connection between the two cases can be established
since they differ in too many important respects.
Once these eigenmodes and eigenvalues were available,
we took advantage of the recently developed general 
formalism from Refs.~\cite{rei24,rei25} to arrive 
at explicit expressions 
for the time-dependent expectation values.
Along these lines we analytically demonstrated 
absence of thermalization
in the case of an end-impurity, provided its strength
exceeds some finite (non-zero) threshold value.
On the other hand, in the case of an impurity in the
middle of the chain, the corresponding threshold 
value was found to be zero, i.e., an arbitrarily weak 
impurity is sufficient for nonthermalization.
More precisely speaking, in both cases
the threshold value refers to the 
postquench impurity strength, 
while the difference between the pre- and post-quench 
impurity strengths is even admitted to assume arbitrarily 
small (non-zero) values.
Moreover, our analytical approximations 
actually become asymptotically exact 
when this difference 
becomes small.

Essentially, the absence of thermalization was found to be 
rooted in the appearance of a localized mode in the 
Jordan-Wigner transformed model.
While this notion of localization -- originally introduced 
in Ref.~\cite{mur19} -- is of a quite formal character
with respect to the original XX model in which we are
actually interested, we found that its main
directly observable manifestation is a localization 
of nonthermalization, meaning that a local quench
may permanently keep the system
out of equilibrium near the location of the quench,
but not far away from it.

Turning to the XXZ model, analogous explorations
could only be accomplished by numerical means.
In the case of an end-impurity, we recovered
a quite similar behavior to that of the XX model,
but not anymore in the case of an impurity
in the middle of the chain.
As detailed in Sec.~\ref{s62},
the basic reason seems to be that the XX model
remains integrable in the presence of impurities
at the end as well as in the middle of the chain,
while the XXZ model remains integrable for end-impurities,
but not anymore for impurities in the middle of the chain.

Straightforward generalizations of our analytical explorations
include periodic instead of open boundary conditions
(see also Sec.~\ref{s611}), the XY instead of the XX model,
or more general types of impurities and observables,
to name but a few.
Working out all the technical details is, however, 
quite a tedious task which goes beyond the scope of
our present work.
The same applies to a more detailed numerical
exploration of the XXZ model (see Sec.~\ref{s6}),
and likewise for the XYZ model,
which may be considered as a generalization 
of either the XY or the XXZ model.

As explained at the end of Sec.~\ref{s24}, 
in cases where a system does {\em not} 
exhibit our present signature of 
non-thermalization,
this is not yet sufficient to 
answer the question of whether or not the
system actually does thermalize.
Instead, a solution of 
this task has been worked out 
in our companion paper \cite{rei25}.
Most importantly, the occurrence of thermalization
after a local quench has been rigorously established 
therein for the same XX model as in 
Secs.~\ref{s4} and \ref{s6},
however under the extra condition that $g=-\gamma$ 
and thus $p=0$ in Eq.~(\ref{17}), meaning that the
prequench system in Eq.~(\ref{2}) exhibits an impurity,
which is switched-off at time $t=0$, resulting
in a postquench system in Eq.~(\ref{5}) 
without any impurity \cite{foot5}.
Together with the findings of our present work this implies that
small details such as whether 
an impurity is switched on or off,
may be of great importance 
with respect to the question of 
whether a system thermalizes 
or not.
Moreover, to obtain truly reliable answers, 
a careful analytical 
exploration of any given specific model
may be practically indispensable 
in general \cite{shr21}.

We finally note that there are also a number of
similarities as well as differences between our
present investigations and the topic of Hilbert space
fragmentation \cite{hsf}.
Notable similarities are that autocorrelation functions,
localization, and 
a strong violation of the ETH
play important 
roles in both cases.
An essential difference regarding the issue of (non)thermalization
is that in the context of Hilbert space fragmentation the focus
is predominantly on global quenches, or on the infinite temperature 
limit and some not explicitly specified nonequilibrium initial states.
More detailed investigations along these lines are the subject
of our ongoing work.

\begin{acknowledgments}
This work was supported by the 
Deutsche Forschungsgemeinschaft (DFG, German Research Foundation)
under Grant No. 355031190 
within the Research Unit FOR 2692
and under Grant No. 
502254252, 
and
by the Paderborn Center for Parallel 
Computing (PC$^2$) within the project 
HPC-PRF-UBI2.
\end{acknowledgments}

\section*{Data availability}
Source data for all the figures are available at Ref. \cite{data}.

\vspace*{0.5cm}
\appendix
\section{Same $\beta$ in Eqs. (\ref{3}) and (\ref{10})}
\label{appA}

In this Appendix we provide a detailed justification of why
$\beta$ can be chosen equal in Eqs. (\ref{3}) and (\ref{10})
for sufficiently large $L$ [see also the discussion above Eq.~(\ref{9})].

We adopt the same definition of the 
system's initial state $\rho(0)$ as in Eq. (\ref{3})
with an arbitrary but fixed value of $\beta$.
However, instead of $\rho_{can}$ in Eq. (\ref{10}), 
we now consider more general canonical ensembles of the form
\begin{eqnarray}
\tilde \rho_{can}(\tilde\beta) &:=&e^{-\tilde{\beta} H}/\tr\{e^{-\tilde{\beta} H} \}
\ ,
\label{a1}
\end{eqnarray}
where $\tilde\beta$ is for the moment still arbitrary.
According to textbook statistical mechanics, 
given a system with Hamiltonian $H$ and some initial state $\rho(0)$,
the pertinent canonical ensemble is of the general form (\ref{a1}).
Moreover, the value of $\tilde\beta$ must obviously be chosen such 
that the system's energy (which is a conserved quantity) 
is correctly reproduced.
Denoting this particular $\tilde\beta$ value as $\beta'$, 
and employing the definition
\begin{eqnarray}
\Delta E(\tilde{\beta})
&:=&
\tr \{\rho(0)H\}-\tr\{\tilde \rho_{can}(\tilde\beta) H\}
\ ,
\label{a2}
\end{eqnarray}
we thus require that $\beta'$ satisfies
\begin{eqnarray}
\Delta E(\beta')=0
\ .
\label{a3}
\end{eqnarray}
Setting
\begin{eqnarray}
\delta\beta
&:=&
\beta' - \beta
\ ,
\label{a4}
\end{eqnarray}
the objective in the remainder of this appendix is 
to show that 
\begin{eqnarray}
\delta\beta=\mathcal{O}(1/L)
\ ,
\label{a5}
\end{eqnarray}
which is equivalent to 
our original claim
at the beginning of this Appendix.

More precisely speaking, for our general XXZ model 
of the form (\ref{1}), (\ref{2}), (\ref{5}), and (\ref{8}),
we will derive (\ref{a5}) under the physically very reasonable extra 
assumption that the system's heat capacity is extensive.
Moreover, for our XX model from (\ref{24}) we will 
derive (\ref{a4}) even without this extra assumption.

To begin with, 
let us 
define a set of auxiliary Hamiltonians $H_r$ and
concomitant canonical ensembles $\rho_r$, 
\begin{eqnarray}
H_r&:=&H_0+r(H-H_0)
\ ,
\label{a6}
\\
\rho_r&:=&e^{-\beta H_r}/Z_r
\ ,
\label{a7}
\\
Z_r&:=&\tr\{e^{-\beta H_r}\}
\ ,
\label{a8}
\end{eqnarray}
where $r\in [0,1]$ is a parameter.
Recalling the basic relations
$-\frac{\partial}{\partial\beta}\ln Z_r=\tr\{\rho_rH_r\}$
and $H=H_0+gV$ [see Eq.~(\ref{5})], we can 
conclude from Eq. (\ref{a2}) that
\begin{eqnarray}
\Delta E(\beta)=g\tr\{\rho(0)V\}+\frac{\partial}{\partial\beta}\ln\frac{Z_1}{Z_0}
\ .
\label{a9}
\end{eqnarray}
Furthermore, by exploiting Eqs. (C.9) and (C.11) in the supplemental 
material of Ref.~ \cite{far17} and
the cyclic invariance of the trace one can infer the relation
\begin{eqnarray}
\ln\frac{Z_1}{Z_0}=-g\beta\int_0^1\textrm{d}r\tr\{\rho_rV\}
\ ,
\label{a10}
\end{eqnarray}
yielding with (\ref{a9}) the result
\begin{eqnarray}
\Delta E(\beta)&=&g[C_1(\beta)-C_2(\beta)- \beta C_3(\beta)]
\ ,
\label{a11}
\\
C_1(\beta)&:=&\tr\{\rho(0)V\}
\ ,
\label{a12}
\\
C_2(\beta)&:=&\int_0^1\textrm{d}r \tr\{\rho_rV\}
\ ,
\label{a13}
\\
C_3(\beta)&:=&\int_0^1\textrm{d}r\frac{\partial}{\partial\beta} \tr\{\rho_rV\}
\ .
\label{a14}
\end{eqnarray}
Observing that $\left|\tr\{\rho_rV\}\right|\leq \opnorm{V}$ 
for all $r$ (where $\opnorm{\cdot}$ denotes the operator norm, see also below Eq. (\ref{16})),
and utilizing the abbreviations
\begin{eqnarray}
v&:=& \opnorm{V}
\ ,
\label{a15}
\\
D&:=& \frac{1}{v} \max_{0\leq r\leq 1}
\left|\frac{\partial}{\partial\beta}\tr\{\rho_rV\}\right|
\ ,
\label{a16}
\end{eqnarray}
we can upper bound the modulus of the three quantities in Eqs. (\ref{a12})-(\ref{a14}) and 
then conclude with Eq. (\ref{a11}) that
\begin{eqnarray}
\left|\Delta E(\beta)\right|\leq |g|v(2+ |\beta| D)
\ .
\label{a17}
\end{eqnarray}

Next, we can infer from (\ref{a3}) that
\begin{eqnarray}
\Delta E(\beta)
&=&
\int_{\beta'}^{\beta}\textrm{d}\tilde\beta \, W(\tilde\beta)
\ ,
\label{a18}
\\
W(\tilde \beta) 
&:=&
\frac{\partial}{\partial\tilde\beta}\Delta E(\tilde \beta)
\ ,
\label{a19}
\end{eqnarray}
and with (\ref{a2}) that
\begin{eqnarray}
W(\tilde\beta) 
&=&
-\frac{\partial}{\partial\tilde{\beta}}\tr\{\tilde{\rho}_{can}(\tilde\beta)\, H\}
\ .
\label{a20}
\end{eqnarray}
Invoking (\ref{a4}), we can evaluate (\ref{a18})
 by the mean-value theorem and introduce the result into (\ref{a17}), implying
\begin{eqnarray}
|\delta\beta|\leq \frac{|g|v(2+ | \beta | D)}{|W(\beta_m)|}
\ ,
\label{a21}
\end{eqnarray}
for some $\beta_m$ with the property 
$\beta-|\delta\beta|\leq\beta_m\leq\beta+|\delta\beta|$.

\subsection{Evaluation of $D$ and $W(\beta_m)$}
\label{appA1}

Let us first remark that the derivation of (\ref{a21})
is still valid for entirely general Hamiltonians $H$ and $H_0$.
Furthermore, our 
final result (\ref{a5}) follows
immediately from (\ref{a21}) under the following three
conditions:
(i) $v$ in (\ref{a15}) remains bounded for $L\to\infty$,
(ii) $D$ in (\ref{a16}) remains bounded for $L\to\infty$,
(iii) $|W(\beta_m)|$ in (\ref{a19}) grows asymptotically linearly with $L$.

Condition (i) is trivially fulfilled for any local operator 
$V$. In particular, for our standard example 
from Eq. (\ref{8}) one finds that
\begin{eqnarray}
v=\opnorm{s^z_{\nu}}
=1/2
\ ,
\label{a22}
\end{eqnarray}
see also below Eq. (\ref{16}).
Conditions (ii) and (iii) are in general more difficult
to verify mathematically, but physically speaking it seems
very reasonable to expect that they will be fulfilled in 
many cases.
For instance, 
the quantity $W(\beta_m)$ in Eq. (\ref{a20})
can be immediately related to the system's heat 
capacity.
Hence, it is expected to be
an extensive quantity in many cases, including
our XXZ model of the form (\ref{1}), (\ref{2}), (\ref{5}), and (\ref{8}).
Similarly, (ii) is expected to hold as long as there 
is no (first-order) phase transition at inverse 
temperature $\beta$ and with an order parameter 
related to $V$, which again can be excluded for 
our present one-dimensional models.

Moreover, in cases where the Hamiltonian $H_r$ is given 
by a sum of local operators of the general structure $H_r=\sum_l h_l$
(which applies for instance for our present XXZ model),
one readily verifies that
\begin{eqnarray}
\frac{\partial}{\partial\beta} \tr\{\rho_r V\}
= - \sum_l  G (V,h_l)
\ ,
\label{a23}
\end{eqnarray}
where
\begin{eqnarray}
G(A,B) :=
\tr\{\rho_r A B \}-\tr\{\rho_r A\}-\tr\{\rho_r B \}
\ \ \ 
\label{a24}
\end{eqnarray}
is the correlation of
the operators $A$ and $B$.
For the canonical ensembles $\rho_r$ and local perturbations $V$
we are considering here, one can furthermore
show that the correlations $G(V,h_l)$
exhibit a so-called cluster decomposition property 
\cite{cdp}, 
from which one
can deduce that the sum in (\ref{a23}) can be upper bounded
(in modulus) by a finite value when $L\to\infty$.
Note that this argument applies separately to any given
parameter value $r$, hence it also applies to that $r$-value
which maximises the right-hand side  of (\ref{a16})
in the limit $L\to\infty$.
Incidentally, one might think that similar arguments may
also be applicable with respect to the above condition (iii).
The problem is that, in contrast to (ii), it is necessary to
establish some suitable kind of {\em lower} bound in order 
to guarantee (iii).

Finally, we provide a more rigorous justification of (ii) and (iii)
for the XX model within the setup considered in Secs.~\ref{s3}
and \ref{s61}.
Regarding (ii), we can identify $\tr\{\rho_rV\}$ with the left-hand 
side of Eq.~(\ref{g1}) with $\alpha=\nu$, while the $r$-dependence of $\tr\{\rho_rV\}$ 
is buried in the quantities $U_{k\nu}$ and $E_k$ 
on the right-hand side of Eq.~(\ref{g1}). 
Utilizing $\frac{\textrm{d}}{\textrm{d}x}\tanh(x)=1/\cosh^2(x)$
one thus obtains
\begin{eqnarray}
\frac{\partial}{\partial\beta}\tr\{\rho_rV\}=-\frac{1}{4}\sum_{k=1}^L|U_{k\nu}|^2\frac{E_k}{\cosh^2(\beta E_k/2)}
. \ \ \ 
\label{a25}
\end{eqnarray}
Furthermore, one can readily infer from the discussions in Sec.~\ref{s42} and 
Appendix \ref{appH} that
\begin{eqnarray}
|E_k|\leq h(p)
\label{a26}
\end{eqnarray}
for all admissible $k$ and $L$, where $h(p)$ is an $L$-independent
function with $h(0)=1$ and $h(p)\approx\mathcal{O}(|p|)$ for $p\gg1$. 
Recalling that $\sum_{k=1}^L |U_{kl}|^2 =1$ (see below Eq. (\ref{39})) 
and that $\cosh(x)\geq1$, we thus can conclude that
\begin{eqnarray}
\left|\frac{\partial}{\partial\beta}\tr\{\rho_rV\}\right|\leq \frac{h(p)}{4}
\label{a27}
\end{eqnarray}
independent of $r$, and with (\ref{a17}) that
\begin{eqnarray}
D\leq\frac{h(p)}{4}
\label{a28}
\end{eqnarray}
independent of $L$,
i.e., condition (ii) is fulfilled.

Turning to (iii),
our first observation is that the pertinent quantity $|W(\beta)|$ can be
identified with $\frac{\partial^2}{\partial \beta^2} \ln Z_1$ 
[see also below Eq.~(\ref{a8})].
Our next observation is that $\ln Z_1$ can be readily 
evaluated by means of the eigenbasis $|\vec b\rangle$ of $H$
as specified in and around Eqs.~(\ref{30})-(\ref{32})
and recalling that $H_1=H$ according to Eq. (\ref{a6}).
The detailed calculations are straightforward and 
very similar as, for instance,  those in Appendix B.1 of Ref.~\cite{rei25}.
Along these lines, one finally arrives at the result
\begin{eqnarray}
|W(\beta_m)|
&=&
\sum_{l=1}^L  \frac{E_l^2}{4\cosh^2(\beta_m E_l/2)} 
\ .
\label{a29}
\end{eqnarray}
Upon utilizing (\ref{a26}) we thus can conclude that
\begin{eqnarray}
|W(\beta_m)|\geq \frac{1}{4\cosh^2(\beta_m h(p)/2)} 
\sum_{l=1}^L E_l^2
\ .
\label{a30}
\end{eqnarray}
Denoting by $N_L$ the number of indices $l \in\{1,...,L\}$
with the property
$|E_l|\geq\frac{1}{2}$ it follows that
\begin{eqnarray}
|W(\beta_m)|\geq 
 \frac{N_L}{16\cosh^2(\beta_m h(p)/2)} 
\ .
\label{a31}
\end{eqnarray}
Focusing on energies $E_l$ of the type discussed in 
Sec.~\ref{s421} and Appendix \ref{appH}, 
one readily sees that for large $L$ at least half of 
them satisfy the condition $|E_l|\geq\frac{1}{2}$, i.e.,
$N_L\geq L/2$.
Together with Eqs. (\ref{a21}), (\ref{a22}), (\ref{a28}), and (\ref{a31})
we thus can conclude
that
\begin{eqnarray}
|\delta\beta|
\leq
\frac{4 |g|(8+|\beta |h(p)) \cosh^2(\beta_m h(p)/2)}{L}
\ . \ \ 
\label{a32}
\end{eqnarray}
The remaining task is to show that the right-hand side approaches 
zero for $L\to\infty$.
Recalling that $g$, $\beta$, and $p$ are considered as arbitrary but fixed,
it is thus sufficient that $\beta_m$ remains finite for $L\to\infty$.
According to (\ref{a4}) and the specification of $\beta_m$ below (\ref{a21})
the latter in turn is guaranteed if the solution $\beta'$ 
of the equation (\ref{a3}) remains bounded for $L\to\infty$.
It seems reasonable to expect that the latter, very weak condition 
can safely be taken for granted in most cases of physical interest.
Indeed, an unlimited growth of $\beta'$ essentially corresponds 
to the zero temperature limit and would mean that the system 
assumes an extremely low energy after a local quantum quench
(and likewise if $\beta'\to-\infty$).
Tacitly ignoring such cases, the numerator in (\ref{a32}) 
stays bounded for $L\to\infty$, and we recover our  
final result (\ref{a5}). 
[Incidentally, once we know that $\delta\beta$ is small, we can
approximate $\beta_m$ in (\ref{a32}) by $\beta$ according to 
the specification of $\beta_m$ below Eq. (\ref{a21}).]

\section{Monotonicity}
\label{appB}

The objective is to show that for any 
$n\in\NN$,
the function
\begin{eqnarray}
g(x) & := & 
\sin(x)\,\cot(n x)
\label{b1}
\end{eqnarray}
is strictly monotonically decreasing
for all $x\in(0,\pi)$ with $n x/\pi\not\in\ZZ$.
[Note that $g(x)$ is not defined (exhibits poles) for
$n x/\pi \in\ZZ$.]
Since $\cos(x)$ is monotonically decreasing
for all $x\in(0,\pi)$, it readily follows that also the function
\begin{eqnarray}
f(x) & := & 
\cos(x) + \sin(x)\,\cot(Lx)
\label{b2}
\end{eqnarray}
is strictly monotonically decreasing
for all
$x\in(0,\pi)$ with $Lx/\pi\not\in\ZZ$,
as claimed in item (iii) above Eq.~(\ref{63}).

We therefore rewrite the derivative of $g(x)$ as
\begin{eqnarray}
g'(x) & = & - h(x)/\sin^2(n x) \ ,
\label{b3}
\\
h(x)  & := & - \cos(x) \cos(n x)\sin(n x) +n \sin(x)
\nonumber
\\
& = & n \sin(x) - \cos(x) \sin(2n x)/2
\ .
\label{b4}
\end{eqnarray}
It is thus sufficient to show that $h(x)>0$ for all $x\in(0,\pi)$.
Observing that $h(\pi-x)=h(x)$, it is furthermore sufficient to
focus on $x\in(0,\pi/2]$.
Since $|\!\cos(x) \sin(2n x)/2|< 1/2$ for all those $x$, 
it readily follows that $h(x)>0$ for all  $x\in(0,\pi/2]$ 
with the property $\sin(x)\geq 1/2n $.
We are thus left with those $x\in(0,\pi/2]$ with the property
$\sin(x) < 1/2n $.
Since $\cos(x)<1$ for all those $x$, it is sufficient to show that
\begin{eqnarray}
2n \, \sin(x) \geq  \sin(2n x)
\label{b5}
\end{eqnarray}
for all $x\in(0,\pi/2]$ with $\sin(x) < 1/2n $.

Our next observation is that $\cos(\alpha y)\geq\cos(y)$ for all
$y\in[0,\pi]$ and arbitrary $\alpha\in(0,1]$, implying
\begin{eqnarray}
\frac{\sin(\alpha y)}{\alpha} = \int_0^y \! dx \cos(\alpha x)\geq 
 \int_0^y \! dy  \cos( x) = \sin(y) .
 \ \ \ \ \ \
 \label{b6}
 \end{eqnarray}
 Choosing $\alpha=1/2n $ and $y=x/\alpha=2nx$  it follows that
 Eq.~(\ref{b5}) is fulfilled
 for all $x\in[0,\pi/2n ]$. 
 Observing that the solution $x\in(0,\pi/2]$ of $\sin(x)=1/2n $
 satisfies $x\leq \pi/2n $ for any 
$n\in\NN$
 we can conclude that, indeed,
 Eq.~(\ref{b5}) is fulfilled for all $x\in(0,\pi/2]$
with $\sin(x) < 1/2n $.

\section{Derivation of Eq.~(\ref{83})}
\label{appC}

Focusing first on $p=1$, our goal is to find $L$ solutions $r_k$ of 
Eq.~(\ref{78}) under the extra constraint $r_k\in(-\pi,0)$,
 see Eqs.~(\ref{60}), (\ref{64}), and below (\ref{74}).
Introducing 
\begin{eqnarray}
x_k & := & -r_k \ ,
\label{c1}
\\
y_k & := & \frac{\pi k}{L+1}+\frac{L}{L+1}x_k \ ,
\label{c2}
\end{eqnarray}
it follows that $x_k\in(0,\pi)$ and $y_k\in(0,2\pi)$
for all $k\in\{1,...,L\}$, while Eq.~(\ref{78}) together with Eq.~(\ref{63}) can be
rewritten as
\begin{eqnarray}
\sin(x_k)=\sin(y_k)
\ .
\label{c3}
\end{eqnarray}
One readily sees that this equation admits at most two possible
solutions under the extra constraints $x_k\in(0,\pi)$ 
and $y_k\in(0,2\pi)$, namely
(i) $x_k=y_k$ and (ii) $x_k-\pi/2=\pi/2-y_k$.
In the case (i) one can conclude with Eq.~(\ref{c2}) that $x_k=k\pi$,
thus violating one of the constraints.
Likewise, in the case (ii) one recovers with Eqs.~(\ref{c1}) and (\ref{c2}) 
the result from Eq.~(\ref{83}).

Turning to the case $p=-1$, one recovers again Eq.~(\ref{c3})
if one defines
\begin{eqnarray}
x_k & := & r_k,
\label{c4}
\\
y_k & := & \frac{\pi k}{L+1}-\frac{L}{L+1}x_k,
\label{c5}
\end{eqnarray}
satisfying $x_k \in (0,\pi)$ and 
$y_k \in(-\pi,\pi)$, and implying again the two possible 
solutions (i) and (ii). 
But now, (i) implies Eq.~(\ref{83}), while (ii)
violates the constraints.

\section{Derivation of Eq.~(\ref{93})}
\label{appD}

Our objective is to show that Eq.~(\ref{93})  amounts to an
approximation whose relative error decreases
as $1/L$ for $L\to\infty$.
In particular, we always take for granted that $L$ is large.
More precisely speaking (see around Eq.~(\ref{93})), 
the quantity $\kappa_k $ in Eq.~(\ref{93})
may either be defined via Eq.~(\ref{68}) 
with $k\in\{1,...,L-1\}$ 
(corresponding to the case $|p|>p_c$)
or via Eq.~(\ref{77}) with  $k\in\{1,...,L\}$
(corresponding to the case $|p|<p_c$).

We first consider the case 
$|p|>p_c$.
Since $\sin(0)=0$ we can replace the lower summation limit
$l=1$ in Eq.~(\ref{68}) by $l=0$.
Observing
$\sin^2(x)=[1-\cos(2x)]/2$, we thus can infer from Eq.~(\ref{68}) that
\begin{eqnarray}
\kappa _k^2
& = & [L+1 - R]/2
\ ,
\label{d1}
\\
R & := & \sum_{l=0}^L \cos(\psi l)
\ ,
\label{d2}
\\
\psi & := & 2\varphi_k\ ,
\label{d3}
\end{eqnarray}
where $\varphi_k$ is defined in Eq.~(\ref{64}).
Since $\varphi_k\in(0,\pi)$ for all $k$, see Eq.~(\ref{61}),
it follows that $e^{i\psi}\not=1$ and thus
\begin{eqnarray}
R & = & \Re(Q)
\ ,
\label{d4}
\\
Q & := & \sum_{l=0}^L e^{i\psi l} =\frac{1-e^{i\chi}}{1-e^{i\psi}}
\ ,
\label{d5}
\end{eqnarray}
where $\chi := (L+1)\psi$.
Together with Eqs.~(\ref{d3}) and (\ref{64}) it follows that
$\chi= \psi+2(\pi k  + r_k)$.
As far as the quantity $e^{i\chi}$ in Eq.~(\ref{d5})
is concerned, we therefore
can equally well employ the definition
\begin{eqnarray}
\chi & := & \psi + 2r_k 
\ .
\label{d6}
\end{eqnarray}

Let us first focus on $k$-values with $k\ll L$, 
say $k/L\leq\epsilon$ with some small $\epsilon>0$.
It follows with Eqs.~(\ref{64}) and (\ref{71}) that 
$\psi$ in Eq.~(\ref{d3}) as well as $\chi$ in Eq.~(\ref{d6})
are small quantities (of order $\epsilon$ or smaller).
Hence $Q$ in Eq.~(\ref{d5}) can be
approximated as
\begin{eqnarray}
Q\simeq \frac{\chi}{\psi}
=
1 + \frac{r_k}{\varphi_k } 
=
1 + \frac{L r_k}{\pi k + r_k}
\ ,
\label{d7}
\end{eqnarray}
where we used  Eqs.~(\ref{d3}), (\ref{d6}) in the second step and
Eq.~(\ref{64}) in the last step.
Observing Eq.~(\ref{71}) it follows that 
the last term in Eq.~ (\ref{d7}) can be 
approximated by $1/(p-p_c+1/L)$
and thus by $1/(p-p_c)$ for 
sufficiently large $L$.
Together with (\ref{d4}) and (\ref{d7})
this yields $R\simeq 1+1/(p-p_c)$.
Rewriting Eq.~(\ref{d1}) as
\begin{eqnarray}
\kappa_k^2
& = &
\frac{L}{2}(1+ a) \ ,
\label{d8}
\\
a & := & \frac{1-R}{L} \ ,
\label{d9}
\end{eqnarray}
we thus recover the approximation (\ref{93}) 
with a relative error $a$ which decreases 
as $1/L$ for large $L$.
Analogous considerations apply for all $k$ 
with $\bar k/L\leq\epsilon$, where $\bar k:= L-k$.
Next we turn to the remaining $k$-values
with the properties $k/L> \epsilon$ and 
$\bar k/L>\epsilon$.
Taking into account Eqs.~(\ref{64}) and (\ref{65}) it follows that
$\psi$ from (\ref{d3}) must be larger than 
$\tilde \epsilon:=\pi\epsilon-\pi/L$
and smaller than $2\pi-\tilde\epsilon$.
For sufficiently large $L$ we thus can conclude
that $\psi\geq \epsilon$ and $\psi<2\pi-\epsilon$.
Since $\epsilon$ is small  
it follows that $|1-e^{i\psi}|$ can be lower 
bounded by $\epsilon$ apart from 
corrections of order $\epsilon^2$.
Hence, $|Q|$ in Eq.~(\ref{d5}) and
$|R|$ in Eq.~(\ref{d4}) can be upper 
bounded by $2/\epsilon$.
Together with Eqs.~(\ref{d8}) and (\ref{d9})
this yield once again the approximation (\ref{93}) 
with a relative error $a\propto 1/L$.

Finally, we turn to the case 
$|p|<p_c$
(see also the first paragraph).
One readily sees that essentially the same line of reasoning as
before can be employed, except that one now has to work with 
Eqs.~(\ref{74}), (\ref{80}), and (\ref{81})
instead of Eqs.~(\ref{64}), (\ref{71}), and (\ref{72}),
respectively.
Moreover, in the special case $p=0$ one finds
the {\em exact} result $\kappa_k=\sqrt{(L+1)/2}$ 
(for arbitrary $L$) as an improvement of
the large-$L$ approximation (\ref{93}).

\section{Strong violation of the ETH for $|p|>p_c$}
\label{appE}

In this Appendix we provide a direct proof of the 
strong violation of the ETH
for $|p|>p_c$,
as announced {at the end of} Sec.~\ref{s442}.

To begin with we recall \cite{dal16} that
the ETH essentially means that the diagonal matrix 
elements $\langle n|A|n\rangle$ 
of any (local) observabel $A$
must be similar
for energy eigenstates $|n\rangle$ with similar 
eigenvalues ${\cal E}_n$,
see also in and around Eqs.~(\ref{21}), (\ref{22}).
More precisely speaking \cite{dal16}, it is sufficient to focus
on large system sizes,
and the variations of $\langle n|A|n\rangle$
must be 
negligible
as long as the differences of the energies
${\cal E}_n$ are 
small compared to the typically extensive 
energy scale of the system itself.
The 
conventional (or strong) ETH requires that {\em all} eigenstates $|n\rangle$ 
(with energies ${\cal E}_n$ not too close to the 
edges of the spectrum) exhibit this property \cite{dal16}.
On the other hand a few exceptional eigenstates
are still admitted in the case of the 
weak ETH (wETH) \cite{weth}, but
their fraction is required to approach zero
with increasing system size.

As already noted in brackets below Eq.~(\ref{32}),
instead of the 
eigenvectors $|n\rangle$ and
eigenvalues ${\cal E}_n$ of $H$
as introduced above Eq.~(\ref{21}), 
we can equally well
work with the eigenvectors $|\vec \xx\rangle$ 
and eigenstates  $E(\vec \xx)$ from 
Eqs.~(\ref{31}) and (\ref{32}), respectively.
Taking into account Eqs.~(\ref{30}), (\ref{33}), (\ref{34})
and the fermionic anticommutation relations for 
$f_k^\dagger$ and $f_k$, a straightforward calculation 
then yields for the diagonal matrix elements 
of the observable $A=s_\alpha^z$ [see also Eq.~(\ref{11})]
the result
\begin{eqnarray}
\langle \vec\xx |s_{\alpha}^z|\vec \xx \rangle=\bigg(\sum_{k=1}^L b_k\, |U_{k \alpha } |^2\bigg)  -1/2
\ ,
\label{e1}
\end{eqnarray}
where $\xx_k$ are the binary components of $\vec \xx$ [see above Eq.~(\ref{30})].

Let us choose an arbitrary but fixed eigenstate $|\vec\xx \rangle$
and define a second eigenstate $|\vec\xx' \rangle$ 
with $\xx'_k=\xx_k$ for all $k\in\{1,...,L-1\}$
and $\xx'_{L}\not=\xx_L$ 
(i.e., $\xx'_L=1$ if $\xx_L=0$ and vice versa).
Hence, the difference of the two energies is, 
according to Eq.~(\ref{32}),
\begin{eqnarray}
|E(\vec\xx')-E(\vec\xx)|=|E_L| 
\ .
\label{e2}
\end{eqnarray}
Focusing on the case $|p|>p_c$ and exploiting the large-$L$ 
approximation (\ref{88}) it follows that
\begin{eqnarray}
|E(\vec\xx')-E(\vec\xx)|=|p+1/p|/2
\ .
\label{e3}
\end{eqnarray}
Likewise, the difference of the corresponding diagonal matrix elements
is, according to Eq.~(\ref{e1}),
\begin{eqnarray}
|\langle \vec\xx' |s_{\alpha }^z|\vec \xx' \rangle-\langle \vec\xx |s_{\alpha }^z|\vec \xx \rangle|
= |U_{L{\alpha }}|^2 = (1-1/p^2) (1/p)^{2 \bar\alpha}
,\ \ \ \ \ \ 
\label{e4}
\end{eqnarray}
where we exploited 
Eqs.~(\ref{89}) and (\ref{90}) in the last step.
Hence, the difference in Eq.~(\ref{e4}) is non-negligible 
compared to the two possible measurement outcomes $\pm1/2$
of the observable $A=s_{\alpha }^z$ \cite{foot1}, for instance, if $\alpha =L$ and 
$|p|$ is not too close to $p_c\simeq 1$.
Likewise, the energy difference in Eq.~(\ref{e3}) might appear
not to be really small.
However, as said in the {second paragraph of 
the present Appendix},
it is not
this energy difference itself which counts, but rather the
difference is required to grow not too fast (usually sublinearly)
with the system size $L$, which is clearly the case.

In summary, for every given energy eigenstate, there exists
a ``partner state'' which is energetically close while the
diagonal matrix elements are notably different at least
for one local observable $A=s_L^z$.
It follows that both the ETH and the wETH are violated,
i.e., we recover the announced 
strong violation of the ETH
for $|p|>p_c$.

\section{Numerical evaluation of the exact quench dynamics}
\label{appF}

This Appendix provides the details of our optimized
numerical treatment of long XX-chains,
as announced in Sec.~\ref{s45}.

As a slightly more general starting point we consider cases where
$H$ in Eq.~(\ref{5}) as well as $H_0$ in Eq.~(\ref{2})
can be written 
in the form
\begin{eqnarray}
H&=&\sum_{k,l=1}^{L} B_{kl}c_k^\dagger c_l,
\label{f1}
\\
H_0&=&\sum_{k,l=1}^{L} \tilde{B}_{kl}c_k^\dagger c_l,
\label{f2}
\end{eqnarray}
with fermionic creation and annihilation operators $c_l^\dagger$ and $c_l$, respectively,
and arbitrary coefficients $B_{kl}$ and $\tilde{B}_{kl}$ apart from the hermiticity
conditions $B_{lk}=B^\ast_{kl}$ and $\tilde B_{lk}=\tilde B^\ast_{kl}$.
In particular, XX models with Hamiltonians $H$ as in Eq.~(\ref{24}) can be 
recast by means of a Jordan-Wigner transformation into the form (\ref{26}) 
and thus (\ref{f1}), and likewise for $H_0$.
Similarly as in Eqs.~(\ref{27}) and (\ref{28}), there furthermore must exist unitary $L\times L$ 
matrices $U$ and $\tilde{U}$ with entries $U_{kl}$ and $\tilde{U}_{kl}$, respectively, 
so that
\begin{eqnarray}
H&=&\sum_{k=1}^{L} E_kf_k^\dagger f_k,
\label{f3}
\\
H_0&=&\sum_{k=1}^{L} \tilde{E}_k\tilde{f}_k^\dagger \tilde{f}_k,\label{f4}
\end{eqnarray}
where
\begin{eqnarray}
f_k& :=&\sum_{l=1}^L U_{kl}c_l,\label{f5}\\
\tilde{f}_k& :=&\sum_{l=1}^L\tilde{U}_{kl}c_l,
\label{f6}
\end{eqnarray}
and their adjoints  $f_k^\dagger$ and $\tilde{f}_k^\dagger$
are again fermionic annihilation and creation operators.
In other words, $U$ is the unitary matrix diagonalizing the matrix $B$ with entries $B_{kl}$, 
while $\tilde{U}$ diagonalizes $\tilde{B}$ with entries $\tilde{B}_{kl}$.
Inverting Eq.~\eqref{f6} and introducing the resulting expression into Eq.~\eqref{f5} yields
\begin{eqnarray}
f_k&=&\sum_{l=1}^LS_{kl}\tilde{f}_l\label{f7},\\
S_{kl}:&=&\sum_{m=1}^L U_{km}\tilde{U}_{lm}^*
\label{f8}.
\end{eqnarray}
Adopting the same definition as in Eq.~\eqref{13},
\begin{eqnarray}
A(t):=e^{iHt}Ae^{-iHt}
\ ,
\label{f9}
\end{eqnarray}
but now for arbitrary (not necessarily Hermitian) operators $A$, 
and utilizing Eq.~(B14) from \cite{rei25}, which 
takes the form
\begin{eqnarray}
f_k(t)=e^{-iE_kt}f_k
\label{f10}
\end{eqnarray}
in our present notation, we can conclude that
\begin{eqnarray}
c_l(t)&=&\sum_{k=1}^LU_{kl}^*f_k(t)
\label{f11}
\\
&=&\sum_{k=1}^LU_{kl}^*e^{-iE_kt}\sum_{m=1}^LS_{km}\tilde{f}_m.
\label{f12}
\end{eqnarray}
Exploiting Eq.~\eqref{f9} it follows that $[c_l^\dagger c_l](t)=[c_l(t)]^\dagger c_l(t)$, 
and together with Eq.~\eqref{12} that
\begin{eqnarray}
\langle c_l^\dagger c_l \rangle_t
&=&
\sum_{j,k=1}^LU_{jl}U_{kl}^*e^{i(E_{j}-E_{k})t}\, T_{jk},
\label{f13}
\\
T_{jk}
&:=&
\sum_{m,n=1}^L S_{jm}^*S_{kn}\, p_{mn},
\label{f14}
\\
p_{mn}
&:=&
\tr\{\rho(0)\tilde{f}_{m}^\dagger\tilde{f}_{n}\}
\ .
\label{f15}
\end{eqnarray}
To further evaluate the right-hand side of Eq.~(\ref{f15}),
we focus on our specific initial conditions from Eq.~(\ref{3}).
By exploiting (\ref{f4}) one thus finds after
a straightforward calculation (see e.g. Eqs. (B21)-(B29)
in Ref. \cite{rei25} for the details)
\begin{eqnarray}
p_{mn}=\frac{1}{1+e^{\beta E_m}} \, \delta_{mn}
\ .
\label{f16}
\end{eqnarray}
Hence, Eq.~(\ref{f14}) simplifies to
\begin{eqnarray}
T_{jk}=\sum_{n=1}^L\frac{S^*_{jn}S_{kn}}{1+e^{\beta E_n}}
\ ,
\label{f17}
\end{eqnarray}
and with Eqs.~(\ref{33}) we arrive at our final result
\begin{eqnarray}
\langle s^z_l \rangle_t&=&\langle c_l^\dagger c_l \rangle_t -1/2
\ ,
\label{f18}
\end{eqnarray}
which can be evaluated by employing Eqs.~\eqref{f8}, \eqref{f13} and \eqref{f17}.

Overall, the numerical effort to evaluate the expectation value
(\ref{f18}) along these lines only grows algebraically with the
system size $L$, while it grows exponentially when employing standard
methods based on a direct diagonalization of the Hamiltonian $H$
in Eq.~(\ref{24}), see also Sec.~\ref{s45}.

\section{Evaluation of $\langle s_\alpha^z\rangle_{\!\rm th}$}
\label{appG}

The thermal expectation value in Eq.~(\ref{9}) for our 
standard observables $A=s_\alpha^z$ [see Eq.~(\ref{11})]
and Hamiltonians $H$ [see Eq.~(\ref{24})]
can be analytically evaluated without 
any approximation, 
as shown in Ref.~\cite{rei25} [see Eq. (B41) therein],
yielding
\begin{eqnarray}
\langle s_\alpha^z \rangle_{\rm th}
= - \frac{1}{2}\sum_{k=1}^L |U_{k\alpha}|^2 \, \tanh(\beta E_k/2)
\ .
\label{g1}
\end{eqnarray}
We recall that Eq.~(\ref{g1}) must be an odd of function $p$
[see below Eq.~(\ref{17})].
Furthermore, Eq.~(\ref{g1}) is obviously an odd 
function of $\beta$.

In order to obtain more detailed quantitative results, we
henceforth focus -- as in Sec.~\ref{s4} of the main paper -- 
on impurities at the right chain-end
[i.e., with $\nu=L$, see in and around Eq.~(\ref{42})].
Hence,
we can exploit our results for $U_{kl}$ and $E_k$ from 
Sec.~\ref{s42}.

Let us first address the case $|p|<p_c$.
Exploiting Eqs.~(\ref{64}), (\ref{66}), and (\ref{67})
we obtain
\begin{eqnarray}
U_{k\alpha}
& = & \kappa^{-1} \sin(\varphi_k\alpha)
\ ,
\label{g2}
\\
E_k
& = &
\cos(\varphi_k)
\ .
\label{g3}
\end{eqnarray}
Since the $U_{k\alpha}$ are the elements of a unitary
matrix [see below Eq.~(\ref{26})] we can conclude with Eq.~(\ref{g2}) that
\begin{eqnarray}
\sum_{k=1}^L \kappa_k^{-2} \sin(\varphi_k m)  \sin(\varphi_k n)
= \sum_{k=1}^L U^\ast _{km}U_{kn} =
\delta_{mn}
\ \ \
\label{g4}
\end{eqnarray}
for all $m,n\in\{1,...,L\}$.

To make further progress we 
adopt the same assumption as in Eq.~(\ref{103})
and approximate
the $\tanh$ function in Eq.~(\ref{g1}) by its leading order approximation in $\beta$.
Together with Eqs.~(\ref{g2}) and (\ref{g3}) we thus obtain
\begin{eqnarray}
\langle s_\alpha^z \rangle_{\rm th}
= - \frac{\beta}{4}\sum_{k=1}^L  \kappa_k^{-2}  \sin^2(\varphi_k\alpha) \cos(\varphi_k)
\ .
\label{g5}
\end{eqnarray}
Observing that $\sin(x)\cos(y)=[\sin(x+y)+\sin(x-y)]/2$ it follows 
that
\begin{eqnarray}
\langle s_\alpha^z \rangle_{\rm th}
& = &
- \frac{\beta}{8}\, (S_\alpha^+ + S_\alpha^-)
\label{g6}
\ ,
\\
S_\alpha^\pm
& := &
\sum_{k=1}^L    \kappa_k^{-2}  \sin(\varphi_k\alpha)\sin(\varphi_k(\alpha\pm 1)) 
\ .
\label{g7}
\end{eqnarray}
With Eq.~(\ref{g2}) we can conclude that $S_\alpha^-=0$
for all $\alpha\in\{2,...,L\}$. 
For $\alpha=1$ the same follows immediately from Eq.~(\ref{g7}).
Likewise, one sees that $S_\alpha^+=0$
for all $\alpha\in\{1,...,L-1\}$, yielding
\begin{eqnarray}
\langle s_\alpha^z \rangle_{\rm th}
& = &
- \delta_{\alpha L}\frac{\beta}{8} \, S
\ ,
\label{g8}
\\
S
& := &
\sum_{k=1}^L   \kappa_k^{-2}  \sin(\varphi_k L)\sin(\varphi_k(L+1)) 
\ .
\label{g9}
\end{eqnarray}
Finally, we recall that all the $\varphi_k$ are solutions of 
Eqs.~(\ref{61}) and (\ref{62}), i.e.,
$\sin(\varphi_k(L+1)) = p \sin(\varphi_kL)$, implying
\begin{eqnarray}
S
& := &
p \sum_{k=1}^L  \kappa_k^{-2} \sin^2(\varphi_k L) 
= p
\ ,
\label{g10}
\end{eqnarray}
where we exploited Eq.~(\ref{g4}) in the last step. 
With Eq.~(\ref{g8}) we thus obtain in leading order $\beta$ the result
\begin{eqnarray}
\langle s_\alpha^z \rangle_{\rm th}
& = & 
- \delta_{\alpha L}\frac{\beta\, p}{8}
\ .
\label{g11}
\end{eqnarray}

Most importantly, Eq.~(\ref{g11}) tells us that,
in leading order $\beta$, only the spin 
at the right chain-end $\alpha=L$ is affected by an impurity at
the right chain-end $\nu=L$. 
Furthermore, Eq.~(\ref{g11}) is in agreement with the 
prediction below Eq.~(\ref{g1})
that $\langle s_\alpha^z \rangle_{\rm th}$ must be 
an odd function of $p$ and of $\beta$.
Moreover, it follows that the neglected corrections on the 
right-hand side of Eq.~(\ref{g11}) must be (at least) of 
order $\beta^3$.
On the other hand, we did {\em not} assume that $p$ is 
a small quantity, i.e., the linearity of Eq.~(\ref{g11}) in $p$
applies for all $|p|<p_c$ 
(at least in leading order $\beta$.)

From the structure of Eqs.~(\ref{g1}) and (\ref{g5})-(\ref{g10}) it is quite 
obvious that keeping higher and higher powers of 
$\beta$ in the expansion of the $\tanh$-function in Eq.~(\ref{g1})
has two main effects.
(i) On the right-hand side of Eq.~(\ref{g11}) there will appear
additional summands with  higher and higher (odd) 
powers of $\beta$, each multiplied with an 
(odd) polynomial of $p$ of the same order as $\beta$.
(ii)~Analogously to the Kronecker $\delta$ in Eq.~(\ref{g11}), 
the coefficients of these polynomials will depend on $\alpha$
such that more and more spins near the right chain-end 
assume nonvanishing values
(which however are expected to still decrease 
with increasing distance from the chain-end). 
Working out the details is straightforward but
not further pursued here.

Next we turn to the case $|p|>p_c$. 
The salient point consist in the following observation: 
If we do {\em not} replace $\varphi$ by $i\varphi$,
as it is done in case B below Eq.~(\ref{56}), then those $\varphi$'s will
become purely imaginary quantities, but for the rest they still satisfy
exactly the same equations as in case A, most importantly
Eqs.~(\ref{48}), (\ref{49}), and (\ref{51}).
As a consequence, also Eqs.~(\ref{g2})-(\ref{g11}) can be taken over,
except that the first factor $\sin(\varphi_k m)$ in Eq.~(\ref{g4}) has to be 
replaced by its complex conjugate, 
and analogously in Eqs.~(\ref{g5}), (\ref{g7}), (\ref{g9}), and (\ref{g10}).

Altogether, the main result (\ref{g11}) is thus found to be 
valid both for $|p|<p_c$ and $|p|>p_c$.
The validity in the case $|p|=p_c$ follows by continuity.
In conclusion, Eq.~(\ref{g11}) as well as the discussion below
(\ref{g11}) is valid for arbitrary $p$. 
Since the left-hand side of Eq.~(\ref{g11}) is bounded in modulus
by $1/2$ it follows that the neglected higher order corrections 
on the right-hand side will become more and more important 
with increasing values of $p$.
In particular, the same (odd) power series in $\beta$
mentioned below Eq.~(\ref{g11}) is expected to apply (and converge)
for arbitrary $p$.
Hence the qualitative properties of the thermal expectation values 
on the left-hand side of Eq.~(\ref{g11}) do {\em not} exhibit any kind of 
``transition'' when $|p|$ crosses the critical 
value $p_c$.

Incidentally, the result (\ref{g11}) and its generalization
to arbitrary $\nu$
can be obtained even more easily by exploiting Eq.~(\ref{15}) 
for $t=0$ and the small-$\beta$ asymptotics 
of Eqs.~(\ref{35})-(\ref{37}), yielding
\begin{eqnarray}
 \langle A\rangle_0 - \langle A\rangle _{\!\rm th}
 & = &  
\frac{g \beta}{4} \left|\sum_{k=1}^L U^\ast_{k\nu} U_{k\alpha}\right|^2
= \frac{g \beta}{4}  \delta_{\alpha \nu}
\ .
\label{g12}
\end{eqnarray}
Apart from the small-$\beta$ approximation 
[see also Eq.~(\ref{103})],
this result becomes
asymptotically exact for small $g$ according to the discussion
below Eq.~(\ref{16}). 
In view of Eqs.~(\ref{3}) and (\ref{4}), the first term $\langle A\rangle_0$ 
in Eq.~(\ref{g12}) is the thermal expectation value for the prequench 
Hamiltonian (\ref{2}) with an arbitrary but fixed value of $\gamma$.
Observing Eqs.~(\ref{5}), (\ref{9}), and (\ref{10}), 
the left-hand side of Eq.~(\ref{g12}) can be identified with the
(negative) change of the thermal expectation value
when increasing $\gamma$ in Eq.~(\ref{2}) by $g$.
According to Eqs.~(\ref{17}) and (\ref{24}) we thus can increase $p$
by means of many small steps in $g$, and add up the 
corresponding small (negative) changes of the thermal expectation 
values on the left-hand side of Eq.~(\ref{g12}). 
Moreover, from the symmetry 
properties
below Eq.~(\ref{g1}) we can conclude that the thermal
expectation value vanishes for $\gamma=0$.
Starting with $\gamma=0$ and letting the step size go 
to zero one thus 
obtains
\begin{eqnarray}
\langle s_\alpha^z \rangle_{\rm th}
& = & 
- \delta_{\alpha \nu}\frac{\beta\, p}{8}
\ .
\label{g13}
\end{eqnarray}
[The extra factor of $1/2$ in (\ref{g13})
compared to Eq.~(\ref{g12})
has its origin in the factor $2$ on the right-hand side of Eq.~(\ref{17}).]
While this way of deriving the leading-order-in-$\beta$ 
behavior is even easier than in our previous calculations,
the generalization to higher orders in $\beta$ turns out to be
more involved.

\section{Analytical solution for a central impurity}
\label{appH}

For the central impurities from Eq.~(\ref{115}) we can rewrite Eq.~(\ref{40}) as
\begin{eqnarray}
a_{l+1}+a_{l-1} & = & 2 E a_l \ \, \mbox{for $l \in\{2,...,L-1\}\! \setminus \! \{\nu\}$,}
\ \ \ 
\label{h1}
\\
a_{\nu+1}+a_{\nu-1} & = & (2 E - p) a_\nu 
\ ,
\label{h2}
\\
a_2 & = & 2 E a_1 
\ ,
\label{h3}
\\
a_{L-1} & = & 2E a_L 
\ .
\label{h4}
\end{eqnarray}
Similarly as in Eqs.~(\ref{46})-(\ref{49}) this suggest the ansatz
\begin{eqnarray}
a_l & = & \alpha_l \ \mbox{for $\l\in\{1,...,\nu-1\}$,}
\label{h5}
\\
a_l & = & \beta_l \ \mbox{for $\l\in\{\nu+1,...,L\}$,}
\label{h6}
\\
\alpha_l & := & \sin(\varphi l) \ \mbox{for all $\l\in\ZZ$,}
\label{h7}
\\
\beta_l & := & B\, \sin(\varphi (L+1 - l)) \ \mbox{for all $\l\in\ZZ$,}
\label{h8}
\end{eqnarray}
involving three, generally complex valued parameters $\varphi, B$, and $a_\nu$.
Observing that
\begin{eqnarray}
\alpha_{l+1}+\alpha_{l-1} & = & 2 \cos(\varphi)\, \alpha_l \ \, \mbox{for all $l\in\ZZ $,}
\label{h9}
\\
\beta_{l+1}+\beta_{l-1} & = &2 \cos(\varphi)\, \beta_l \ \, \mbox{for all $l\in\ZZ $,}
\label{h10}
\end{eqnarray}
it follows that Eqs.~(\ref{h5}) and (\ref{h6}) indeed solve Eq.~(\ref{h1}) for all $l=2,...,\nu-2$
and  $l=\nu+2,...,L-1$, respectively, if and only if 
\begin{eqnarray}
E = \cos(\varphi)
\ ,
\label{h11}
\end{eqnarray}
independently of the actual choice of the parameters 
$\varphi, B\in \CC$. 
Likewise, Eqs.~(\ref{h3}) and (\ref{h4}) are seen to be fulfilled.

The parameter $a_\nu$ [see below Eq.~(\ref{h8})] can be
fixed by choosing $l=\nu-1$ in Eq.~(\ref{h1}), implying
\begin{eqnarray}
a_\nu\!=\!-a_{\nu-2} + 2Ea_{\nu-1}\!=\!-\alpha_{\nu-2}+2E\alpha_{\nu-1}\!=\!\alpha_{\nu}
,
\ \ \ \  \ \ \ 
\label{h12}
\end{eqnarray}
where we exploited Eq.~(\ref{h5}) in the second step and Eqs.~(\ref{h9}) and (\ref{h11}) 
in the last step.
Analogously, one verifies by choosing $l=\nu+1$ in Eq.~(\ref{h1})
that $a_{\nu}=\beta_{\nu}$.
Together with Eqs.~(\ref{h5})-(\ref{h8}) we thus obtain
\begin{eqnarray}
a_l & = & \alpha_l =\sin(\varphi l)\ \, \mbox{for $l\in\{1,...,\nu\}$,}
\label{h13}
\\
a_l & = & \beta_l =B \sin(\varphi (L+1 - l))\ \, \mbox{for $l\in\{\nu,...,L\}$.}
\ \ \ \
\label{h14}
\end{eqnarray}
Choosing $l=\nu$ in Eqs.~(\ref{h13}) and (\ref{h14}) yields the first relation which
the two remaining parameters $\varphi$ and $B$ have to fulfill, namely
\begin{eqnarray}
\sin(\varphi \nu) = B \sin(\varphi (L+1  - \nu)) = B \sin(\varphi \nu) 
\ ,
\label{h15}
\end{eqnarray}
where we exploited Eq.~(\ref{115}) in the last step.
The second relation can be deduced from 
Eq.~(\ref{h2}) by rewriting it with Eqs.~(\ref{h13}) and (\ref{h14}) in the form
$\beta_{\nu+1}=-\alpha_{\nu-1} + (2E\alpha_m-p)\alpha_\nu$.
Exploiting Eqs.~(\ref{h9}) and (\ref{h11}), the right-hand side can be identified
with $\alpha_{\nu+1}-p\alpha_\nu$. Observing Eqs.~(\ref{h7}) and (\ref{h8}) 
we thus obtain
\begin{eqnarray}
B\sin(\varphi(L- \nu)) = \sin(\varphi(\nu+1))-p\sin(\varphi \nu)
\, . \ \ \ \ \ \ \
\label{h16}
\end{eqnarray}

Our next goal is to show $B$ must be non-zero by means of a
proof by contradiction.
Let us therefore assume that $B=0$.
With Eq.~(\ref{h15}) it follows that $\sin(\varphi \nu)=0$ and then with
(\ref{h16}) that $\sin(\varphi(\nu+1))=0$.
A straightforward calculation shows that 
\begin{eqnarray}
\mbox{$\sin(z)=0$ with $z\in\CC$ 
implies $z=\pi n$ with $n \in\ZZ$.}
\ \ \ \ \ \ 
\label{h17}
\end{eqnarray}
We thus can conclude 
that $\varphi(\nu+1)=\pi n$ and $\varphi n=\pi n'$
with $n,n'\in\ZZ$, implying $\varphi=\pi m$ with $m\in\ZZ$.
With Eqs.~(\ref{h13}) and (\ref{h14}) this yields $a_l=0$ for all
$l\in\{1,...,L\}$. According to the discussion below Eq.~(\ref{39}) 
this is not an admissible solution, thus completing our proof by 
contradiction.

In a next step, we take $B\not=0$ for granted 
and separately discuss the cases 
$\sin(\varphi \nu)=0$ and $\sin(\varphi \nu)\not=0$.
Considering first the case $\sin(\varphi \nu)=0$, 
we can conclude from Eq.~(\ref{h17}) that
$\varphi \nu=\pi n$ with $n \in\ZZ$.
In particular, $\varphi$ must be purely real.
Similarly as before, $\varphi=0$ would imply $a_l=0$ for 
all $l$ and can thus be excluded. 
Moreover, $\varphi$ and $-\varphi$ lead to equivalent solutions 
in Eqs.~(\ref{h11}), (\ref{h13}), and (\ref{h14})
[see below Eq.~(\ref{39})].
Likewise $\varphi$ and $\varphi+2\pi$ are equivalent.
Hence, we can restrict ourselves to $\varphi\in(0,\pi)$.
Altogether, the admissible $\varphi$ values 
then must be of the form 
\begin{eqnarray}
\varphi=\frac{\pi k}{\nu} = \frac{2\pi k}{L+1}
\  \mbox{with $k\in\{1,...,\nu-1=\frac{L-1}{2}\}$,}
\ \ \ \ 
\label{h18}
\end{eqnarray}
where we exploited Eq.~(\ref{116}) in the last equalities.
Finally,  we can conclude from
(\ref{h16}) by rewriting $L-\nu$ 
as $L+1-(\nu+1)$ and observing Eq.~(\ref{h18}) and 
$\sin(\varphi \nu)=0$
that
\begin{eqnarray}
B\sin(-\varphi (\nu+1)) = \sin(\varphi(\nu+1))
\, . \ \ \ \ \ \ \
\label{h19}
\end{eqnarray}
Similarly as below Eq.~(\ref{h17}), or directly from Eq.~(\ref{h18}) one finds 
that $\sin(\varphi(\nu+1))\not = 0$, hence Eq.~(\ref{h19}) implies
\begin{eqnarray}
B=-1
\ .
\label{h20}
\end{eqnarray}
Altogether, we thus obtained $(L-1)/2$ solutions of the form 
(\ref{h11}), (\ref{h13}), (\ref{h14}), (\ref{h18}), (\ref{h20}).
A particularly important property of all those solutions is
that they are of odd symmetry about $l=\nu$,
meaning that $a_{\nu-j}=-a_{\nu+j}$ for all $j\in\{0,...,\nu-1\}$.
Also quite remarkable is that the impurity strength $p$ 
does not play any role. 
The reason is that $a_\nu=0$ for all those solutions,
hence the value of $p$ in Eq.~(\ref{h2}) does not matter.

Next we turn the case $\sin(\varphi \nu)\not=0$,
implying with Eq.~(\ref{h15}) that
\begin{eqnarray}
B & = & 1
\ .
\label{h21}
\end{eqnarray}
Together with Eq.~(\ref{h16}) this yields after a short calculation the 
following equation for $\varphi$:
\begin{eqnarray}
2 \sin(\varphi ) \cot(\varphi  \nu) & = & p \ .
\label{h22}
\end{eqnarray}
In a next step one can show similarly as in Sec.~\ref{s41} that
with respect to the solutions of Eq.~(\ref{h22})
it is sufficient to focus on the following two cases.
Case A: $\varphi\in(0,\pi)$. 
Case B: $\varphi = iy+n\pi$ with $y\in\RR^+$ and $n\in\{0,1\}$.
In case A one encounters, similarly as in Sec.~\ref{s421},
a critical impurity strength of the form (\ref{116}),
and one finds that
whenever
$|p|>p_c$ there 
are
$\nu-1$ 
solutions of Eq.~(\ref{h22}) of the form
\begin{eqnarray}
\varphi = \frac{\pi (k+1/2) +r_k}{\nu} \ \mbox{with $k \in\{ 1,...,\nu\!-\!1=\frac{L-1}{2}\}$,}
\ \ \ \ \ \ 
\label{h23}
\end{eqnarray}
where the ``remainders''  $r_k$ are 
bounded according to $|r_k|<\pi/2$ 
[cf.  Eq.~(\ref{65})].
Compared to Eq.~(\ref{63}), the main new feature of
(\ref{116}) is that $p_c$ approaches zero
for large $L$.
Hence, the case $|p|<p_c$ is of minor interest
and not further pursued here.
Likewise, a more detailed discussion of $r_k$
is quite similar to that in Sec.~\ref{s421}
and therefore omitted.

Similarly as below Eq.~(\ref{h20}), the $(L-1)/2$ solutions 
from Eqs.~(\ref{h21}) and (\ref{h23}) are of even symmetry 
about $l=\nu$, meaning that $a_{\nu-j}=a_{\nu+j}$ 
for all $j\in\{0,...,\nu-1\}$.
Altogether, we thus obtained $L-1$ solutions so far,
half of them of odd and half of even symmetry.
As in Sec.~\ref{s422}, the still ``missing'' $L$ th
solution is obtained by exploring the above 
mentioned case B.
Omitting the details, one arrives along these lines at
\begin{eqnarray}
E & = & \pm\cosh(y)
\ ,
\label{h24}
\\
a_l & = & (\pm 1)^l\sinh(y l)\ \, \mbox{for $l\in\{1,...,\nu$\},}
\label{h25}
\\
a_l & = & (\pm 1)^{l} \sinh(y (L+1 - l))\ \, \mbox{for $l\in\{\nu,...,L\}$,}
\ \ \
\label{h26}
\end{eqnarray}
where the plus sign 
applies
when $p>p_c$ and the minus sign when $p<-p_c$.
Moreover, $y$ is defined as the (unique) solution of the equation
\begin{eqnarray}
2 \sinh(y) \coth(y\nu) =  |p| 
\label{h27}
\end{eqnarray}
with the property $y\in\RR^+$.
In particular, this solution (\ref{h25}), (\ref{h26}) is once again
of even symmetry about $l=\nu$ \cite{foot3}.

Under the very weak extra assumption that $|p|\gg 1/L$ it 
follows with Eq.~(\ref{115}) that the solution of Eq.~(\ref{h27}) 
must satisfy $y \nu \gg 1$. Hence, the factor $\coth(y\nu)$
in Eq.~(\ref{h27}) can be approximated by unity, yielding
\begin{eqnarray}
y = \mbox{arsinh}(|p|/2)
\ .
\label{h28}
\end{eqnarray}
Under the even stronger assumption
$1\gg |p|\gg 1/L$ 
this yield the even simpler approximation
\begin{eqnarray}
y =|p|/2
\ .
\label{h29}
\end{eqnarray}
On the other hand, for $p=p_c$ one readily verifies
[similarly as in Eq.~(\ref{91})] that 
\begin{eqnarray}
\mbox{$E=1$ and $a_l = l=a_{L+1-l}$ for $l=1,...,\nu$}
\label{h30}
\end{eqnarray}
indeed satisfies Eqs.~(\ref{h1})-(\ref{h4}) even exactly,
and analogously for $p=-p_c$ \cite{foot3}.

Finally, by employing the approximation (\ref{h29}) one recovers Eq.~(\ref{117})
by means of analogous calculations as in Sec.~\ref{s44}.




\begin{thebibliography}{0}%
\makeatletter
\providecommand \@ifxundefined [1]{%
 \@ifx{#1\undefined}
}%
\providecommand \@ifnum [1]{%
 \ifnum #1\expandafter \@firstoftwo
 \else \expandafter \@secondoftwo
 \fi
}%
\providecommand \@ifx [1]{%
 \ifx #1\expandafter \@firstoftwo
 \else \expandafter \@secondoftwo
 \fi
}%
\providecommand \natexlab [1]{#1}%
\providecommand \enquote  [1]{``#1''}%
\providecommand \bibnamefont  [1]{#1}%
\providecommand \bibfnamefont [1]{#1}%
\providecommand \citenamefont [1]{#1}%
\providecommand \href@noop [0]{\@secondoftwo}%
\providecommand \href [0]{\begingroup \@sanitize@url \@href}%
\providecommand \@href[1]{\@@startlink{#1}\@@href}%
\providecommand \@@href[1]{\endgroup#1\@@endlink}%
\providecommand \@sanitize@url [0]{\catcode `\\12\catcode `\$12\catcode
  `\&12\catcode `\#12\catcode `\^12\catcode `\_12\catcode `\%12\relax}%
\providecommand \@@startlink[1]{}%
\providecommand \@@endlink[0]{}%
\providecommand \url  [0]{\begingroup\@sanitize@url \@url }%
\providecommand \@url [1]{\endgroup\@href {#1}{\urlprefix }}%
\providecommand \urlprefix  [0]{URL }%
\providecommand \Eprint [0]{\href }%
\providecommand \doibase [0]{https://doi.org/}%
\providecommand \selectlanguage [0]{\@gobble}%
\providecommand \bibinfo  [0]{\@secondoftwo}%
\providecommand \bibfield  [0]{\@secondoftwo}%
\providecommand \translation [1]{[#1]}%
\providecommand \BibitemOpen [0]{}%
\providecommand \bibitemStop [0]{}%
\providecommand \bibitemNoStop [0]{.\EOS\space}%
\providecommand \EOS [0]{\spacefactor3000\relax}%
\providecommand \BibitemShut  [1]{\csname bibitem#1\endcsname}%
\let\auto@bib@innerbib\@empty
\end{thebibliography}%


\begin{thebibliography}{71}

\bibitem{gog16}
C. Gogolin and J. Eisert,
Equilibration, thermalization, and the emergence
of statistical mechanics in closed quantum systems,
Rep. Prog. Phys. {\bf 79}, 056001 (2016).

\bibitem{mor18}
T. Mori, T. N. Ikeda, E. Kaminishi, and M. Ueda,
Thermalization and prethermalization 
in isolated quantum systems: a theoretical overview,
J. Phys. B  {\bf 51}, 112001 (2018).

\bibitem{tas24}
N. Shiraishi and H. Tasaki,
Nature abhors a vacuum: A simple rigorous example of 
thermalization in an isolated macroscopic quantum system,
J. Stat. Phys. {\bf 191}, 82 (2024).

\bibitem{deu91}
J. M. Deutsch,  
Quantum statistical mechanics in a closed system,
Phys. Rev. A {\bf 43}, 2046
(1991).

\bibitem{sre94}
M. Srednicki,
Chaos and quantum thermalization,
Phys. Rev. E {\bf 50}, 888
(1994).

\bibitem{rig08}
M. Rigol, V. Dunjko, and M. Olshanii,
Thermalization and its mechanism for generic
isolated quantum systems,
Nature {\bf 452}, 854
(2008).

\bibitem{dal16}
L. D'Alessio, Y. Kafri, A. Polkovnikov, and M. Rigol,
From Quantum Chaos and Eigenstate Thermalization
to Statistical Mechanics and Thermodynamics,
Adv. Phys.  {\bf 65}, 239 (2016).

\bibitem{ess16}
F. H. L. Essler and M. Fagotti,
Quench dynamics and relaxation in isolated integrable quantum spin chains,
J. Stat. Mech. {\bf 6}, 064002 (2016).

\bibitem{bar70}
E. Barouch, B. M. McCoy, and M. Dresden,
Statistical Mechanics of the XY Model. I,
Phys. Rev. A {\bf 2}, 1075 (1970);
E. Barouch and B. M. McCoy, 
Statistical Mechanics of the XY Model. III,
Phys. Rev. A {\bf 3}, 2137 (1971).

\bibitem{weth}
G. Biroli, C. Kollath, and A. M. L\"auchli, 
Effect of Rare Fluctuations on the Thermalization of Isolated Quantum Systems, 
Phys. Rev. Lett. {\bf 105}, 250401 (2010);
V. Alba, 
Eigenstate thermalization hypothesis and integrability in quantum spin chains, 
Phys. Rev. B {\bf 91}, 155123 (2015);
T. Mori,
Weak eigenstate thermalization with large deviation bound,
arXiv:1609.09776;
E. Iyoda, K. Kaneko, and T. Sagawa,
Fluctuation Theorem for Many-Body Pure Quantum States,
Phys. Rev. Lett. {\bf 119}, 100601 (2017);
T. Kuwahara and K. Saito,
Eigenstate thermalization from the clustering property of correlations,
Phys. Rev. Lett. {\bf 124}, 200604 (2020).

\bibitem{far17}
T. Farrelly, F. G. S. L. Brand\~ao,   and M. Cramer,
Thermalization and return to equilibrium on finite quantum lattice systems,
Phys. Rev. Lett. {\bf 118}, 140601 (2017).

\bibitem{dab22}
L. Dabelow, P. Vorndamme, and P. Reimann,
Thermalization of locally perturbed many-body quantum systems,
Phys. Rev. B {\bf 105}, 024310 (2022).

\bibitem{rei25}
P. Reimann and C. Eidecker-Dunkel,
Thermalization in a simple spin-chain model
Phys. Rev. B {\bf 111}, 054312 (2025).

\bibitem{bre20}
M. Brenes, T. LeBlond, J. Goold, and M. Rigol,
Eigenstate Thermalization in a Locally Perturbed Integrable System,
Phys. Rev. Lett. {\bf 125}, 070605 (2020).

\bibitem{pc}
M. Rigol, private communication regarding Ref.~\cite{bre20}.

\bibitem{san04}
L. F. Santos,
Integrability of a disordered Heisenberg spin-1/2 chain,
J. Phys. A  {\bf 37}, 4723 (2004).

\bibitem{bar09}
O. S. Barisic, P. Prelovsek, A. Metavitsiadis, and X. Zotos,
Incoherent transport induced by a single static impurity in a Heisenberg chain,
Phys. Rev. B {\bf 80}, 125118 (2009).

\bibitem{tor14}
E. J. Torres-Herrera and L. F. Santos,
Local quenches with global effects in interacting quantum systems,
Phys. Rev. E {\bf 89}, 062110 (2014).

\bibitem{pan20}
M. Pandey, P. W. Claeys, D. K. Campbell, A. Polkovnikov, and D. Sels,
Adiabatic deformations as a sensitive probe for quantum chaos,
Phys. Rev. X {\bf 10}, 041017 (2020)

\bibitem{san20}
L. F. Santos, F. P\'erez-Bernal, and E. J. Torres-Herrera,
Speck of chaos,
Phys. Rev. Research {\bf 2}, 043034 (2020).

\bibitem{foot1}
As usual, the units of the magnetic moments are chosen
such that the spin-1/2 operators $s_l^a$ can be expressed
in the eigenbasis of $s_l^z$ by the corresponding Pauli 
matrices $\sigma^a_l$ as $s_l^z=\sigma^a_l/2$.

\bibitem{tou15}
H. Touchette,
Equivalence and nonequivalence of ensembles: thermodynamic, macrostate, and measure levels,
J. Stat. Phys. {\bf 159}, 987 (2015).

\bibitem{bra15}
F. G. S. L. Brandao and M. Cramer,
Equivalence of Statistical Mechanical Ensembles for Non-Critical Quantum Systems,
arXiv:1502.03263

\bibitem{tas18}
H. Tasaki,
On the local equivalence between the canonical and the 
microcanonical ensembles for quantum spin systems,
J. Stat. Phys. {\bf 172}, 905 (2018).

\bibitem{kuw20}
T. Kuwahara and K. Saito,
Gaussian concentration bound and ensemble equivalence in generic quantum 
many-body systems including long-range interactions,
Ann. Phys. {\bf 421}, 168278 (2020).

\bibitem{rei24}
P. Reimann and C. Eidecker-Dunkel,
Onsager's regression hypothesis adjusted to quantum systems,
Phys. Rev. B {\bf 101}, 014306 (2024).

\bibitem{fn}
Our present notation is similar but not exactly identical to the 
one in Ref. \cite{rei24}. In particular, our present Eq. (\ref{5}) 
corresponds to Eq. (11) in Ref. \cite{rei24}.

\bibitem{eid23}
C. Eidecker-Dunkel and P. Reimann,
Allosteric impurity effects in long spin chains,
Phys. Rev. B {\bf 108}, 054407 (2023).

\bibitem{uhr14}
G. S. Uhrig, J. Hackmann, D. Stanek, J. Stolze, and F.~B.~Anders,
Conservation laws protect dynamic spin correlations from decay: 
Limited role of integrability in the central spin model,
Phys. Rev. B {\bf 90}, 060301(R) (2014).

\bibitem{alh20}
A. M. Alhambra, J. Riddell, and L. P. Garcia-Pintos,
Time evolution of correlation functions in quantum many-body systems,
Phys. Rev. Lett. {\bf 124}, 110605 (2020).

\bibitem{spo18}
H. Spohn,
Interacting and noninteracting integrable systems,
J. Math. Phys. {\bf 59}, 091402 (2018).

\bibitem{lie61}
E. Lieb, T. Schultz, and D. Mattis,
Two Soluble Models of an Antiferromagnetic Chain,
Ann. Phys. {\bf 16}, 407 (1961).

\bibitem{cru81}
H. B. Cruz and L. L. Goncalves,
Time-dependent correlations of the one-dimensional isotropic XY model,
J. Phys. C: Solid State Phys. {\bf 14}, 2785 (1981).

\bibitem{sto92}
J. Stolze, V. S. Viswanath, and G. M\"uller,
Dynamics of semi-infinite quantum spin chains at $T=\infty$,
Z. Phys. B {\bf 89}, 45 (1992).

\bibitem{sto95}
J. Stolze, A. N\"oppert, and G. M\"uller,
Gaussian, exponential, and power-law decay of time-dependent correlation functions in quantum spin chains,
Phys. Rev. B {\bf 52}, 4319 (1995).

\bibitem{der00}
O. Derzhko, T. Krokhmalskii, and J. Stolze,
Dynamics of the spin-1/2 isotropic XY chain in a transverse field,
J. Phys. A: Math. Gen. {\bf 33}, 3063 (2000).

\bibitem{foot3}
See also the general symmetry considerations 
around Eq.~(\ref{41}).

\bibitem{tas24b}
H. Tasaki,
Heat flows from hot to cold: A simple rigorous example 
of thermalization in an isolated macroscopic quantum system,
arXiv:2404.04533

\bibitem{mur19}
C. Murthy and M. Srednicki,
Relaxation to Gaussian and generalized Gibbs states in systems of particles with quadratic Hamiltonians,
Phys. Rev. E {\bf 100}, 012146 (2019).

\bibitem{foot4}
The latter finding is strictly speaking not yet sufficient 
to guarantee the occurrence of thermalization,
see item (ii) at the end of Sec.~\ref{s24}.

\bibitem{foot6}
Note that the mean-values of the long-time fluctuations are
again quantitatively quite similar in Figs.~\ref{fig9} and \ref{fig5},
as can be deduced from the similarity of the red crosses in 
Figs.~\ref{fig3}(b) and \ref{6}(b).

\bibitem{skl88}
E. K. Sklyanin,
Boundary conditions for integrable quantum systems,
J. Phys. A {\bf 21}, 2375 (1988).

\bibitem{alc87}
F. C. Alcaraz, M. N. Barber, M. T. Batchelor, R. J. Baxter, and G. R. W. Quispel,
Surface exponents of the quantum XXZ, Ashkin-Teller and Potts models,
J. Phys. A {\bf 20}, 6397 (1987).

\bibitem{bei13}
N. Beisert, L. Fievet, M. de Leeuw, and F. Loebbert,
Integrable deformations of the XXZ spin chain,
J. Stat. Mech. {\bf  P09028} (2013).

\bibitem{hok25}
A. Hokkyo, Rigorous test for quantum integrability and nonintegrability,
arXiv:2501.18400

\bibitem{fra16}
G. Francica, T. J. G. Appolaro, N. Lo Gullo, and F. Plastina,
Local quench, Majorana zero modes, and disturbance propagation in the Ising chain,
Phys. Rev. B {\bf 94}, 245103 (2016).

\bibitem{foot5}
More precisely speaking, for the end-impurities
from Sec.~\ref{s4}, thermalization has been
verified in Ref.~\cite{rei25} for all single-site observables
of the form (\ref{11}), while for the inner impurities 
from Sec.~\ref{s6}, only a subset of such single-site
observables was covered.
In both cases, the generalization from single-site 
to arbitrary local 
(multiple-site) observables readily follows by
exploiting the ``Gaussification''-results from 
Refs.~\cite{mur19}.

\bibitem{shr21}
N. Shiraishi and K. Matsumoto, 
Undecidability in quantum thermalization,
Nat. Commun. {\bf 12}, 5084 (2021).

\bibitem{hsf}
P. Sala, T. Rakovszky, R. Verresen, M. Knap, and F. Pollmann,
Ergodicity Breaking Arising from Hilbert Space Fragmentation in Dipole-Conserving Hamiltonians,
Phys. Rev X {\bf 10}, 011047 (2020);
T. Rakovszky, P. Sala, R. Verresen, M. Knap, and F. Pollmann,
Statistical localization: From strong fragmentation to strong edge modes
Phys. Rev B {\bf 101}, 125126 (2020);
S. Moudgalya, B. A. Bernevig, and N. Regnault,
Quantum many-body scars and Hilbert space fragmentation: a review of exact results,
Rep. Prog. Phys. {\bf 85}, 086501 (2022).


\bibitem{data}
https://doi.org/10.6084/m9.figshare.29900183

\bibitem{cdp}
M Kliesch, C. Gogolin, M. J. Kastoryano, A. Riera, and J. Eisert,
Locality of temperature,
Phys. Rev. X {\bf 4}, 031019 (2014);
H. Araki,
Gibbs states of a one dimensional quantum lattice,
Commun. Math. Phys. {\bf 14}, 120 (1969);
Y. M. Park,
The cluster expansion for classical and quantum lattice systems,
J. Stat. Phys. {\bf 27}, 553 (1982);
Y. M. Park and H. J. Yoo,
Uniqueness and clustering properties of Gibbs states 
for classical and quantum unbounded spin systems,
J. Stat. Phys. {\bf 80}, 223 (1995).






\end{thebibliography}
\end{document}